\documentclass[iop,numberedappendix]{emulateapj} \def \apjloutput {}

\usepackage{color}
\usepackage{hyperref}
\hypersetup{colorlinks,citecolor=blue}
\usepackage{graphicx}
\usepackage{amssymb}
\usepackage{amsmath}
\usepackage{rotating}
\usepackage{multirow}
\usepackage{float}

\def\HST{\textit{HST}}

\def\wise{\textit{WISE}}
\def\WISE{\textit{WISE}}
\def\W1{\textit{W1}}
\def\W2{\textit{W2}}
\def\W3{\textit{W3}}
\def\W4{\textit{W4}}
\def\GALEX{\textit{GALEX}}

\def\Herschel{\textit{Herschel}}
\def\AKARI{\textit{Akari}}
\def\Spitzer{\textit{Spitzer}}

\def\uJy{$\mu$Jy}

\def\w1248{W1248$-$2154}

\def\ehylirg{ELIRG}
\def\ehylirgs{ELIRGs}
\def\elirg{ELIRG}
\def\elirgs{ELIRGs}
\def\Lbol{$L_{\rm bol}$}
\def\hotdog{Hot DOG}
\def\hotdogs{Hot DOGs}
\def\thetaE{$\theta_{\rm E}$}

\ifx \apjloutput \undefined
\else
\slugcomment{{The Astrophysical Journal, in Press}}
\fi

\begin{document}

\title{The Most Luminous Galaxies Discovered by WISE}
\author{
Chao-Wei Tsai\altaffilmark{1,2},
Peter R. M. Eisenhardt\altaffilmark{1}, 
Jingwen Wu\altaffilmark{3},
Daniel Stern\altaffilmark{1},
Roberto J. Assef\altaffilmark{4}, \\
Andrew W. Blain\altaffilmark{5}, 
Carrie R. Bridge\altaffilmark{6}, 
Dominic J. Benford\altaffilmark{7}, 
Roc M. Cutri\altaffilmark{8}, 
Roger L. Griffith\altaffilmark{9}, 
Thomas H. Jarrett\altaffilmark{10}, 
Carol J. Lonsdale\altaffilmark{11}, 
Frank J. Masci\altaffilmark{8}, 
Leonidas A. Moustakas\altaffilmark{1}, 
Sara M. Petty\altaffilmark{12}, 
Jack Sayers\altaffilmark{6}, 
S. Adam Stanford\altaffilmark{13}, 
Edward L. Wright\altaffilmark{3}, 
Lin Yan\altaffilmark{8}, 
David T. Leisawitz\altaffilmark{7}, 
Fengchuan Liu\altaffilmark{1}, 
Amy K. Mainzer\altaffilmark{1},
Ian S. McLean\altaffilmark{3}, 
Deborah L. Padgett\altaffilmark{7}, 
Michael F. Skrutskie\altaffilmark{14}, 
Christopher R. Gelino\altaffilmark{8}, 
Charles A. Beichman\altaffilmark{8}, 
St\'{e}phanie Juneau\altaffilmark{15}
}
\altaffiltext{1}{Jet Propulsion Laboratory, California Institute of Technology, 4800 Oak Grove Dr., Pasadena, CA 91109, USA}
\altaffiltext{2}{NASA Postdoctoral Program Fellow; [email: Chao-Wei.Tsai@jpl.nasa.gov]}
\altaffiltext{3}{Department of Physics and Astronomy, UCLA, Los Angeles, CA 90095-1547}
\altaffiltext{4}{N\'ucleo de Astronom\'ia de la Facultad deIngenier\'ia, Universidad Diego Portales, Av. Ej\'ercito Libertador 441, Santiago, Chile.}
\altaffiltext{5}{Department of Physics \& Astronomy, University of Leicester, 1 University Road, Leicester, LE1 7RH, UK}
\altaffiltext{6}{Division of Physics, Math, and Astronomy, California Institute of Technology, Pasadena, CA 91125, USA}
\altaffiltext{7}{NASA Goddard Space Flight Center, Greenbelt, MD 20771, USA}
\altaffiltext{8}{Infrared Processing and Analysis Center, California Institute of Technology, Pasadena, CA 91125, USA}
\altaffiltext{9}{Department of Astronomy and Astrophysics, The Pennsylvania State University, 525 Davey Lab, University Park, PA 16802, USA}
\altaffiltext{10}{Astronomy Department, University of Cape Town, Private Bag X3, Rondebosch 7701, South Africa}
\altaffiltext{11}{National Radio Astronomy Observatory, 520 Edgemont Road, Charlottesville, VA 22903, USA}
\altaffiltext{12}{Department of Physics, Virginia Tech, Blacksburg, VA, 24061, USA}
\altaffiltext{13}{Department of Physics, University of California Davis, One Shields Avenue, Davis, CA 95616, USA}
\altaffiltext{14}{Department of Astronomy, University of Virginia, Charlottesville, VA 22903, USA}
\altaffiltext{15}{CEA-Saclay, DSM/IRFU/SAp, F-91191 Gif-sur-Yvette, France}

\begin{abstract} 
We present 20 \textit{Wide-field Infrared Survey Explorer} (\wise)-selected galaxies with bolometric luminosities $L_{\rm bol} > 10^{14}\,L_{\sun}$, including five with infrared luminosities $L_{\rm IR} \equiv L_{\rm (rest\ 8-1000\,\mu \rm m)} > 10^{14}\,L_{\sun}$. These ``extremely luminous infrared galaxies,'' or ELIRGs, were discovered using the ``$W1$$W2$-dropout'' selection criteria \citep{2012ApJ...755..173E} which requires marginal or non-detections at 3.4 and 4.6 \micron\ ($W1$ and $W2$, respectively) but strong detections at 12 and 22 \micron\ in the \wise\ survey. Their spectral energy distributions are dominated by emission at rest-frame 4--10 \micron, suggesting that hot dust with $T_{d}\sim 450\,K$ is responsible for the high luminosities. These galaxies are likely powered by highly obscured active galactic nuclei (AGNs), and there is no evidence suggesting these systems are beamed or lensed. We compare this \wise-selected sample with 116 optically selected quasars that reach the same $L_{\rm bol}$ level, corresponding to the most luminous unobscured quasars in the literature. We find that the rest-frame 5.8 and 7.8 \micron\ luminosities of the \wise-selected \elirgs\ can be 30--80\% higher than that of the unobscured quasars. The existence of AGNs with $L_{\rm bol} > 10^{14}\,L_{\sun}$ at $z > 3$ suggests that these supermassive black holes are born with large mass, or have very rapid mass assembly. For black hole seed masses $\sim 10^3\,M_{\sun}$, either sustained super-Eddington accretion is needed, or the radiative efficiency must be $<$\,15\%, implying a black hole with slow spin, possibly due to chaotic accretion.

\end{abstract}


\section{Introduction}\label{sec:intro}
Hyperluminous infrared galaxies \citep{1994ApJ...424L..65C}, or HyLIRGs, are galaxies whose infrared luminosity ($L_{\rm IR}$) exceeds $10^{13}\,L_{\sun}$ \citep{1996ARA&A..34..749S}. They have generally been discovered due to their substantial IR emission from far-IR surveys such as those with the \textit{Infrared Astronomical Satellite} (\textit{IRAS}), \citep{1984ApJ...278L...1N}, the Submillimetre Common-User Bolometer Array (SCUBA) at James Clerk Maxwell Telescope, or, more recently, the \textit{Herschel Space Telescope} (\citealt[][and references therein]{1994ApJ...424L..65C,1998ApJ...506L...7F,2000MNRAS.316..885R}; \citealt{2012ApJ...761..140C}). Infrared emission dominates the energy output of these hyperluminous systems, so their $L_{\rm IR}$ is approximately equal to their bolometric luminosity $L_{\rm bol}$. Galaxies with such high luminosity, usually powered by AGN \citep{2012ApJ...761..184W}, intense starbursts \citep{2000MNRAS.316..885R}, or a mixture of both, represent a rapid growth phase of the supermassive black holes (SMBHs) and/or the stellar mass portfolio of the host galaxies. It has been suggested that this phase at $1< z < 3$ dominates both the stellar mass assembly of massive galaxies and the mass accretion of SMBHs \citep{2006ApJS..163....1H,2008ApJS..175..356H}.

Both observations and simulations suggest that the high accretion rate phase of AGNs occurs after the major merging event of two large galaxies \citep{1988ApJ...325...74S,2008ApJS..175..356H}. At that stage, gas from the two parent galaxies loses angular momentum due to cloud--cloud collisions, quickly sinks to the center of the coalescing galaxy, and fuels the AGN which is still shrouded by a dusty cocoon. As the AGN accretes, its luminosity increases dramatically, and it becomes a quasar. At early stages of AGN accretion, the infalling dust and gas may cause severe obscuration toward the quasar, making it difficult to identify at optical wavelengths. The intense starburst induced by the gas cloud collisions is quickly followed by an optically luminous quasar phase, and eventually suppressed by feedback from the quasar \citep{1998A&A...331L...1S,2005MNRAS.361..776S}. 

Some optical quasars have comparable intrinsic $L_{\rm bol}$ to the extreme luminosity end of the HyLIRG population ($> 10^{14}\,L_{\sun}$), such as S5 0014+81 \citep{1994ApJ...436..678O}, SDSS J1701+6412 \citep{2010MNRAS.405.2302H}, and HS 1946+7658 \citep{1996ApJS..107..475L}. These quasars have SMBH masses of $\sim$ few $\times 10^{9}\,M_{\sun}$ or higher, if they are emitting at or close to the Eddington limit. At the Eddington limit, SMBH mass grows on the Salpeter $e$-folding time scale of 45 Myr \citep{1964ApJ...140..796S}, and the most massive SMBHs can reach $\sim 3 \times 10^{10}\,M_{\sun}$ at $z>2$ \citep{2010ApJ...719.1315K} in the broad-line QSO phase. The discovery of hyper-luminous quasars at  $z>6$ \citep{2001AJ....122.2833F} further suggests that SMBHs can grow to $10^{9}\,M_{\sun}$ \citep{2003ApJ...594L..95B,2005ApJ...633..630W,2011Natur.474..616M} by the time the universe is only $\sim1$ Gyr old. The existence of such luminous quasars at early times constrains SMBH seed masses and their growth history \citep{2006ApJ...650..669V}, implying a high accretion rate at high redshift, rather than slower accretion over a Hubble time \citep{2006ApJ...643..641H}.

By selection, optical quasars have relatively low extinction at visible wavelengths, suggesting that feedback to the ambient material may have cleared out the surrounding dust cocoon and terminated further accretion or star formation in the inner regions of the host galaxy. Several lines of evidence suggest that quasars must have spent significant time growing in the obscured phase \citep{2010ApJ...719.1315K,assef_opt}, and the 20--30 keV peak of the cosmic X-ray background implies that most black hole growth is obscured \citep{2007A&A...463...79G}. Key open questions for understanding quasar evolution include: What was the SMBH activity just prior to the quasar ``blowout'' phase, when the SMBH was still highly embedded in dust and gas from the parent galaxy coalescence event? Were the SMBHs accreting as rapidly as optical quasars, faster than quasars because of the infalling material, or was the accretion suppressed by the dynamical interaction? Answers to these questions may be hidden in highly obscured, but still powerful quasars.  

The \textit{Wide-field Infrared Survey Explorer} (\wise), which surveyed the entire sky at 3.4, 4.6, 12, and 22 \micron, was designed to identify nearby cool brown dwarfs and the most luminous dusty galaxies in the universe \citep{2010AJ....140.1868W}. By selecting objects with marginal or no detection in the \wise\ 3.4 and 4.6 \micron\ bands and strong detections in the 12 and 22 \micron\ bands, we have discovered a population of hyperluminous galaxies with $L_{\rm bol} > 10^{13}\,L_{\sun}$ \citep{2012ApJ...755..173E,2012ApJ...756...96W,2013ApJ...769...91B}. Spectroscopy reveals that these ``$W1$$W2$-dropouts'' are predominantly systems at redshift $1.6<z<4.6$ \citep{assef_opt,eisenhardt_spec}. Extended Ly$\alpha$ emission is observed in a large fraction of these systems, and may be the result of strong AGN feedback \citep{2013ApJ...769...91B}. Their steeply rising spectral energy distributions (SEDs) from rest frame 1--10\micron\, and decreasing luminosity contribution at longer wavelengths imply that the bulk of the energy in these galaxies is radiated by hot dust \citep{2012ApJ...756...96W}. They meet the selection criteria for dust-obscured galaxies \citep[DOGs; $F_{24\,\mu \rm m} > 0.3$\,mJy and $F_{24\,\mu \rm m}/F_{R} > 1000$;][]{2008ApJ...677..943D}, but have hotter dust temperatures \citep[$> 60$ K;][]{2012ApJ...756...96W,2013ApJ...769...91B,jones_scuba2} than DOGs \citep[30K--40K; ][]{2008ApJ...689..127P,2012AJ....143..125M}. Thus, we also refer to this population as ``\hotdogs'' \citep{2012ApJ...756...96W}. 

Here we examine the most luminous \hotdogs\ identified, corresponding to galaxies with $L_{\rm bol}> 10^{14}\,L_{\sun}$. Luminosities this high correspond to a star formation rate of many thousands of solar masses per year, or to a SMBH accretion rate of tens of solar masses per year. If this luminosity is maintained for $\sim 10^{8}$\,yrs, these high luminosity sources represent the main growth phase for stellar mass if they are powered by starbursts, or of SMBH mass if they are powered by AGNs. From spectroscopic and far-infrared followup observations of over 200 \hotdogs\, we have identified 20 that meet this $L_{\rm bol}$ threshold. Among these 20 \hotdogs\, five have intrinsic $L_{\rm IR} \equiv L_{\rm (rest\ 8-1000\,\mu \rm m)} > 10^{14}\,L_{\sun}$, an order magnitude higher than the HyLIRG luminosity threshold. We refer to such systems as ``extremely luminous infrared galaxies,'' or \ehylirgs. The rest of our sample has $L_{\rm IR}> 5 \times 10^{13}\,L_{\sun}$, which should be considered a conservative lower limit due to our luminosity estimate approach (see \S\ \ref{sec:luminosity_cal}). For convenience, we refer to these sightly less luminous objects as \ehylirgs\ as well throughout this paper. For comparison, we also present 116 optically selected quasars from the literature with $L_{\rm bol} > 10^{14}\,L_{\sun}$. 

We present the sample and mid-IR and far-IR observations in Section \ref{sec:observation}. Our luminosity estimates are detailed in Section \ref{sec:results_and_analyses}. The implications of the high luminosities are discussed in Section \ref{sec:discussion}, followed by a summary in Section \ref{sec:summary}. We adopt a cosmology with $H_0=70$\,km\,s$^{-1}$\,Mpc$^{-1}$, $\Omega_{m} = 0.3$, and $\Omega_{\Lambda} = 0.7$.

\section{Sample and Observations}\label{sec:observation}

The \wise\ \ehylirgs\ presented in this paper are from the subset of \hotdogs\ selected from the \wise\ All-Sky Source Catalog \citep{2012wise.rept....1C} with spectroscopic redshifts \citep{eisenhardt_spec} and far-infrared photometry. The redshift and $L_{\rm bol}$ distribution of the current \hotdog\ sample is shown in Figure \ref{fig:hotdog_z_nuLnu}. The sample of 20 \hotdogs\ with $L_{\rm bol} \geq 10^{14}\, L_{\sun}$ corresponds to approximately 15\% of the current sample with spectroscopic redshifts and multi-wavelength followup observations. The coordinates and redshifts of the 20 sources are listed in Table \ref{table:properties_hotdogs}. The redshift quality flag ``A'' in Table \ref{table:properties_hotdogs} indicates unambiguous redshift typically determined from multiple emission or absorption features. The flag ``B'' signifies a less secure redshift determined from a robustly detected line but with uncertain identification of the line \citep{2002AJ....123.2223S}. The typical uncertainty in the redshift in Table \ref{table:properties_hotdogs} is $\Delta z \sim 0.002$.

The photometric measurements used in this paper are listed in Table \ref{table:photometry}. We include measurements of optical $r^{\prime}$-band and selected near-IR bands from ground-based follow-up observations, mid-IR photometry from \wise\ and the \textit{Spitzer Space Telescope}, and far-IR photometry from the \textit{Herschel Space Telescope}\footnote{\Herschel\ is an ESA space observatory with science instruments provided by European-led Principal Investigator consortia and with important participation from NASA.}.

\ifx \apjloutput \undefined
\else
\begin{deluxetable}{lcccc}  
\tabletypesize{\scriptsize}
\tablewidth{0in}
\tablecaption{Properties of \wise\ \elirgs\label{table:properties_hotdogs}}
\tablehead{
\colhead{Source} &  
\multicolumn{2}{c}{\WISE\ Coordinates} &
 \colhead{$z$} & 
 \colhead{Q$_{z}$} 
 \\
\colhead{} & 
  \colhead{R.A. (J2000)} & 
  \colhead{Decl. (J2000)} & 
  \colhead{} & 
  \colhead{}
}
\startdata
W0116$-$0505 	&	  01:16:01.42 	&		$-$05:05:04.2	&	3.173\tablenotemark{a} 	&	A		\\
W0126$-$0529 	&	  01:26:11.96  	&		$-$05:29:09.6	&	2.937 	&	B		\\
W0134$-$2922 	&	  01:34:35.71  	&		$-$29:22:45.4	&	3.047 	&	A		\\
W0149$+$2350 	&	  01:49:46.18  	&		$+$23:50:14.6	&	3.228 	&	A		\\
W0220$+$0137 	&	  02:20:52.13  	&		$+$01:37:11.4	&	3.122\tablenotemark{a} 	&	A		\\
W0255$+$3345 	&	  02:55:34.90  	&		$+$33:45:57.8	&	2.668 	&	A		\\
W0410$-$0913 	&	  04:10:10.61  	&		$-$09:13:05.2	&	3.592\tablenotemark{a} 	&	A		\\
W0533$-$3401 	&	  05:33:58.44  	&		$-$34:01:34.5	&	2.904 	&	A		\\
W0615$-$5716 	&	  06:15:11.07  	&		$-$57:16:14.6	&	3.399 	&	B		\\
W0831$+$0140 	&	  08:31:53.26  	&		$+$01:40:10.8	&	3.888 	&	A		\\
W0859$+$4823	&	  08:59:29.93	&		$+$48:23:02.0	&	3.245\tablenotemark{a}	&	A		\\
W1248$-$2154 	&	  12:48:15.21  	&		$-$21:54:20.4	&	3.318 	&	A		\\
W1322$-$0328 	&	  13:22:32.57  	&		$-$03:28:42.2	&	3.043 	&	A		\\
W1838$+$3429 	&	  18:38:09.16  	&		$+$34:29:25.9	&	3.205 	&	B		\\
W2042$-$3245 	&	  20:42:49.28  	&		$-$32:45:17.9	&	3.963 	&	B		\\
W2201$+$0226 	&	  22:01:23.39  	&		$+$02:26:21.8	&	2.877 	&	A		\\
W2210$-$3507 	&	  22:10:11.87  	&		$-$35:07:20.0	&	2.814 	&	B		\\
W2246$-$0526 	&	  22:46:07.57  	&		$-$05:26:35.0	&	4.593 	&	A		\\
W2246$-$7143 	&	  22:46:12.07  	&		$-$71:44:01.3	&	3.458 	&	A		\\
W2305$-$0039 	&	  23:05:25.88  	&		$-$00:39:25.7	&	3.106 	&	A		
\enddata
\tablecomments{The \wise\ coordinates are from the AllWISE database. The ``Q$_{z}$'' flag indicates the quality of the redshift (see Section \ref{sec:observation} for details).}
\tablenotetext{a}{Redshift from \cite{2012ApJ...756...96W}}.
\end{deluxetable}  

\fi

\ifx \apjloutput \undefined
\else
\begin{figure}
\epsscale{1.0}
\begin{center}
\plotone{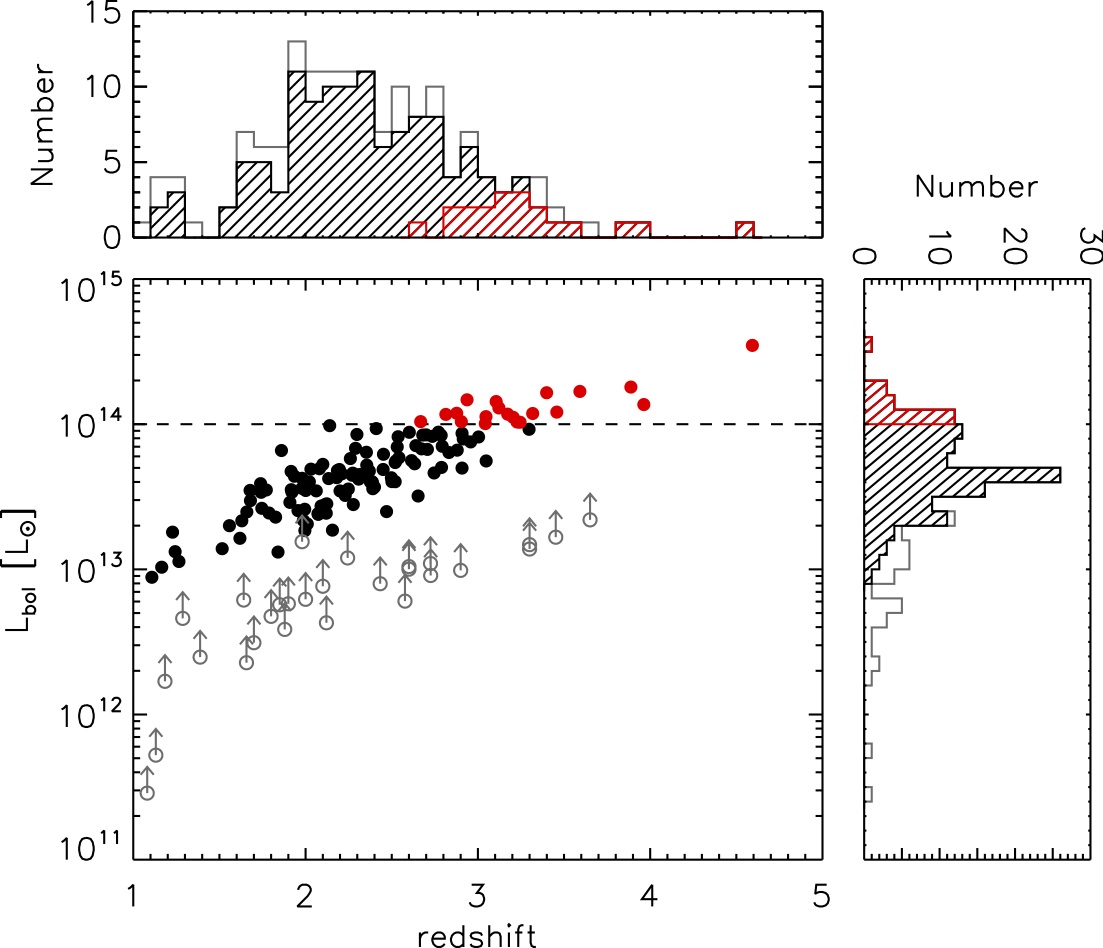}
\end{center}
\caption{The redshift and bolometric luminosity, $L_{\rm bol}$,
of \wise\ Hot DOGs at $1 < z < 5$. The luminosity distribution of the confirmed \hotdogs\ is on the right, and the redshift distribution is on the top. The black dots and black shaded regions represent \hotdogs\ with \Herschel\ measurements, while the gray open circles and open histograms indicate the lower luminosity limits for \hotdogs\ without far-IR data. The red points and histogram are for the sample of objects presented in this paper which exceed the $L_{\rm bol} > 10^{14}\,L_{\sun}$ threshold, shown by the horizontal dashed line.}\label{fig:hotdog_z_nuLnu}
\end{figure}

\fi

\subsection{Mid-infrared Observations}

The mid-IR photometry of the \wise\ \elirgs\ is listed in Table \ref{table:photometry}. \wise\ photometry is from the AllWISE Data Release \citep{2013wise.rept....1C}, which contains enhanced data products relative to the \wise\ All-Sky Source Catalog \citep{2012wise.rept....1C} from improved data processing pipelines on the full 7 months of cryogenic data at 12 and 22 \micron, and 12 months of both cryogenic and post-cryogenic data at 3.4 and 4.6 \micron. By selection, the \hotdogs\ are not well detected at \wise\ 3.4 and 4.6 \micron\ in the 7-month \wise\ All-Sky Source Catalog. However, more than half of them are detected at $\gtrsim 5 \sigma$ using the deeper 3.4 and 4.6 \micron\ data in the AllWISE Source Catalog. The [3.6] and [4.5] photometry for the $W1$$W2$-non-detected sources are from \Spitzer\ IRAC obtained during the \Spitzer\ warm mission phase, as reported by \cite{2012AJ....144..148G}. For sources with AllWISE [3.4] and [4.6] detections, we convert the data to IRAC [3.6] and [4.5] using the color correction $[3.6] = {W1} - 0.29 \times (W1 - W2)$\footnote{based on Figure 2, 3, and 4 of \url{http://wise2.ipac.caltech.edu/docs/release/allsky/expsup/sec6_3a.html}\label{footnote:colors}}. The anticipated color difference in between IRAC [4.5] and \WISE\ [4.6] is less than 0.1 magnitude\altaffilmark{\ref{footnote:colors}}, or about 10\% in flux density, thus no color correction has been applied for that band.

\subsection{Far-IR and Submillimeter Observations}
The far-IR and submillimeter photometry of the \WISE\ selected \ehylirgs, listed in Table \ref{table:photometry}, was acquired with \Herschel. The \Herschel\ data (PI: P. Eisenhardt, Proposal ID: OT1\_peisenha\_1 and OT2\_peisenha\_2) include both PACS \citep{2010A&A...518L...2P} and SPIRE \citep{2010A&A...518L...3G} observations. The SPIRE maps were made using small jiggle map mode, with a total 487 s integration time per source. The PACS images were obtained with two concatenated mini-scans for a total of 679 s on each source. The data were processed and analyzed with \textsc{hipe} v11.1.0. For W0831$+$0140, which was not included in the \Herschel\ program but was covered by the \Herschel\ ATLAS survey \citep{2010PASP..122..499E}, the far-IR photometry was taken from the public \Herschel\ archive. 

\ifx \apjloutput \undefined
\else
\ifx \apjloutput \undefined
\begin{deluxetable}{lllllllllll}  
\else
\begin{deluxetable*}{lllllllllll}  
\fi
\tabletypesize{\scriptsize}
\tablewidth{0in}
\tablecaption{Photometry of \wise\ \elirgs\label{table:photometry}}
\tablehead{
\colhead{Source} &  
 \colhead{$r^{\prime}$-band} & 
 \colhead{3.6$\mu$m} & 
 \colhead{4.5$\mu$m} &
 \colhead{12$\mu$m} & 
 \colhead{22$\mu$m} &
 \colhead{70$\mu$m} &
 \colhead{160$\mu$m} &
 \colhead{250$\mu$m} &
 \colhead{350$\mu$m} &
 \colhead{500$\mu$m} 
\\
\colhead{} & 
  \colhead{(\uJy)} & 
  \colhead{(\uJy)} & 
  \colhead{(\uJy)} &
  \colhead{(mJy)} & 
  \colhead{(mJy)} & 
  \colhead{(mJy)} &
  \colhead{(mJy)} & 
  \colhead{(mJy)} & 
  \colhead{(mJy)} & 
  \colhead{(mJy)}
}
\startdata
W0116$-$0505		&	10.2(0.5)					&	51(2)				&	89(1)				&	2.4(0.2)	&	12.1(1.1)	&	50(3)	&	93(6)	&	42(11)	&	$<$30				&	$<$42	\\
W0126$-$0529 	&  	4.1(0.3)      				& 	33(2)      				&  	37(1)    				&  	1.0(0.2) 	&  	27.5(1.3) 	&  	29(2)    	&  	219(6)      &  	213(10) 	&  	137(11) 				&  	71(14) 	\\
W0134$-$2922		&	\nodata					&	38(8)\tablenotemark{b}	&	99(11)\tablenotemark{b}	&	5.0(0.2)	&	19.7(1.4)	&	36(3)	&	40(6)	&	46(12)	&	41(10)				&	50(11)	\\
W0149$+$2350	&	$<$1.4					&	20(2)				&	35(1)				&	2.1(0.1)	&	9.8(0.8)	&	35(3)	&	91(4)	&	48(11)	&	89(16)				&	$<$57	\\
W0220$+$0137	&	6.7(0.2)\tablenotemark{a}		&	25(2)				&	38(1)				&	2.0(0.1)	&	12.4(1.0)	&	55(3)	&	120(6)	&	64(10)	&	56(11)				&	$<$42	\\
W0255$+$3345	&	1.5(0.2)\tablenotemark{a}		&	39(6)\tablenotemark{b}	&	36(10)\tablenotemark{b}	&	2.3(0.2)	&	16.5(1.2)	&	86(2)	&	73(7)	&	52(10)	&	42(10)				&	$<$42	\\
W0410$-$0913		&	2.0(0.2)\tablenotemark{a}		&	27(2)				&	46(1)				&	2.9(0.2)	&	13.4(1.2)	&	28(3)	&	110(6)	&	122(10)	&	117(11)				&	97(15)	\\
W0533$-$3401		&	7.0(0.2)\tablenotemark{a}		&	36(2)				&	73(1)				&	3.2(0.1)	&	12.0(1.0)	&	39(2)	&	98(10)	&	124(10)	&	85(10)				&	50(15)	\\
W0615$-$5716 	&  	\nodata					&  	32(2) 				& 	49(1)   				&  	2.4(0.1)  	&  	15.0(0.8) 	&  	58(3)  	&  	110(6)  	&  	53(10)  	&  	37(11) 				&  	$<$42 	\\
W0831$+$0140	&	5.7(0.2)\tablenotemark{a}		&	31(8)\tablenotemark{b}	&	63(11)\tablenotemark{b}	&	2.8(0.2)	&	10.3(1.1)	&	$<$35	&	$<$60	&	114(12)	&	93(10)		&	81(11)	\\
W0858$+4823$	&	5.4(0.2)\tablenotemark{a}		&	16(2)				&	45(1)				&	2.6(0.2)	&	12.2(1.3)	&	29(3)	&	63(10)	&	55(10)	&	57(11)				&	48(14)	\\
W1248$-$2154		&	2.7(0.2)\tablenotemark{a}		&	46(5)\tablenotemark{b}	&	36(10)\tablenotemark{b}	&	2.8(0.1)	&	13.1(0.9)	&	37(3)	&	67(2)	&	53(11)	&	36(10)				&	$<$42	\\
W1322$-$0328		&	2.6(0.2)\tablenotemark{a}		&	29(2)				&	60(1)				&	2.5(0.1)	&	11.5(1.1)	&	47(3)	&	64(7)	&	67(10)	&	47(11)				&	$<$39	\\
W1838$+$3429	&	\nodata					&	31(2)				&	35(1)				&	0.7(0.1)	&	8.4(0.9)	&	94(2)	&	38(7)	&	$<$27	&	$<$30				&	$<$42	\\
W2042$-$3245		&	2.6(0.3)\tablenotemark{a}		&	15(2)				&	19(1)				&	2.7(0.2)	&	16.4(1.3)	&	20(3)	&	30(5)	&	44(10)	&	$<$30				&	22(15)	\\
W2201$+$0226	&	0.9(0.2)\tablenotemark{a}		&	42(8)\tablenotemark{b}	&	92(11)\tablenotemark{b}	&	4.9(0.2)	&	18.1(1.4)	&	27(3)	&	141(7)	&	135(11)	&	138(12)				&	82(15)	\\
W2210$-$3507		&	1.3(0.1)\tablenotemark{a}		&	32(6)\tablenotemark{b}	&	36(12)\tablenotemark{b}	&	2.3(0.1)	&	16.5(1.0)	&	51(3)	&	140(6)	&	86(10)	&	95(11)				&	77(15)	\\
W2246$-$0526		&	$<$3.9					&	28(2)				&	27(1)				&	2.5(0.2)	&	15.9(1.6)	&	37(3)	&	192(5)	&	89(9)	&	81(12)				&	44(15)	\\
W2246$-$7143		&	\nodata					&	22(4)\tablenotemark{b}	&	17(6)\tablenotemark{b}	&	1.4(0.1)	&	12.6(1.0)	&	29(3)	&	87(6)	&	71(9)	&	62(11)				&	31(15)	\\
W2305$-$0039		&	0.6(0.2)\tablenotemark{a}		&	58(6)\tablenotemark{b}	&	67(11)\tablenotemark{b}	&	3.4(0.2)	&	24.6(1.4)	&	\nodata	&	\nodata	&	83(10)	&	59(11)				&	44(15)
\enddata
\tablecomments{The numbers in parentheses are the 1--$\sigma$ uncertainty in photometry. The upper limits are at 3--$\sigma$.}
\tablenotetext{a}{Ground-based $r^{\prime}$-band photometry from \cite{eisenhardt_spec}.} 
\tablenotetext{b}{Data from \WISE\ 3.4$\mu$m or 4.6$\mu$m measurements.}
\ifx \apjloutput \undefined
\end{deluxetable}  
\else
\end{deluxetable*}  
\fi
\fi

\subsection{Extremely Luminous Optically Selected Quasars from the Literature}\label{sec:opt_qso}
 
As a comparison sample, we identified known quasars with $L_{\rm bol} > 10^{14} L_{\sun}$ from the following large-scale quasar catalogs: (i) the 13th edition of the Catalogue of Quasars and Active Nuclei \citep{2010A&A...518A..10V}, (ii) the 2dF QSO Redshift Survey \citep{2004MNRAS.349.1397C}, (iii) the 2dF-SDSS LRG and QSO Survey \citep{2009MNRAS.392...19C}, (iv) the Sloan Digital Sky Survey (SDSS) Quasar Catalog V from the 7th SDSS data release \citep{2010AJ....139.2360S}, and (v) the Sloan Digital Sky Survey Quasar Catalog from SDSS 9th data release \citep{2012A&A...548A..66P}. In addition, we considered objects with the spectroscopic class of ``QSO'' in the SDSS 10th Data Release \citep[DR10;][]{2013arXiv1307.7735A}. For the luminous SDSS DR10 quasar sample, we visually checked for mis-identified spectral features or artifacts. We also included 46 objects that are listed as ``HyLIRGs'' in the NASA/IPAC Extragalactic Database (2013 August 27th version). We utilized the redshift information of quasars reported in these catalogs, and estimated their bolometric luminosities using photometric data from \GALEX\ GR7 \citep{2005ApJ...619L...1M}, SuperCosmos \citep{2001MNRAS.326.1295H}, SDSS DR10 \citep{2013arXiv1307.7735A}, 2MASS \citep{2006AJ....131.1163S}, UKIDSS DR9 \citep{2007MNRAS.379.1599L}, the AllWISE Data Release \citep{2013wise.rept....1C}, \textit{IRAS} \citep{1984ApJ...278L...1N} and \AKARI\ \citep{2007PASJ...59S.369M}. 

We then visually inspected the SEDs and images of $\sim$ 1300 sources with estimated $L_{\rm bol} > 5 \times 10^{13} L_{\sun}$ in the optical, near-IR, and mid-IR to identify possible cases where the photometry used for the luminosity calculation was confused by nearby objects. Some sources from the low spectral resolution surveys \citep[e.g.][]{1996A&AS..119..265I} showed \WISE\ colors close to zero, much bluer than typical for quasars \citep{2012ApJ...753...30S,2013ApJ...772...26A}, and their SEDs resemble the thermal emission of stellar objects. Furthermore, the objects clustered at $z=1.97-2.20$, triggering suspicions that their redshifts might be incorrect. Other quasars with unusual blue mid-IR colors such as J071046.20$+$473211.0 and J072810.14$+$393027.7 were removed due to known photometric contamination from nearby stars \citep{1983A&AS...51...41M,1997A&AS..123..219V}. Some of the sources have proper motions detected between the 2MASS and \wise\ observations, and their SEDs suggest they are likely late-type dwarf stars \citep[e.g., J003332.60$-$392245.0, an M-type dwarf star;][]{2008ApJS..175..191P} or known brown dwarfs \citep[e.g., J144825.70$+$103158.0, an L3.5 brown dwarf; ][]{2003IAUS..211..197W}. After this culling from visual inspection, a total of 140 optically selected quasars reach the luminosity cut of $10^{14} L_{\sun}$, assuming their emission is isotropic.

To ensure that the intrinsic luminosities of the optically selected quasars are greater than the $10^{14} L_{\sun}$ threshold, we removed known gravitationally lensed systems and blazars. We invoked the catalog of strong gravitational lensing systems from ``the Master Lens Database\footnote{\url{http://www.masterlens.org/}}'' \citep{Moustakas_orphan}, and the list of blazars from ``the Roma-BZCAT Multi-frequency Catalogue of Blazars'' \citep[][version 4.1.1 -- 2012 August]{2009A&A...495..691M}. Of the 140 luminous quasars, 9 are in known strong gravitational lensing systems, and 15 are known blazars. This leaves a total of 116 hyperluminous quasars with $L_{\rm bol} > 10^{14} L_{\sun}$, including 68 quasars from the SDSS DR7 quasar search \citep{2010AJ....139.2360S}. These quasars are listed in Table \ref{table:properties_qso}. 

To compare the far-IR SEDs of hyperluminous quasars and \hotdogs, we have gathered the available \Herschel\ photometry for our quasar sample. \Herschel\ SPIRE data are available for 15 quasars, and two of them also have PACS measurements. This photometry is listed in Table \ref{table:photometry_qso}.

\ifx \apjloutput \undefined
\else
\ifx \apjloutput \undefined
\begin{deluxetable}{llllcl}  
\else
\begin{deluxetable*}{llllcl}  
\fi
\tabletypesize{\scriptsize}
\tablewidth{0in}
\tablecaption{Properties of Optically Selected Quasars with $L_{\rm bol} > 10^{14} L_{\sun}$ \textbf{(Short Version)}\label{table:properties_qso}}
\tablehead{
\colhead{Source} &  
\multicolumn{2}{c}{\WISE\ Coordinate} &
 \colhead{Redshift} & 
 \colhead{$L_{\rm bol}$\tablenotemark{a}} &
 \colhead{Redshift Ref.}
\\
\colhead{} & 
  \colhead{R.A. (J2000)} & 
  \colhead{Decl. (J2000)} & 
  \colhead{} & 
  \colhead{($10^{14}\,L_{\sun}$)} & 
  \colhead{}
}
\startdata
J000322.91$-$260316.8 &  00:03:22.91 & $-$26:03:16.8 &  4.098   &   1.6   & NED, V10 \\
J001527.40$+$064012.0 &  00:15:27.40 & $+$06:40:12.0 &  3.17    &   1.2   & V10 \\
J004131.50$-$493612.0 &  00:41:31.50 & $-$49:36:12.0 &  3.24    &   1.8   & V10 \\
J010311.30$+$131618.0 &  01:03:11.30 & $+$13:16:18.0 &  2.681   &   1.6   & NED, V10 \\
J012156.04$+$144823.9 &  01:21:56.03 & $+$14:48:23.9 &  2.870    &   1.1   & S10, V10 \\
J012412.47$-$010049.8 &  01:24:12.47 & $-$01:00:49.7 &  2.826   &   1.0   & S10, P12, V10 \\
J013301.90$-$400628.0 &  01:33:01.90 & $-$40:06:28.0 &  3.023   &   1.0   & V10 \\
J015636.00$+$044528.0 &  01:56:36.00 & $+$04:45:28.0 &  2.993   &   1.0   & V10 \\
J020727.20$-$374156.0 &  02:07:27.20 & $-$37:41:56.0 &  2.404   &   1.2   & V10 \\
J020950.70$-$000506.0 &  02:09:50.71 & $-$00:05:06.4 &  2.850    &   1.2   & V10, S10, P12 \\
J024008.10$-$230915.0 &  02:40:08.10 & $-$23:09:15.0 &  2.225   &   1.4   & V10 \\
J024854.30$+$180250.0 &  02:48:54.30 & $+$18:02:50.0 &  4.42    &   1.0   & V10 \\
J025240.10$-$553832.0 &  02:52:40.10 & $-$55:38:32.0 &  2.35    &   1.2   & V10 \\
J030722.80$-$494548.0 &  03:07:22.80 & $-$49:45:48.0 &  4.728   &   1.3   & V10 \\
J032108.45$+$413220.9 &  03:21:08.45 & $+$41:32:20.8 &  2.467   &   1.1   & S10
\enddata
\tablecomments{Redshifts from: V10 \citep{2010A&A...518A..10V}; C04 \citep{2004MNRAS.349.1397C}, C09 \citep{2009MNRAS.392...19C}; S10 \citep{2010AJ....139.2360S}; P12 \citep{2012A&A...548A..66P}, DR10 \citep{2013arXiv1307.7735A}, and NED (2013 April version of HyLIRG list from NASA/IPAC Extragalactic Database). Only the first 15 sources are listed here. The complete electronic table of 116 sources is available online at the journal website.
}
\tablenotetext{a}{See Section \ref{sec:luminosity_cal} for definition.}
\ifx \apjloutput \undefined
\end{deluxetable}  
\else
\end{deluxetable*}  
\fi

\ifx \apjloutput \undefined
\begin{deluxetable}{lllllllllll}  
\else
\begin{deluxetable*}{lllllllllll}  
\fi
\tabletypesize{\scriptsize}
\tablewidth{0in}
\tablecaption{Photometry of Optically Selected Quasars with $L_{\rm bol} > 10^{14}\,L_{\sun}$ \textbf{(Short Version)}\label{table:photometry_qso}}
\tablehead{
\colhead{Source} &  
 \colhead{$R$-band} & 
 \colhead{3.4$\mu$m} & 
 \colhead{4.6$\mu$m} &
 \colhead{12$\mu$m} & 
 \colhead{22$\mu$m} &
 \colhead{70$\mu$m} &
 \colhead{160$\mu$m} &
 \colhead{250$\mu$m} &
 \colhead{350$\mu$m} &
 \colhead{500$\mu$m} 
\\
\colhead{} & 
  \colhead{(mJy)} & 
  \colhead{(mJy)} & 
  \colhead{(mJy)} &
  \colhead{(mJy)} & 
  \colhead{(mJy)} & 
  \colhead{(mJy)} &
  \colhead{(mJy)} & 
  \colhead{(mJy)} & 
  \colhead{(mJy)} & 
  \colhead{(mJy)}
}
\startdata
\ifx \apjloutput \undefined
J000322.910$-$260316.80 & 0.40(0.11) & 0.86(0.02) & 0.69(0.02) & 2.2(0.1) & 8.5(0.9) &  \nodata &  \nodata &  \nodata &  \nodata &  \nodata \\ 
J001527.400$+$064012.00 & 0.43(0.12) & 0.93(0.02) & 1.09(0.03) & 5.3(0.2) & 10.9(1.1) &  \nodata &  \nodata &  \nodata &  \nodata &  \nodata \\ 
J004131.500$-$493612.00 & 0.79(0.22) & 1.61(0.04) & 1.62(0.04) & 4.3(0.1) & 9.1(0.8) &  \nodata &  \nodata &  \nodata &  \nodata &  \nodata \\ 
J010311.300$+$131618.00 & 1.09(0.30) & 1.08(0.03) & 1.51(0.04) & 9.0(0.2) & 19.4(1.0) &  \nodata &  \nodata &  \nodata &  \nodata &  \nodata \\ 
J012156.038$+$144823.94 & 0.31(0.09) & 0.93(0.03) & 1.16(0.03) & 3.6(0.2) & 9.2(1.0) &  \nodata &  \nodata &  \nodata &  \nodata &  \nodata \\ 
J012412.470$-$010049.76 & 0.45(0.12) & 1.28(0.03) & 1.40(0.03) & 3.6(0.1) & 6.7(1.0) &  \nodata &  \nodata &  \nodata &  \nodata &  \nodata \\ 
J013301.900$-$400628.00 & 0.49(0.13) & 0.65(0.02) & 0.89(0.03) & 3.4(0.1) & 9.4(0.9) &  \nodata &  \nodata &  \nodata &  \nodata &  \nodata \\ 
J015636.000$+$044528.00 & 0.43(0.12) & 0.60(0.02) & 0.67(0.02) & 2.2(0.1) & 3.1(0.8) &  \nodata &  \nodata &  \nodata &  \nodata &  \nodata \\ 
J020727.200$-$374156.00 & 0.85(0.23) & 1.49(0.03) & 1.71(0.04) & 6.0(0.1) & 12.3(0.8) &  \nodata &  \nodata &  \nodata &  \nodata &  \nodata \\ 
J020950.712$-$000506.49 & 0.63(0.17) & 0.93(0.02) & 1.36(0.03) & 6.1(0.2) & 15.0(0.8) &  \nodata &  \nodata & 66(6) & 48(6) & 22(7) \\ 
J024008.100$-$230915.00 & 1.01(0.28) & 1.92(0.04) & 3.23(0.07) & 11.1(0.2) & 21.1(1.0) &  \nodata &  \nodata &  \nodata &  \nodata &  \nodata \\ 
J024854.300$+$180250.00 & 0.17(0.05) & 0.62(0.02) & 0.55(0.02) & 1.6(0.1) & 4.1(1.1) &  \nodata &  \nodata &  \nodata &  \nodata &  \nodata \\ 
J025240.100$-$553832.00 & 0.88(0.24) & 1.73(0.04) & 1.99(0.04) & 8.2(0.2) & 19.0(0.9) &  \nodata &  \nodata & 46(2) & 39(3) & 14(3) \\ 
J030722.800$-$494548.00 & 0.06(0.02) & 0.54(0.01) & 0.51(0.02) & 1.1(0.1) & 4.0(0.7) &  \nodata &  \nodata &  \nodata &  \nodata &  \nodata \\ 
J032108.450$+$413220.87 & 0.56(0.15) & 1.39(0.03) & 1.94(0.05) & 6.8(0.2) & 12.7(1.1) &  \nodata &  \nodata &  \nodata &  \nodata &  \nodata
\else

\fi
\enddata
\tablecomments{Only photometry of the first 15 sources are listed here. The complete electronic table with photometry of 116 sources is available online at the journal website. 
The numbers in parentheses are the 1--$\sigma$ uncertainty in photometry.}
\ifx \apjloutput \undefined
\end{deluxetable}  
\else
\end{deluxetable*}  
\fi
\fi

\section{Results and Analyses}\label{sec:results_and_analyses}

\subsection{Color-Color Diagram}

The \WISE\ mid-IR color-color diagram at [3.4], [4.6], and [12] is shown in Figure \ref{fig:color_color}. The \WISE-selected \ehylirg\ \hotdogs\ occupy a wider range of [3.4]$-$[4.6] color than do the hyperluminous quasars, and the \hotdogs\ are $\sim$ 2--3 mag redder in [4.6]$-$[12] color. The \hotdog\ redshifts span $2.8 < z < 4.6$, which is narrower than the quasar redshifts range of $0.9 < z < 4.9$. This is likely due, in part, to a selection effect which biases the \hotdog\ selection to $z\gtrsim 1.5$ \citep{assef_opt}. The large gap between $4<$ [4.6]$-$[12] $<5$ is a result of the $W1$$W2$-dropout selection criteria. Some hyperluminous objects have been discovered in this color region based on different mid-IR color selection criteria accompanied by criteria at other wavelengths \citep[e.g.,][]{2013ApJ...769...91B,lonsdale_NVSS_WISE,stern_w1819}.

\ifx \apjloutput \undefined
\else
\begin{figure}
\epsscale{1.0}
\begin{center}
\plotone{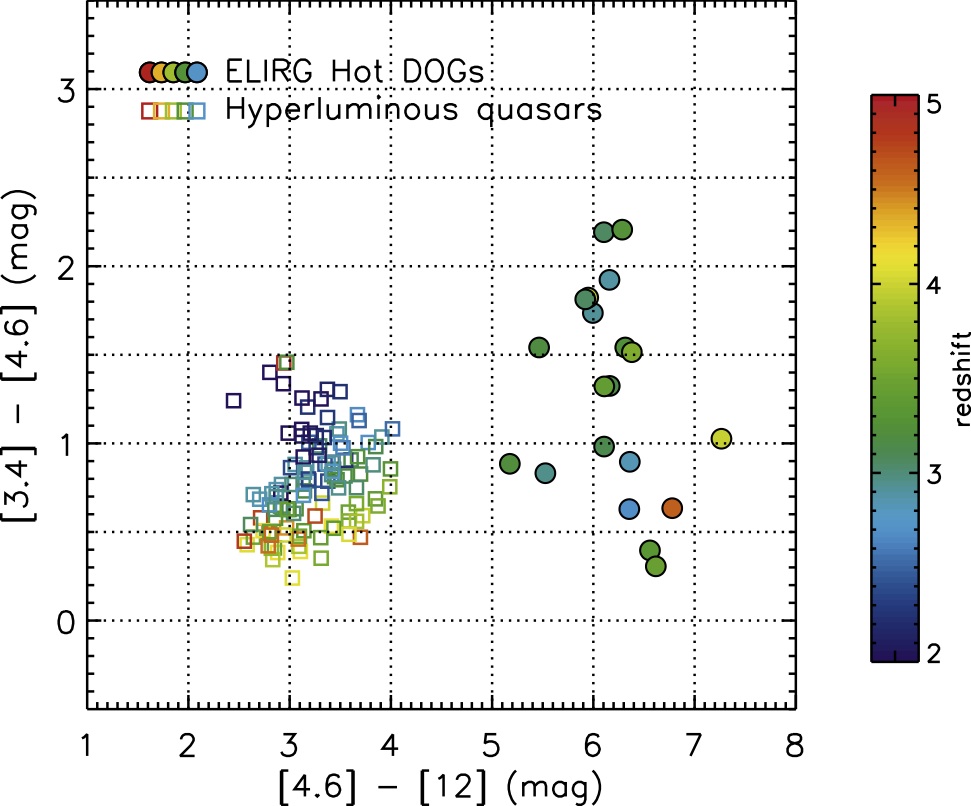}
\end{center}
\caption{Mid-IR colors of the hyperluminous ($L_{\rm bol} > 10^{14}\,L_{\sun}$) \hotdogs\ and optically selected hyperluminous quasars. The open squares show the 116 selected quasars; the filled circles show the 20 \wise-selected \hotdogs. For \hotdogs\ not detected in \WISE\ [3.4] and [4.6], their IRAC measurements are plotted with the color correction from IRAC [3.6] to \WISE\ [3.4] and IRAC [4.5] to \WISE\ [4.6] applied.
}\label{fig:color_color}
\end{figure}

\fi

\subsection{SEDs}\label{sec:sed}

The SEDs of the \wise-selected \ehylirgs\ are shown in Figure \ref{fig:normalized_SEDs}, normalized by the integrated luminosity over the plotted SEDs. The SEDs of the $L_{\rm bol} \geq 10^{14}\,L_{\sun}$ \hotdogs\ are similar to those of their less luminous siblings, which are outlined by the shaded region. The steep rise from rest frame 1--4\micron\ reflects the selection criteria. These SEDs do not match empirical starburst or dusty AGN templates, although they are close to the torus model of \cite{2006ApJ...642..673P}. However, they are steeper than the torus model at $\lambda < 4\,\micron$ and drop faster toward the far-IR at $\lambda > 60\,\micron$. The rest-frame flux density peak is at shorter wavelengths than the peak of the dusty starburst system Arp\,220, which is at about 60\,\micron. This indicates emission from hotter dust in these \wise-selected hyperluminous galaxies. The emission excess around rest frame 6\,\micron\ can be explained by dust emission $T_d \sim 450\,$ K, as shown in the upper panel of Figure \ref{fig:normalized_SEDs}. This suggested temperature does not imply a single-temperature dust system, but is rather a characteristic temperature for the hot dust emission component. Further discussion of the SEDs and implied dust temperatures is included in Section \ref{sec:Tdust_and_SED}.

\ifx \apjloutput \undefined
\else
\ifx \apjloutput \undefined
\begin{figure}
\else
\begin{figure*}
\fi
\ifx \apjloutput \undefined
\epsscale{0.75}
\else
\epsscale{0.75}
\fi
\begin{center}
\plotone{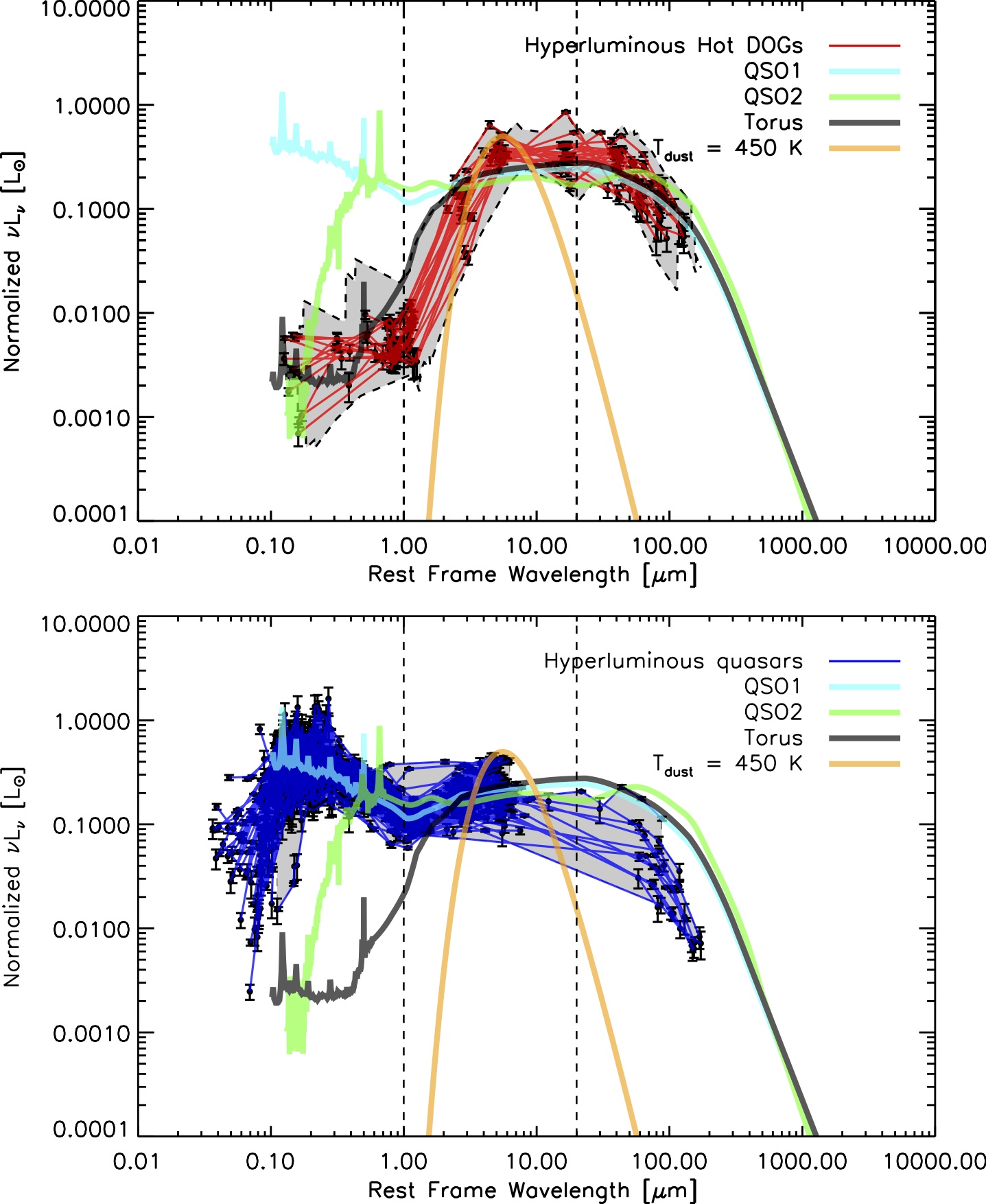}
\end{center}
\caption{Normalized rest-frame SEDs of the hyperluminous ($L_{\rm bol} > 10^{14}\,L_{\sun}$) 
\ehylirg\ \hotdogs\ and optically-selected hyperluminous quasars. The SEDs are normalized to the total bolometric luminosity $L_{\rm bol}$. The shaded region in gray in the upper panel represents the scatter of SEDs for all \hotdogs\ with $z > 1.6$ and $L_{\rm bol} > 10^{13}\,L_{\sun}$, while the gray region in the lower panel covers the scatter of SEDs for the hyperluminous quasar sample. The plotted QSO and torus SED models are adopted from \cite{2006ApJ...642..673P,2007ApJ...663...81P}. The dust model line 
assumes a dust temperature $T_d = 450$ K with emissivity index $\beta = 2.0$. The vertical dashed lines indicate rest-frame wavelengths of 1 and 20\micron.}\label{fig:normalized_SEDs}
\ifx \apjloutput \undefined
\end{figure}
\else
\end{figure*}
\fi
\fi

\subsection{Luminosity Estimates}\label{sec:luminosity_cal}

The bolometric luminosity \Lbol\ is calculated very conservatively by integrating over the photometric data, only considering $> 3\,\sigma$  detections, with a power law interpolated between observed flux density measurements, and extrapolated to 20\% beyond the shortest and longest wavelength bands by assuming no luminosity beyond these wavelengths. We do not incorporate any extinction correction or SED model in our luminosity estimate. The resulting luminosity values from this approach can be considered as conservative lower limits. If the best-fit SED templates or spline-smoothed SEDs are considered, the luminosity values typically increase by a factor of 2.

The SEDs of the quasars do not extend to rest frame wavelengths $> 8$\,\micron\ due to a lack of  comprehensive far-IR data. The contribution to the bolometric luminosity at longer wavelengths is expected to be $< 35\%$ of the $L_{\rm bol}$ based on quasar SED templates \citep{2006ApJ...642..673P,2010ApJ...713..970A}. Using available \Herschel\ archival data, we estimate the contribution to \Lbol\ by far-IR emission is $<$ 20\% for the optically selected quasars in this paper. 

To estimate the luminosity contributed by different components to the SED, we separate the SED into three parts: rest-frame blue emission ($\sim$ 1000 \AA) to 1\micron, emission from 1\micron\ to 20\micron, and emission from 20\micron\ and beyond. The corresponding luminosities from these wavelength ranges are referred to as $L_{\rm 0.1-1\,\mu \rm m}$, $L_{\rm 1-20\,\mu \rm m}$, and $L_{\rm>20\,\mu \rm m}$, respectively. For simplicity, we refer to $L_{\rm 1-20\,\mu \rm m}$ as $L_{\rm MIR}$ hereafter. The $L_{\rm>20\,\mu \rm m}$ should be distinguished from infrared luminosity, $L_{\rm IR}$, which is defined as the accumulated luminosity between 8--1000 \micron, and from the traditional far-infrared luminosity, $L_{\rm FIR}$, which covers emission from 40 to 500\,\micron. The results are listed in Table \ref{table:hotdog_lum}. 

\section{Discussion}\label{sec:discussion}

The luminosity distributions of hyperluminous \hotdogs\ and quasars are shown in Figure \ref{fig:lum_distribution}. The 20 luminous \hotdogs\ have luminosities up to $10^{14.6}\,L_{\sun}$, reaching the same level as the non-lensed quasars, although the numbers are a factor of $\sim$ 3--5 lower without any consideration of selection effects. \cite{assef_opt} find that the space density of \elirg\ \hotdogs\ is comparable to the space density of extremely luminous unobscured quasars from SDSS-III BOSS \citep{2013ApJ...773...14R} after correcting for the selection function used to identify \hotdogs\ from \wise\ photometry and for spectroscopic incompleteness.

\ifx \apjloutput \undefined
\else
\ifx \apjloutput \undefined
\begin{deluxetable}{llrrrrrrrl}  
\else
\begin{deluxetable*}{llrrrrrrrl}  
\fi

\tabletypesize{\scriptsize}
\tablewidth{0in}
\tablecaption{Luminosities of \wise\ \elirgs\label{table:hotdog_lum}}
\tablehead{
\colhead{Source} &  
 \colhead{Redshift} & 
 \colhead{$L_{\rm bol}$} &
 \colhead{$L_{\rm IR}$} &
 \colhead{$L_{\rm 0.1-1\,\mu \rm m}$} & 
 \colhead{$L_{\rm 1-20\,\mu \rm m}$} & 
 \colhead{$L_{\rm >20\,\mu \rm m}$} &
 \colhead{$L_{\rm 5.8\,\mu \rm m}$} &
 \colhead{$L_{\rm 7.8\,\mu \rm m}$} 
\\
\colhead{} & 
\colhead{} & 
  \colhead{($10^{13} L_{\odot}$)} &
  \colhead{($10^{13} L_{\odot}$)} &
  \colhead{($10^{13} L_{\odot}$)} &
  \colhead{($10^{13} L_{\odot}$)} &
  \colhead{($10^{13} L_{\odot}$)} &
  \colhead{($10^{13} L_{\odot}$)} &
  \colhead{($10^{13} L_{\odot}$)} 
}
\startdata

W0116$-$0505 & 3.173 & 11.7 & 8.2 & 0.0 & 7.8 & 3.9 & 0.9 & 1.0 \\
W0126$-$0529 & 2.937 & 14.7 & 10.7 & 0.1 & 7.0 & 7.6 & 1.8 & 1.3 \\
W0134$-$2922 & 3.047 & 11.3 & 6.2 & 0.1 & 9.0 & 2.7 & 1.4 & 1.2 \\
W0149$+$2350 & 3.228 & 10.4 & 7.4 & 0.0 & 6.2 & 4.1 & 0.8 & 0.8 \\
W0220$+$0137 & 3.122 & 12.9 & 9.6 & 0.1 & 7.6 & 5.2 & 1.0 & 1.1 \\
W0255$+$3345 & 2.668 & 10.4 & 7.9 & 0.0 & 6.8 & 3.6 & 0.9 & 1.1 \\
W0410$-$0913 & 3.592 & 16.8 & 11.3 & 0.1 & 9.3 & 7.3 & 1.1 & 1.0 \\
W0533$-$3401 & 2.904 & 10.4 & 7.5 & 0.1 & 5.7 & 4.6 & 0.8 & 0.8 \\
W0615$-$5716 & 3.399 & 16.5 & 11.3 & 0.0 & 11.0 & 5.4 & 1.3 & 1.4 \\
W0831$+$0140 & 3.888 & 18.0 & 12.0 & 0.2 & 11.0 & 7.1 & 1.1 & 1.1 \\
W0859$+$4823 & 3.245 & 10.0 & 6.2 & 0.1 & 6.6 & 3.3 & 0.9 & 0.8 \\
W1248$-$2154 & 3.318 & 11.8 & 7.4 & 0.1 & 8.0 & 3.6 & 1.0 & 1.0 \\
W1322$-$0328 & 3.043 & 10.1 & 7.0 & 0.1 & 6.6 & 3.4 & 0.8 & 0.9 \\
W1838$+$3429 & 3.205 & 11.1 & 8.9 & 0.1 & 8.6 & 2.5 & 0.9 & 1.4 \\
W2042$-$3245 & 3.963 & 13.7 & 5.8 & 0.1 & 11.4 & 2.1 & 1.4 & 1.1 \\
W2201$+$0226 & 2.877 & 11.9 & 8.0 & 0.0 & 6.6 & 5.5 & 1.2 & 0.9 \\
W2210$-$3507 & 2.814 & 11.7 & 8.8 & 0.1 & 6.3 & 5.3 & 1.0 & 1.0 \\
W2246$-$0526 & 4.593 & 34.9 & 22.1 & 0.1 & 22.2 & 12.6 & 1.9 & 1.8 \\
W2246$-$7143 & 3.458 & 12.1 & 8.3 & 0.0 & 7.8 & 4.6 & 1.0 & 1.0 \\
W2305$-$0039 & 3.106 & 13.9 & 8.3 & 0.0 & 10.1 & 3.8 & 1.7 & 1.4 
\enddata
\tablecomments{The bolometric luminosity $L_{\rm bol}$ is conservatively estimated using power-laws to interpolate over photometry from $r^{\prime}$-band to \Herschel\ SPIRE [500] \micron\ if applicable. See Section \ref{sec:luminosity_cal} for details. $L_{\rm IR}$ is the conventional infrared luminosity from rest-frame 8--1000 \micron. $L_{\rm 0.1-1\,\mu \rm m}$ is the luminosity from rest-frame 0.1-1\micron. $L_{\rm 1-20\,\mu \rm m}$, which is also refered as $L_{\rm MIR}$ in this paper, covers 1--20\micron. $L_{\rm > 20\,\mu \rm m}$ is the luminosity at wavelengths longer than 20 \micron. $L_{\rm 5.8\,\mu \rm m}$ and $L_{\rm 7.8\,\mu \rm m}$ are the monochromatic luminosities at rest-frame 5.8 and 7.8\micron, respectively, estimated by interpolating the SEDs.}
\ifx \apjloutput \undefined
\end{deluxetable}  
\else
\end{deluxetable*}  
\fi
\fi

\ifx \apjloutput \undefined
\else
\begin{figure}
\ifx \apjloutput \undefined
\epsscale{0.65}
\else
\epsscale{1.0}
\fi
\begin{center}
\plotone{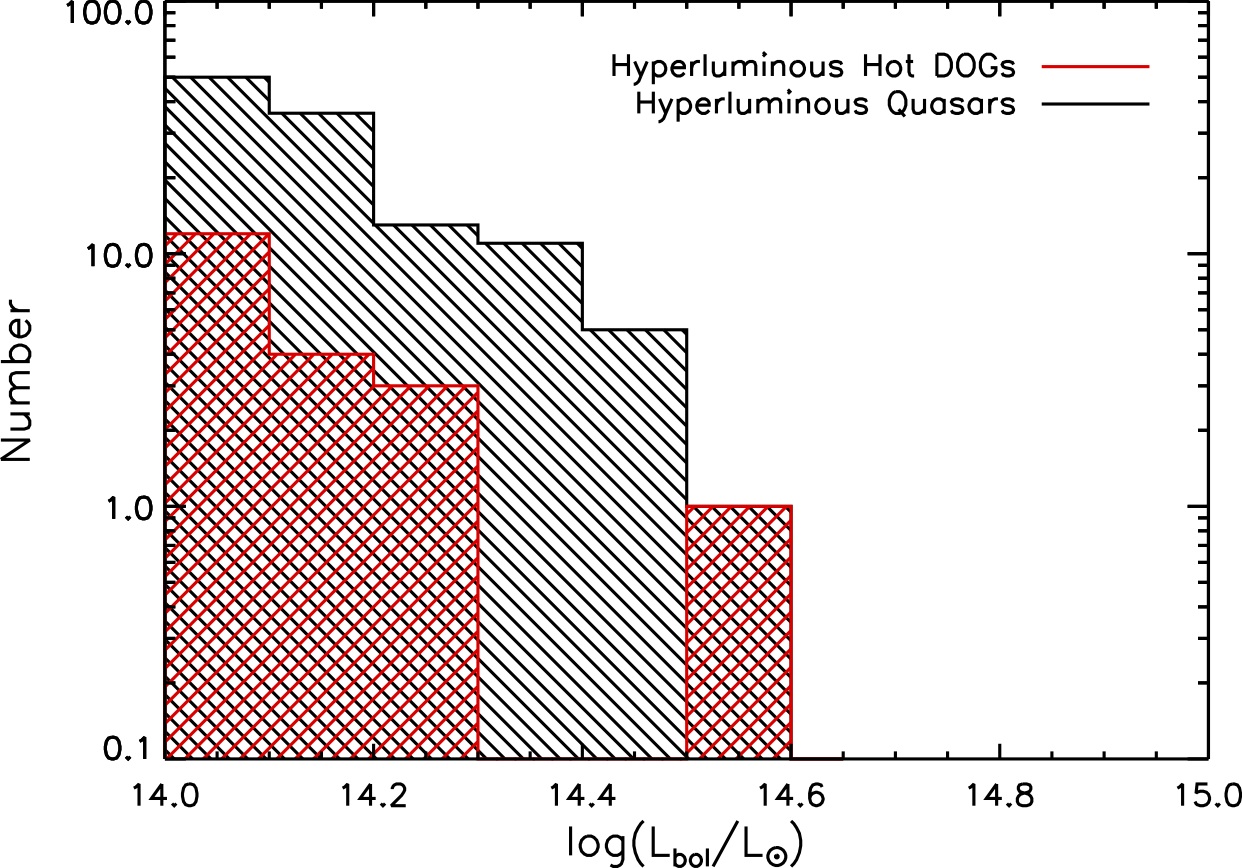}
\end{center}
\caption{Luminosity histogram of $L_{\rm bol} > 10^{14}\,L_{\sun}$  \wise\ \hotdogs\ (in red) and optically selected quasars (in black). There are 20 \hotdogs\ and 116 quasars in this plot. 
}\label{fig:lum_distribution}
\end{figure}

\fi

\subsection{Possible Effects of Beaming?}\label{sec:beaming}
The luminosities reported in this paper are calculated based on the assumption of isotropic emission in the observed wavebands. If the escaped energy is beamed, the intrinsic luminosity could be significantly overestimated. However, beaming, which is observed in blazars, is associated with variable light curves. In addition, \hotdogs\ are only weakly detected or undetected in shallow (1 mJy), wide-area radio surveys \citep{tsai_radio}, unlike beamed objects which are typically radio bright \citep{1995PASP..107..803U}. 

Beaming implies small physical scales, hence the potential for rapid variability. We do not see significant variation in the \WISE\ data. None of the \hotdogs\ varies in $W3$ and $W4$ to a limit of 30\% over 6 months, and none are flagged as significantly varying in the AllWISE catalog. Finally, many \hotdogs\ have emission-line spectra \citep{2012ApJ...756...96W,eisenhardt_spec}, unlike the featureless spectral characteristic of BL Lac objects. These properties distinguish \hotdogs\ from known beamed populations. 

\subsection{Possible Effects of Lensing?}

Another possible explanation for the high luminosity of \hotdogs\ is gravitational lensing by massive foreground systems. The most luminous known quasars, J0831$+$5245 and J1424$+$2256 with apparent $L_{\rm bol} \gtrsim 10^{15}\,L_{\sun}$, are both gravitationally lensed \citep{1992MNRAS.259P...5L,1992MNRAS.259P...1P,1998ApJ...505..529I}. For the \WISE\ \elirgs, while we cannot completely rule out the lensing hypothesis for these hyperluminous \hotdogs, we consider the likelihood of strong lensing to be small based on the following arguments.

First, we consider what may be inferred from the \WISE\ imaging data, estimating upper bounds on the possible magnification. Significant lensing requires the magnified source to be close to the effective Einstein radius (\thetaE) of the lens. In general, \thetaE\ is governed by the redshifts of background target and foreground lens, as well as the mass of the lens. In Case 1, \thetaE\ is larger than the angular resolution of $W1$ (6\arcsec). In Case 2, \thetaE\ is smaller than 6\arcsec\ and the lensed images and foreground lens could be blended.

Case 1 is addressed by Figure \ref{fig:GLens_resolved}, which shows the $W1$ photometry vs. separation for objects in the 20 \elirg\ fields. We use a SWIRE elliptical galaxy SED template and assume a total mass-to-light ratio of $M/L_{B} \sim 5$ in solar units \citep[e.g.][]{1979ARA&A..17..135F,2005MNRAS.357..691N} to show the relationship between \thetaE, $W1$ photometry, and the mass of the lens. The maximum \thetaE\ occurs near a lens redshift $\sim$ 2.5, for a source at $z=3.2$ (the median \elirg\ redshift), and the value of \thetaE\ vs. $W1$ for a lens at this redshift is shown by the solid blue line. Neighboring objects must be above this line to produce high magnification of the \hotdog. No source falls above the line, and only one comes close. At the observed separations of $W1$ objects in the \elirg\ fields, such objects would need masses well above $10^{14}\,M_{\sun}$.

Galaxies with masses $> 10^{13}\,M_{\sun}$ exist, such as ESO 146-5 \citep[$M \sim 10^{13}\,M_{\sun}$,][]{2010ApJ...715L.160C}, which dominates the Abell 3827 cluster of galaxies, but massive galaxy clusters are not in evidence near the \elirgs\ \citep{2012AJ....144..148G,assef_opt}. We conclude that the resolved individual \WISE\ sources that we detect are not likely to be able to cause strong magnification of the \elirgs.

There is the additional possibility that a much larger scale galaxy cluster potential could magnify ELIRGs near a correspondingly larger scale critical curve, which would not necessarily show extreme distortions or local multiple-images, particularly if the sources are quite compact intrinsically. However, the gravitational lensing magnification under this condition is usually small. Although the current optical and near-IR imaging data are not sufficient to fully explore that possibility, the \Spitzer\ data do not show the aggregation of objects within 1\arcmin\ expected for a massive foreground lensing cluster of galaxies \citep{assef_opt}.

\ifx \apjloutput \undefined
\else
\ifx \apjloutput \undefined
\begin{figure}
\epsscale{0.8}
\else
\begin{figure*}
\epsscale{0.7}
\fi
\begin{center}
\plotone{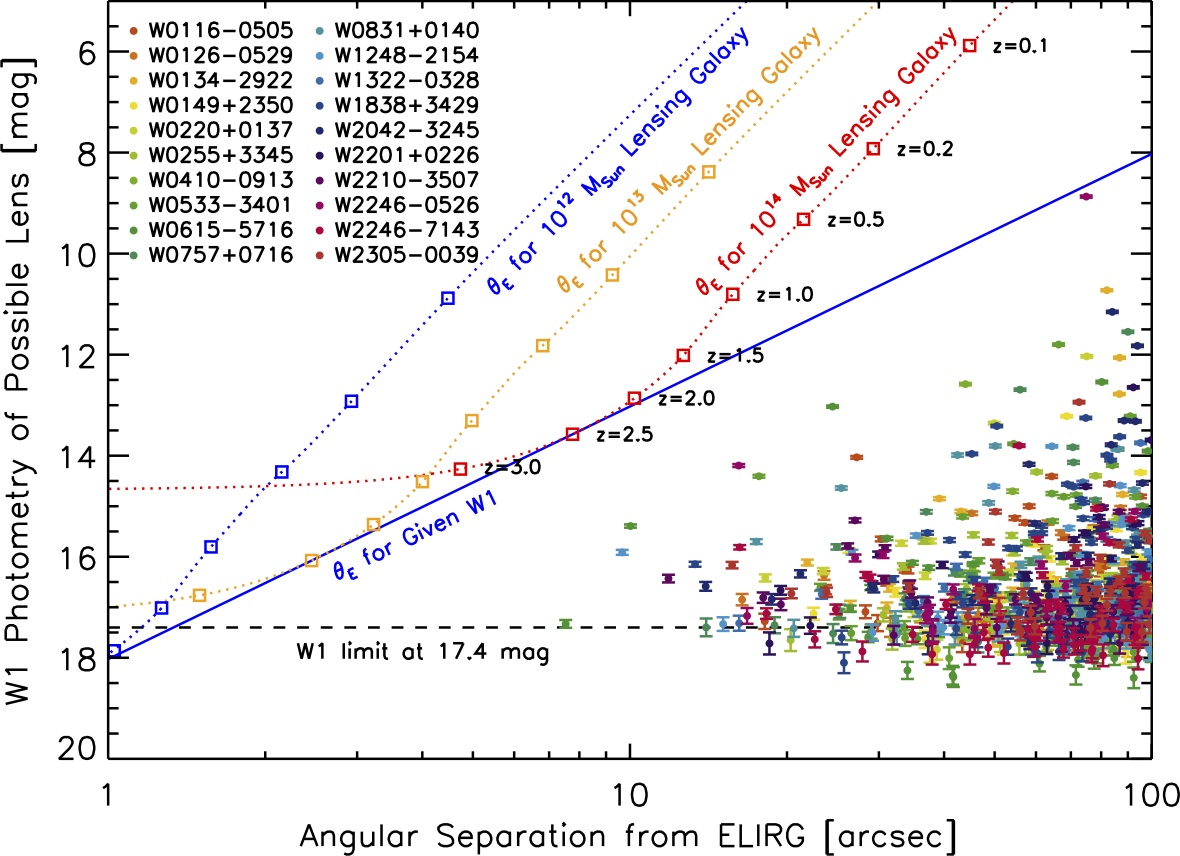}
\end{center}
\caption{
Photometry and angular separation of \WISE\ $W1$ sources within 100\arcsec of the \elirgs. Data points are color-coded by individual \elirg\ as shown in the legend. The dashed black line at $W1 = 17.4$ mag shows the original selection limit for WISE Hot DOGs. The dotted lines show the Einstein radius (\thetaE) and $W1$ magnitude \citep[assuming $M/L_B \sim 5$ and using the elliptical galaxy SED template from][]{2007ApJ...663...81P} for lensing elliptical galaxies with different masses as labeled. Open squares along each dotted line mark representative lensing galaxy redshifts. The solid blue line represents the maximum \thetaE\ vs. $W1$ for a source at $z = 3.2$, the median redshift of the 20 \WISE\ \elirgs, which occurs for a lens redshift of $z \sim 2.5$. \WISE\ sources below the solid blue line are too faint and have too large an angular separation to be lenses. 
}\label{fig:GLens_resolved}
\ifx \apjloutput \undefined
\end{figure}
\else
\end{figure*}
\fi
\fi

Case 2 is addressed by Figure \ref{fig:GLens_unresolved}, which shows the lens mass vs. redshift for a source at $z=3.2$, the median redshift of the \elirgs. In this case the lens would need to have $W1$ $>16.8$ (or $< 58\,\mu$Jy) to be consistent with the observed \elirg\ data (see Table \ref{table:photometry}). As shown by the blue curve, this excludes $10^{12}\,M_{\sun}$ lenses up to $z = 1.5$, assuming $M/L_{B} \sim 5$. Lower mass lenses are possible, of course, but they must reach the critical mass surface density for gravitational lensing \citep[see e.g.][]{1986MNRAS.219..333S}. The remaining parameter space is highlighted as the gray shaded region in Figure \ref{fig:GLens_unresolved}. This parameter space can be investigated where we have high-resolution near-IR imaging.

\ifx \apjloutput \undefined
\else
\begin{figure}
\ifx \apjloutput \undefined
\epsscale{0.7}
\else
\epsscale{1.1}
\fi
\begin{center}
\plotone{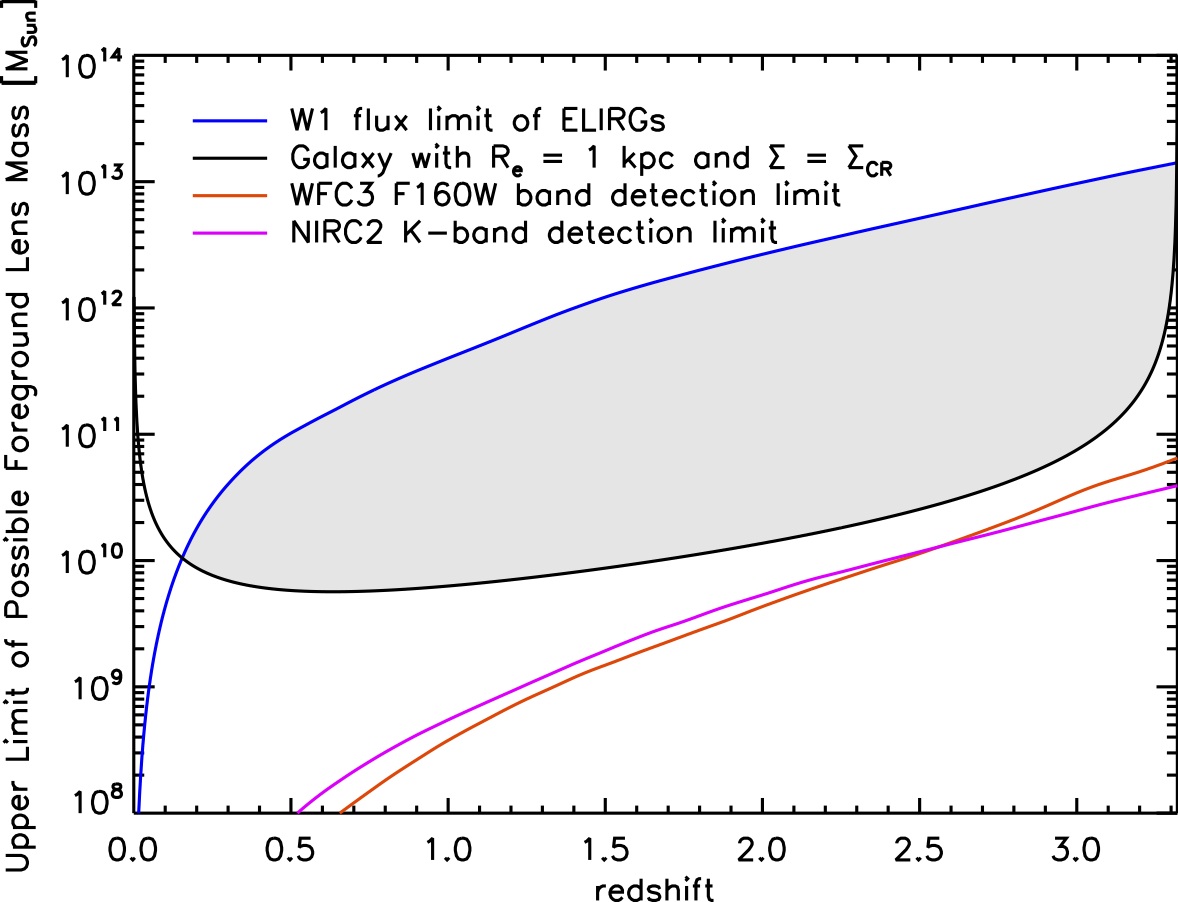}
\end{center}
\caption{
Mass limit of unresolved lenses vs. redshift, assuming a source at $z = 3.2$. The blue curve shows the upper limit due to the brightest $W1$ photometry ($W1 =16.8$) observed for \elirgs. The red-orange and magenta curves show the 3-$\sigma$ detection limit of our high resolution imaging with \HST\ WFC3 in F160W, and Keck NIRC2 with AO in the $K$-band. These curves are calculated using elliptical galaxy SED template from \cite{2007ApJ...663...81P} with assumption of $M/L_{B} \sim 5$. The solid black line shows the mass of a lensing galaxy with size $R_e \sim 1$ kpc and surface mass density $\Sigma$ equals to $\Sigma_{\rm CR}$, the critical surface mass density necessary for gravitational lensing. The shaded region indicates the remaining lens galaxy mass parameter space.
}\label{fig:GLens_unresolved}
\end{figure}

\fi

We have obtained high angular resolution (PSF FWHM $\la 0\farcs15$) near-IR imaging of over 30 \hotdogs, including 6 ELIRGs reported in this paper, with the NIRC2 camera on Keck-II with adaptive optics, and with \HST. These images do not show lensing features such as multiple images or arcs \citep{wu_carma,petty_hst,bridge_hst}. For the 5 \elirgs\ in our sample with high-resolution near-IR imaging data, it is not uncommon to see other objects a few arcseconds from the \elirgs\ in the images from \HST\ or Keck with adaptive optics. However, these objects' morphologies are typical of recent or ongoing mergers, rather than characteristic lensing geometries, based on our experience with strong lensing work \citep{1996ApJ...461...72E,2007ApJ...660L..31M}. There are sources which fall in the gray shaded area in Figure \ref{fig:GLens_unresolved}, but in no case is the \thetaE\ corresponding to the inferred mass as large as their separation from the \elirg. Thus, unless the lensing galaxies are anomalously faint or highly obscured, the high luminosity of \hotdogs\ seems to be intrinsic rather than due to gravitational lensing.

We have also examined all of our 2D spectra. We have closely examined the 4 cases (W1248$-$2154, W2042$-$3245, W2246$-$0526, and W2246$-$7143) where nearby objects appear in the data \citep{eisenhardt_spec}. Other than \w1248, discussed below, we have not identified any cases of two different redshifts superimposed that might be indicative of strong lensing \citep[e.g. SLACS survey sample;][]{2004AJ....127.1860B,2006ApJ...638..703B}. As discussed in detail below, we conclude that gravitational lensing is not causing the high luminosity of \w1248.

\subsubsection{\w1248}

Among the optical spectra of all 20 \elirgs, only \w1248\ suggests lensing. The spectrum of \w1248 shows two sources at $z = 0.339$ and $z = 3.326$ separated by 1\farcs3 \citep{eisenhardt_spec}. To explore the lensing hypothesis, we obtained $K$-band images of \w1248\ using the NIRC2 camera with the Laser Guide Star Adaptive Optics (LGS-AO) system on the Keck II Telescope \citep{2006PASP..118..310V,2006PASP..118..297W}. WISE J1248$-$2154 was observed on the night of 2014 May 18 (UT) under good weather conditions. USNO-B star 0681-0325487 \citep{2003AJ....125..984M} with $R$=16.9 located 49\arcsec from the target was used for the tip-tilt reference. Images were obtained with the MKO $K$ filter with field of view of 40\arcsec$\times$40\arcsec\ per frame and pixel scale of 0\farcs0397/pixel. Thirty $K$-band images (120 sec per image) were obtained using a three-position dither pattern that avoided the noisy, lower-left quadrant. The total effective exposure time was 60 minutes.

\ifx \apjloutput \undefined
\else
\begin{figure}
\ifx \apjloutput \undefined
\epsscale{0.5}
\else
\epsscale{0.9}
\fi
\begin{center}
\plotone{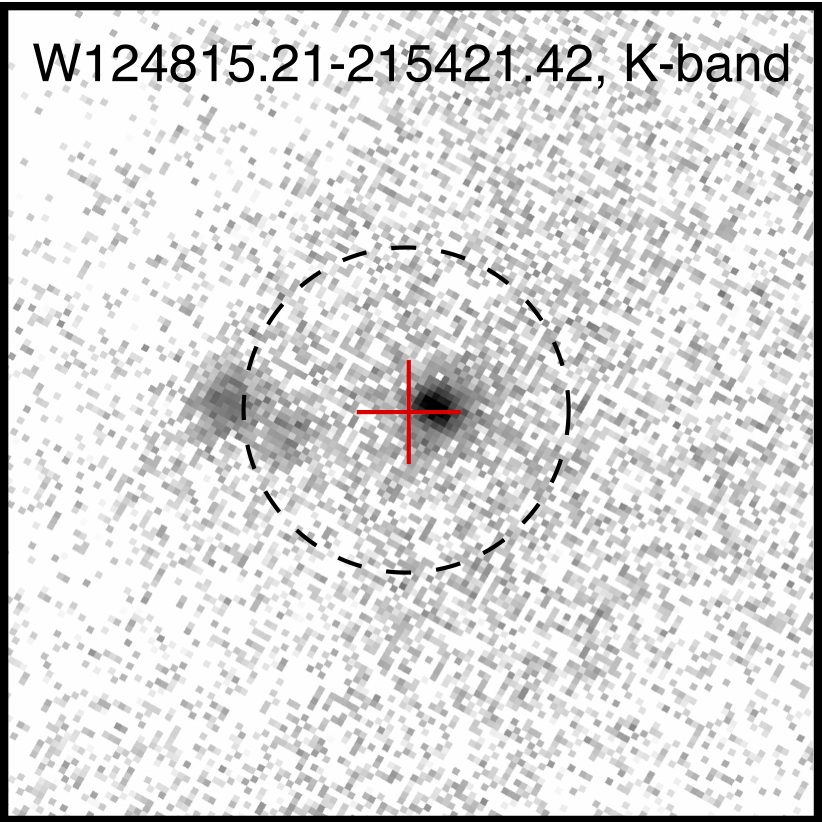}
\end{center}
\caption{
A 5\arcsec $\times$ 5\arcsec\ subsection of the Keck NIRC2 image of W1248$-$2154 in the $K$-band. The PSF FWHM is $\sim$ 0\farcs14. The red cross shows the position of W1248$-$2154 from the AllWISE Catalog and the 0\farcs32 uncertainty, which includes 0\farcs2 registration uncertainty (1\,$\sigma$) of the $K$-band image and the $\sim$ 0\farcs25 position uncertainty of W1248$-$2154 in AllWISE images. The black dashed circle indicates a $r=1\arcsec$ region. The object on the right of the red cross is at $z=3.326$, and the source on the left lies just outside the $r=1\arcsec$ circle is at $z=0.339$. 
}\label{fig:w1248_nirc2}
\end{figure}

\fi

The raw images were dark-subtracted and then sky-subtracted using a sky frame based on the median average of all frames, and a dome flat\footnote{Because of an issue with the lamp used to illuminate the spot for dome flats, we were unable to obtain $K$-band flats that night.  Instead, we used the Keck Observatory Archive (\url{https://koa.ipac.caltech.edu}) to download $K$-band dome flats acquired on 2012 October 26 (PI: A. Cooray).} was used to correct for pixel-to-pixel sensitivity variations. The co-added image (Figure \ref{fig:w1248_nirc2}) is the median average of aligned single frame images based on the pixel location of a star in the field.  The FWHM of point-like sources in the field is $\sim$ 0\farcs14. The final NIRC2 image was registered to the seeing-limited $J$-band image from \cite{assef_opt}, which has its WCS matched to the AllWISE WCS using $W1$ sources in the field of view. 

The red cross in Figure \ref{fig:w1248_nirc2} shows the AllWISE position of \w1248, which has an uncertainty of 0\farcs25. The 0\farcs2 extent of the cross indicates the astrometric uncertainty of the $K$-band image with respect to the AllWISE coordinate system. We do not find any significant gravitational lensing signatures in this image. We identify the brighter $K$-band source on the right of the cross as \w1248\ at $z = 3.326$. The object to the left of the $r = 1\arcsec$ dashed circle is at $z = 0.339$. The source on the left just within the dashed circle is blended with the $z = 0.339$ object, and does not show noticeable spectroscopic features in our spectrum. The $z=0.339$ object's $K$-band magnitude of $\sim$ 22.4 mag corresponds to a lensing mass of $2 \times 10^{9}\,M_{\sun}$, and an Einstein radius of $<$0\farcs12, significantly smaller than the separation between the companion and the \elirg. So this foreground source is not producing strong lensing of the \w1248\ \hotdog. Thus we conclude that \w1248\ is not a lensed system.

We also considered whether source confusion in the AllWISE mid-IR photometry might cause the luminosity for \w1248\ to be overestimated. Like other \hotdogs, the SED for \w1248\ is dominated by the AllWISE photometry at 12 and 22 \micron. The $z=0.34$ galaxy and $z=3.3$ \hotdog\ are blended in the AllWISE Atlas images. However, the separation is significantly larger than the 0\farcs25 positional uncertainty of \w1248\ in the AllWISE catalog. \w1248\ does not have noticeable blending in the AllWISE catalog, suggesting that the foreground galaxy is not detected in the $W1$ and $W2$ bands. While we do not have 3 and 4 \micron\ images with resolution better than the AllWISE $W1$ images, the $z=0.34$ source is significantly bluer than the $z=3.3$ source between the $J$- and $K$-bands. Thus we believe that the foreground galaxy does not contribute significantly to the $W3$ and $W4$ photometry or the $L_{\rm bol}$ estimate for \w1248.  

\ifx \apjloutput \undefined
\else
\begin{figure}
\ifx \apjloutput \undefined
\epsscale{0.65}
\else
\epsscale{1.0}
\fi
\begin{center}
\plotone{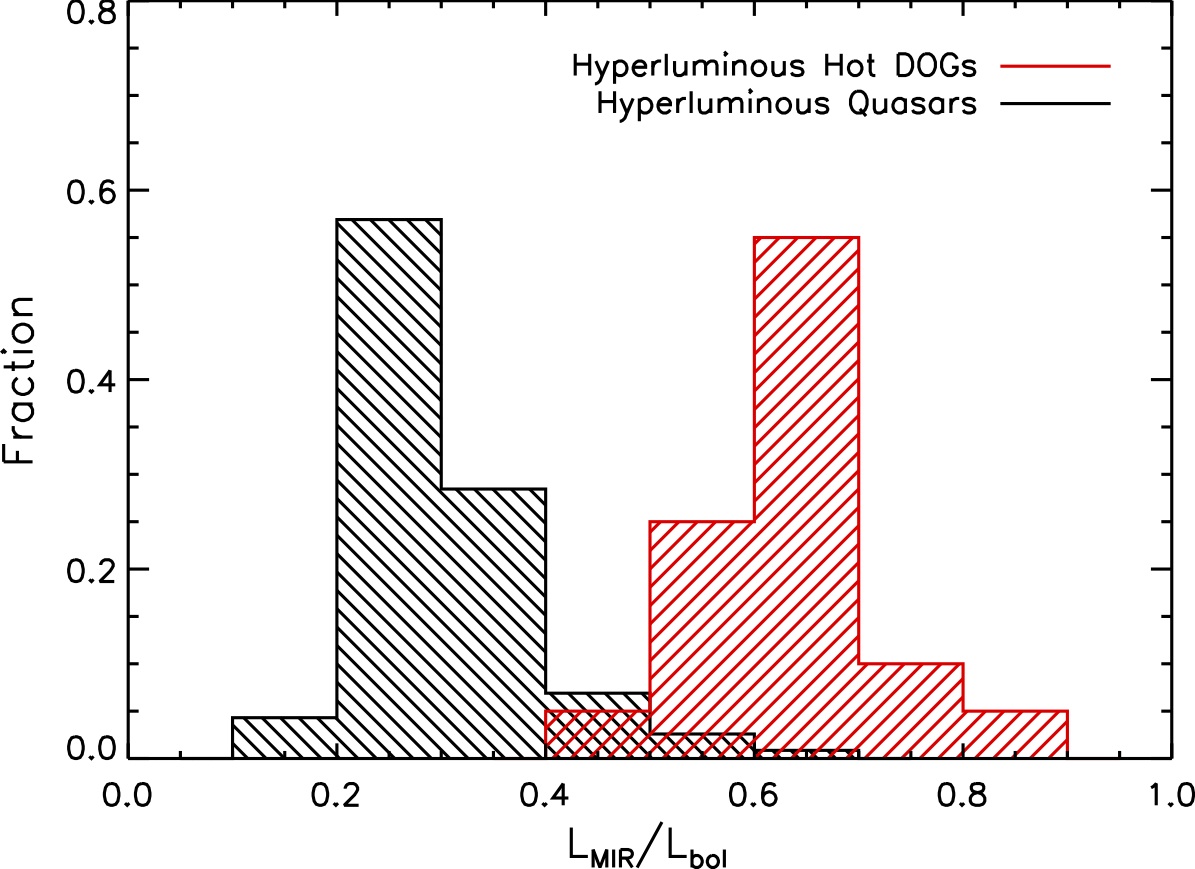}
\end{center}
\caption{Normalized histogram of $L_{\rm MIR} / L_{\rm bol}$ for \wise\ \hotdogs\ and optically selected quasars with $L_{\rm bol} > 10^{14}\,L_{\sun}$, where $L_{\rm bol}$ is the total integrated luminosity calculated using power-law-connected SEDs and $L_{\rm MIR}$ is the integrated luminosity from 1 to 20\,\micron.}\label{fig:Rmir}
\end{figure}

\fi

\ifx \apjloutput \undefined
\else
\ifx \apjloutput \undefined
\begin{figure}
\else
\begin{figure*}
\fi
\epsscale{1.0}
\begin{center}
\plottwo{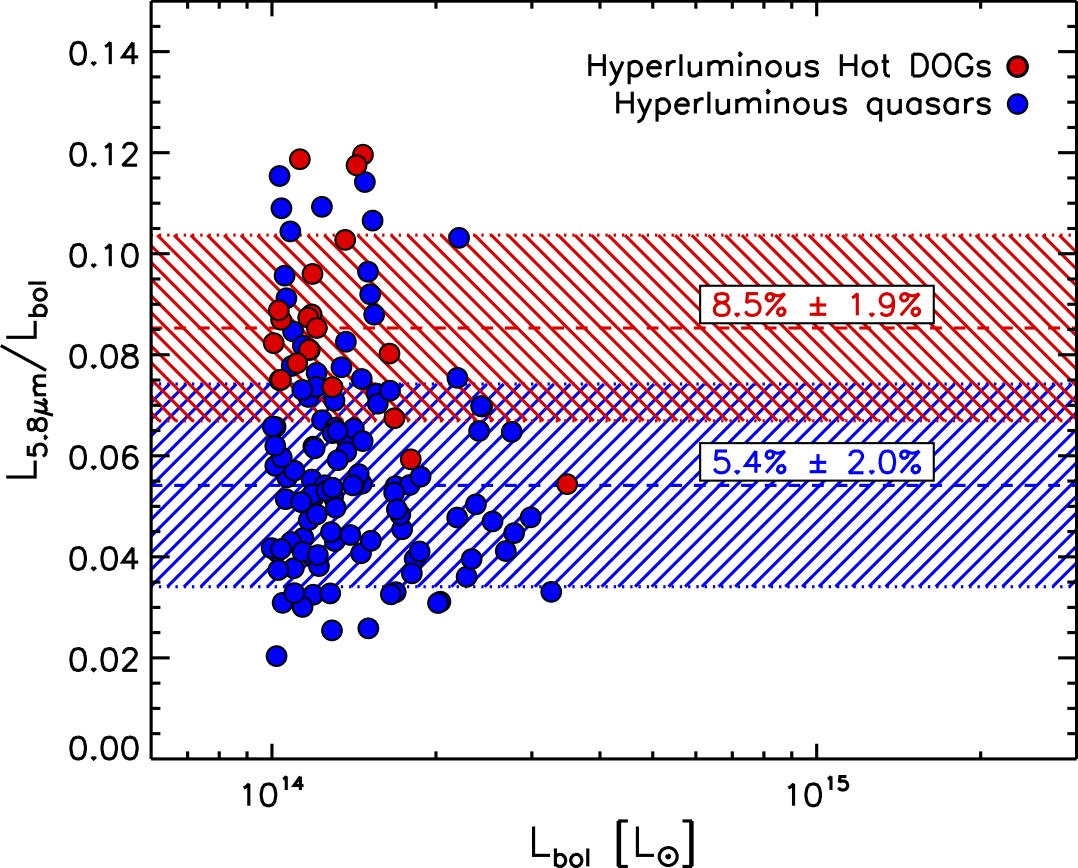}{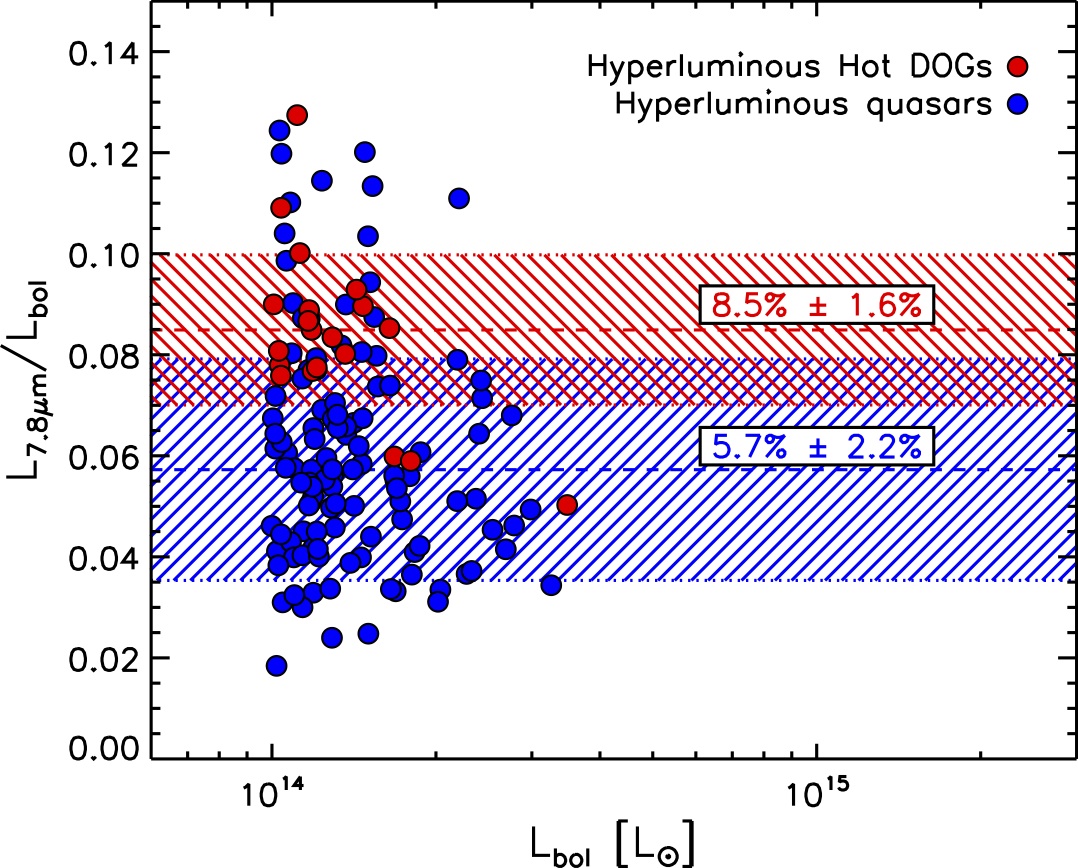}
\end{center}
\caption{Ratio of $L_{\rm 5.8\micron}$ to $L_{\rm bol}$ (\textit{\textbf{left}}) and $L_{\rm 7.8\micron}$ to $L_{\rm bol}$ (\textit{\textbf{right}}) of hyperluminous \hotdogs\ and quasars. The dashed lines represent the median values of luminosity ratios for \hotdogs\ (red), and quasars (blue). The hatched regions show a $1\,\sigma$ range from the median value. 
}\label{fig:Lbol_ratio}
\ifx \apjloutput \undefined
\end{figure}
\else
\end{figure*}
\fi

\fi

\subsection{Dust Temperatures and SED Components}\label{sec:Tdust_and_SED}

The \hotdog\ SEDs generally peak between rest-frame 4 and 10 \micron\ (see Figure \ref{fig:normalized_SEDs}), suggesting that the emitting dust can have temperatures up to $T_{d} \sim 450\,K$. The SED becomes Rayleigh-Jeans around 40--60 \micron, corresponding to $T_{d} \sim 60\,K$ \citep{2012ApJ...756...96W,2013ApJ...769...91B}. Ground-based sub-millimeter follow-up observations of \hotdogs\ indicate the rest-frame far-IR luminosity of \hotdogs\ is about an order of magnitude lower than the rest-frame mid-IR luminosity \citep{2012ApJ...756...96W,jones_scuba2}. Considering the three SED components introduced in section 3.3, the blue component (0.1--1\,\micron) represents direct emission from the host galaxy, as well as direct or scattered emission from the AGN and its accretion disk; the mid-IR component (1--20\,\micron) represents emission from the AGN dust torus, or dust emission from the cocoon of highly obscured AGNs; and the $>$ 20\,\micron\, component represents far-IR emission from dust at the outskirts of the AGN, or starburst-powered dust emission from the host galaxy. 

$L_{\rm 1-20\,\mu \rm m}$, or $L_{\rm MIR}$ as shown in Figure \ref{fig:Rmir}, is a better indicator of the hot dust-dominated luminosity of these hyperluminous systems, in contrast to the traditional infrared luminosity $L_{\rm IR}$ value which is more sensitive to diffuse dust emission powered by a starburst. As shown in Table \ref{table:hotdog_lum}, $L_{\rm MIR} > L_{\rm>20\,\mu \rm m}$ for every \elirg. The bolometric contribution of emission blueward of 1 \micron\ is negligible for the \hotdogs\ and is likely redistributed by dust to rest-frame mid-IR wavelengths, resulting in the high $L_{\rm MIR}/L_{\rm bol} \sim 65\%$ (median value). In optically selected populations, especially the high luminosity quasars, $L_{\rm MIR}$ contributes only $\sim 30$\% (median value) of $L_{\rm bol}$.

Figure \ref{fig:Lbol_ratio} shows $\nu L_{\nu}$ at 5.8 \micron\ ($L_{\rm 5.8\,\mu \rm m}$) and 7.8 \micron\ ($L_{\rm 7.8\,\mu \rm m}$), which have often been used to characterize AGN luminosities in the literature \citep[e.g.][]{2012ApJ...761..184W}. Common practice is to interpolate the SED to obtain these numbers, or to estimate them from an SED model. These numbers are convenient for statistical analyses such as deriving the quasar luminosity function. However, these monochromatic luminosities do not capture the variation in dust temperature distribution in different AGNs, which could be dramatic in obscured systems such as \hotdogs. The ratios of $L_{\rm 5.8\,\mu \rm m} / L_{\rm bol}$ and $L_{\rm 7.8\,\mu \rm m} / L_{\rm bol}$ are substantially offset between \elirg\ quasars and \elirg\ \hotdogs, and the scatter is large for both populations, particularly for quasars. The conversion from these monochromatic values to total $L_{\rm bol}$\ can vary by a factor of $\sim$ 3 (Figure \ref{fig:Lbol_ratio}). Hence we suggest that $\nu L_{\nu}$ at 5.8\,\micron\ or 7.8\,\micron\ as an estimate of total bolometric AGN luminosity should be used with caution. 

\subsection{Spatial Distribution of the Hot Dust}\label{sec:dust_distribution}

The extremely red SEDs of \hotdogs\ suggests the extinction toward their central AGNs is very high, reaching $A_{\rm V} \gtrsim 30$ mag \citep{2012ApJ...755..173E,2014ApJ...794..102S,assef_opt}. The absorbed energy is released at mid-IR wavelengths via thermal dust emission. As discussed in the previous two sections, the reprocessed energy ($L_{\rm MIR}$) output in hyperluminous \hotdogs\ dominates the total luminosity, and matches the luminosity emitted directly from the accretion disk ($L_{\rm 0.1-1\,\mu \rm m}$) in optically selected hyperluminous quasars. This enormous thermal dust luminosity suggests a dust covering fraction close to unity, rather than an edge-on dusty torus.

We estimate a dust sublimation radius \citep{1987ApJ...320..537B} of $\sim 8$\,pc for the $\geq 10^{14}\,L_{\sun}$ \hotdogs\ if they are heavily obscured by graphite-silicate mixed dust grains with a sublimation temperature of $\sim 1500$\,K. Here we assume that the bolometric luminosity is equal to the dust absorbed UV luminosity in these highly obscured systems. Unlike optically selected quasars, in which the variation timescale of optical emission is much shorter because the optical light comes directly from smaller physical scales \citep{1963Natur.197.1040S}, the fluctuations of luminosity over time in \hotdogs\ will be smoothed out by radiation reprocessing by the dust. The 16 pc diameter of the sublimation region sets the shortest variation timescale to be $\sim$ 50 yr in the rest frame. However, the timescale for luminosity changes is more likely to be related to the scale of the dust that produces the peak emission in the SED. In Section \ref{sec:sed}, we show that the highest temperature dust and therefore smallest scale that contributes substantially to the SED is at a temperature of $T_{d} \sim 450\,K$. If we assume  the emitting dust is approximately in thermal equilibrium, the characteristic radius at that temperature is $\sim$ 40 pc. Therefore we do not expect large luminosity variations over a rest-frame timescale less than $\sim200$ yr, or many centuries in the observed frame. 
As noted in Section \ref{sec:beaming}, the flux variation at 12 and 22 \micron\ is less than 30\% over a 6-month period based on \WISE\ observations.

\subsection{Energy Source Other than AGNs?}\label{sec:other_energy_source}

In 200 years at $10^{14}\,L_{\sun}$, the total energy output is $\gtrsim 2.5\times 10^{57}\,{\rm erg}$, or six orders of magnitude higher than the total energy output of a long gamma-ray burst (GRB, $E_{\rm total} \sim 10^{51}\,{\rm erg}$). \cite{2000MNRAS.316..885R} argues that a starburst component is necessary for explaining the far-IR and submillimeter ($\lambda \geq 50 \micron$) SEDs of HyLIRGs. Could an extreme starburst provide this much energy for \elirgs?

Without obvious starburst examples in the \elirg\ luminosity range, we consider He 2-10, a local dwarf galaxy and a highly obscured starburst system, as an analogy to evaluate the possibility that the high luminosity is supported by star formation. In He 2-10, compact starburst regions are highly obscured and show mid-IR emission from hot dust \citep{2001AJ....122.1365B}. \cite{2001AJ....122.1365B} estimate an infrared luminosity $L_{\rm IR} \sim 2 \times 10^{5}\,L_{\sun}$ for the Lyman continuum photon rate $N_{\rm Lyc} = 10^{49}\,{\rm s}^{-1}$ derived from radio measurements. If the obscured starburst case applies to the hyperluminous \hotdogs, $10^{14}\,L_{\sun}$ would correspond to $N_{\rm Lyc} \sim 5 \times 10^{57}\,{\rm s}^{-1}$. 

We use the STARBURST99 simulation \citep{2010ApJS..189..309L} with instantaneous starburst models to estimate the star formation needed to produce such a Lyman continuum photon rate. For a flat IMF \citep[observed in compact starbursts in extreme environments such as young and massive clusters near the Galactic center --][]{1999ApJ...525..750F,2002ApJ...581..258F}, a total star formation of $\sim 2 \times 10^{10}\,M_{\sun}$ is needed, with higher masses for a \cite{1955ApJ...121..161S} IMF. These must be formed within a few Myr -- the lifetime of massive stars which produce most of the Lyman continuum photons.  The implied star formation rate is SFR $> 5 \times 10^3\,M_{\sun}\,{\rm yr}^{-1}$, at least an order of magnitude higher than known extreme starburst systems such as sub-millimeter galaxies \citep{2010A&A...514A..67M,2014MNRAS.438.1267S} or Lyman break galaxies \citep{2005ApJ...626..698S} at high redshift. Large masses of molecular gas and cold dust should accompany this level of star formation, but we observe neither substantial CO emission from these systems \citep{blain_co} nor abundant cold dust \citep{2012ApJ...756...96W,jones_scuba2}. Thus we conclude, as did \cite{2012ApJ...755..173E}, that a starburst is unlikely to be the dominant mechanism driving the high luminosity in \hotdogs.  

\subsection{Black Hole Mass and Accretion History}

Like quasars, \hotdogs\ are likely powered by efficiently accreting SMBHs, albeit with extremely high obscurations. Here we consider the constraints on the mass and growth history of the SMBHs in the extremely luminous \hotdogs\ and quasars based on their luminosities and redshifts. Below we show that the existence of ELIRGs at $z > 3$ implies SMBHs in ELIRGs have (1) a seed mass $\gg 10^{3}\,M_{\sun}$; (2) a sustained super-Eddington accretion phase; or (3) a sustained radiation efficiency of $< 15\%$, producing less radiation feedback to limit accretion.

\subsubsection{Current Eddington Ratio}\label{sec:current_eddington_ration}

The Eddington luminosity corresponds to the total emission from an isotropic accreting AGN when its radiation pressure is balanced by the gravitation of the SMBH. If we assume a hydrogen-dominated plasma, the SMBH mass at observed redshift $z$ is thus
\begin{equation}
M_{\rm Eddington} = M(z) \sim \frac{3 \times 10^{9}\,M_{\sun}}{\epsilon(z)}\frac{L_{\rm bol}}{10^{14}\,L_{\sun}},
\end{equation}
where $\epsilon(z)$ is the Eddington ratio. 

If \hotdogs\ are accreting below the Eddington limit the SMBH masses for \ehylirgs\ are $> 3 \times 10^{9}\,M_{\sun}$. This implies stellar masses $\sim 10^{12}\,M_{\sun}$ if the host galaxies follow the M-$\sigma$ correlation, comparable to local giant elliptical galaxies in clusters. If $\epsilon(z) < 0.1$ for \elirgs, the implied SMBH mass would be $M_{\rm BH} \geq 3 \times 10^{10}\,M_{\sun}$, larger than the most massive SMBHs known in the local universe \citep{2011Natur.480..215M}. The lack of such massive black holes at the present epoch is not easy to explain when the abundance of \hotdogs\ matches that of powerful quasars \citep{assef_opt}, whose abundance is in turn consistent with the distribution of large elliptical galaxies today. Furthermore, if $M_{\rm BH} > 10^{10}\,M_{\sun}$ for \hotdogs, we would expect more massive host galaxies than observed unless \hotdogs\ deviate substantially from the empirical $M-\sigma$ relation \citep{assef_opt}. While the extremely massive black holes and galaxy hosts expected for \ehylirgs\ with low Eddington ratios are not found, there is some evidence that \hotdogs\ are in rich environments. Follow-up observations using the SCUBA-II camera on JCMT show enhanced numbers of 850\micron\ continuum sources within 1\farcm5 of \hotdogs\ \citep{jones_scuba2}, and \Spitzer\ IRAC images also show enhanced densities of sources with red IRAC colors ($[3.6]-[4.5] > 0.37$), indicative of galaxies at $z\gtrsim1$ \citep{assef_opt}. 

Although super-Eddington accretion (i.e. $\epsilon(z) > 1$) is considered to be an unstable phase, it has been suggested to be common at $z> 1.7$ \citep{2010MNRAS.402.2637S}. It is possible that \hotdogs\ are in a transitional super-Eddington phase which produces their extraordinary luminosity. Indeed, super-Eddington accretion is commonly invoked to explain the non-nuclear ``ultraluminous X-ray'' (ULX) source populations seen in local galaxies \citep[e.g.,][]{2014Natur.514..202B}. However, super-Eddington accretion requires a special configuration for the accretion disk, and the timescales for super Eddington accretion in AGN are not thoroughly investigated and understood. \cite{2002ApJ...568L..97B} suggests that the Eddington limit can be exceeded by 10--100 times via small scale inhomogeneities in the thin disk accretion. \cite{2007ApJ...670.1283O} suggest accreting material through the photon trapping regions around the accreting black hole can help the system stably bypass the radiation feedback. Statistical study of SDSS quasars suggests the maximum Eddington ratio for Type 1 quasars is $\epsilon \sim 3$ \citep{2013ApJ...764...45K}, though \hotdogs\ could be in a different accretion phase from Type 1 quasars.

At high Eddington rates, the black hole mass in \ehylirgs\ can grow by an order of magnitude over $10^{7}$\,yr, much faster than the growth of stellar bulges. The relatively tight $M-\sigma$ relation seen in the local universe suggests that high Eddington rate phases do not account for a significant fraction of cosmic SMBH mass growth.

\subsubsection{Time-averaged Eddington Ratios and Radiation Efficiency}

The \ehylirg\ \hotdog\ systems are at $z > 2.5$, leaving the SMBHs in them $< 3$ Gyr to grow to the mass at which we observe them. Based on the arguments of \cite{2005ApJ...620...59S}, we can estimate the time-averaged Eddington ratio in these systems. In the following discussion, we do not consider BH mass growth via BH mergers, although these may play a role at high redshift when the BH density is relatively high. However, the merging timescale, driven by the coalescence timescale of stellar relaxation of host galaxies, is much longer than the Salpeter timescale for BH mass growth by accretion. Thus, compared to mass accretion, mass increase by mergers plays a relatively minor role in BH mass growth history. 

For a black hole with mass $M_{\rm BH}(z)$ at redshift $z$, accretion rate $\dot{M}_{\rm acc}$, black hole mass growth rate $\dot{M}_{\rm BH}$, and radiative efficiency $\eta(z)$, the observed luminosity at redshift $z$ is 
\begin{eqnarray}
L(z) 	= \eta \dot{M}_{\rm acc} c^2 = \frac{\eta  \dot{M}_{\rm BH}}{(1-\eta)} c^2. 
\label{eq:L_accretion}\end{eqnarray}
We can also relate the observed luminosity to the Eddington luminosity, 
\begin{eqnarray}
L(z) 	&=& \epsilon(z) L_{\rm Edd} = \epsilon(z) a M_{\rm BH}(z)
\label{eq:Ledd}\end{eqnarray}
where $\epsilon_{z}$ is the Eddington ratio, and $a$ is a constant associated with the opacity of accreting materials. By combining Eq. \ref{eq:L_accretion} and Eq. \ref{eq:Ledd} with cosmic time $t(z)$, one can derive
\begin{eqnarray}
\frac{{\rm d}M_{\rm BH}}{{\rm d}t} = \epsilon(t) (1-\eta) \frac{a}{\eta c^2} M_{\rm BH} = \frac{\epsilon(t) M_{\rm BH}}{\tau_{\rm Salpeter}}.
\label{eq:dmm_vs_dt}\end{eqnarray}
The factor $\tau_{\rm Salpeter} \equiv \frac{\eta c^2}{(1-\eta) a}$ is the \cite{1964ApJ...140..796S} timescale, which describes the time span for an e-fold mass increase of a black hole accreting at its Eddington limit. For a hydrogen-dominated plasma, $a \simeq 3.3 \times 10^{4}\, L_{\sun}/M_{\sun}$, and with $\eta = 0.057$ for a Schwarzschild black hole \citep{1973CMaPh..31..161B} rather than the more commonly adopted empirical value of 0.1 \citep{2002MNRAS.335..965Y}, $\tau_{\rm Salpeter}$ is $\sim 50$\,Myr. For a Kerr black hole where $\eta = 0.3$ \citep{1974ApJ...191..507T}, $\tau_{\rm Salpeter} \sim 192$\,Myr. In other words, a non-spinning black hole experiences less radiation feedback to accreting material due to its lower radiation efficiency, thus its mass doubling time could be $\gtrsim$ 3 times shorter than a fast spinning black hole.

Assuming black holes of seed mass $M_{\rm seed}$ appear at $z \sim 20$ \citep{1986MNRAS.221...53C,2009Natur.459...49B}, the age of the black holes since their appearance is $T_{\rm age}(z) \equiv \int_{z}^{z\sim20} {\rm d}t$, and the time-averaged Eddington ratio $\bar{\epsilon}(z)$ is
\begin{eqnarray}
\bar{\epsilon}(z) \equiv \frac{\int_{z}^{z\sim20} \epsilon(t) {\rm d}t}{\int_{z}^{z\sim20} {\rm d}t} = \frac{\int_{z}^{z\sim20} \epsilon(t) {\rm d}t}{T_{\rm age}}
\end{eqnarray}
between $z\sim20$ and redshift $z$. Thus, from Eq. \ref{eq:dmm_vs_dt}, we derive that the evolution of black hole mass $M_{\rm BH}$ can be written as
\begin{eqnarray}
\ln\left(\frac{M_{\rm BH}(z)}{M_{\rm seed}}\right) &=& (1-\eta) \frac{a}{\eta c^2} \int_{z}^{z\sim20} \! \epsilon(t) {\rm d}t\nonumber \\
					&=& \bar{\epsilon}(z) \frac{T_{\rm age}(z)}{\tau_{\rm Salpeter}},
\label{eq:Mbh_history_average}
\end{eqnarray}
and
\begin{eqnarray}
\bar{\epsilon}(z) = \ln\left(\frac{L(z)}{\epsilon(z) L_{\rm Edd}(M_{\rm seed})}\right)\left(\frac{T_{\rm age}(z)}{\tau_{\rm Salpeter}}\right)^{-1}.
\label{eq:epsilon_z_final}
\end{eqnarray}

For the \ehylirg\ \hotdogs\ at  $2.8 \la z \la 4.6$, we assume the current value $\epsilon(z) \sim 1$ (see discussion of Section \ref{sec:current_eddington_ration}). The age of the Universe at that redshift is $\sim$ 1.4--2.4\,Gyr. If $M_{\rm seed}$ of the SMBHs in these systems is of order 10--100\,$M_{\sun}$ as suggested by simulations of SMBH seeds from population III stars (\citealt{2008ApJ...679..639Z}; also see review by \citealt{2010A&ARv..18..279V}), the derived time-averaged Eddington ratios of our systems are $\bar{\epsilon} \sim 0.71$ ($z=4.6$) or $\sim 0.46$ ($z=2.8$). We note that the uncertainties in assumptions on $\epsilon(z)$, $M_{\rm seed}$, and $L(z)$ do not affect the $\bar{\epsilon}$ estimate significantly due to the logarithmic scaling (Equation \ref{eq:epsilon_z_final}). An order of magnitude uncertainty in these parameters will change the estimate of $\bar{\epsilon}$ by 0.1 at the most. $T_{\rm age}$ depends on the when SMBH seeds appear ($z \sim 20$), so the maximum possible $T_{\rm age}$ would decrease $\bar{\epsilon}$ by less than 0.15. Differences in $\tau_{\rm Salpeter}$ will affect $\bar{\epsilon}$. The higher $\eta$ of rapidly spinning black hole results in even higher value of $\bar{\epsilon}$. The same arguments apply to the \ehylirg\ quasars in the same redshift range. Thus, the time-averaged Eddington ratios $\bar{\epsilon}$ of the \ehylirg\ systems we discuss here are securely larger than 25\%. In comparison, the most massive black hole known so far with $\sim 2 \times 10^{10}\, M_{\sun}$ \citep{2011Natur.480..215M} would have $\bar{\epsilon} \sim 0.07$ over the Hubble time.

\subsubsection{Seed Black Hole Mass and Black Hole Spin}

The previous section discusses a simple model of $M_{\rm BH}$ growth history (Equation \ref{eq:Mbh_history_average}) in which the variables are the seed black hole mass $M_{\rm seed}$, the radiative efficiency $\eta$, and the Eddington ratio $\epsilon({z})$. 
Current models suggest seed black hole masses ranging from $\sim 10 M_{\sun}$ from simulations of the end products of population III stars \citep[e.g.,][]{2008ApJ...679..639Z} to $\sim 100 M_{\sun}$ from run away collisions between stars in dense clusters \citep[e.g.,][]{1978MNRAS.185..847B} to as much as $10^{6}\,M_{\sun}$ for supermassive stars that quickly accumulate ambient material and collapse into black holes \citep{2008ApJ...682..745W,2009MNRAS.396..343R}.

\ifx \apjloutput \undefined
\else
\ifx \apjloutput \undefined
\begin{figure}
\else
\begin{figure*}
\fi
\epsscale{0.7}
\begin{center}
\plotone{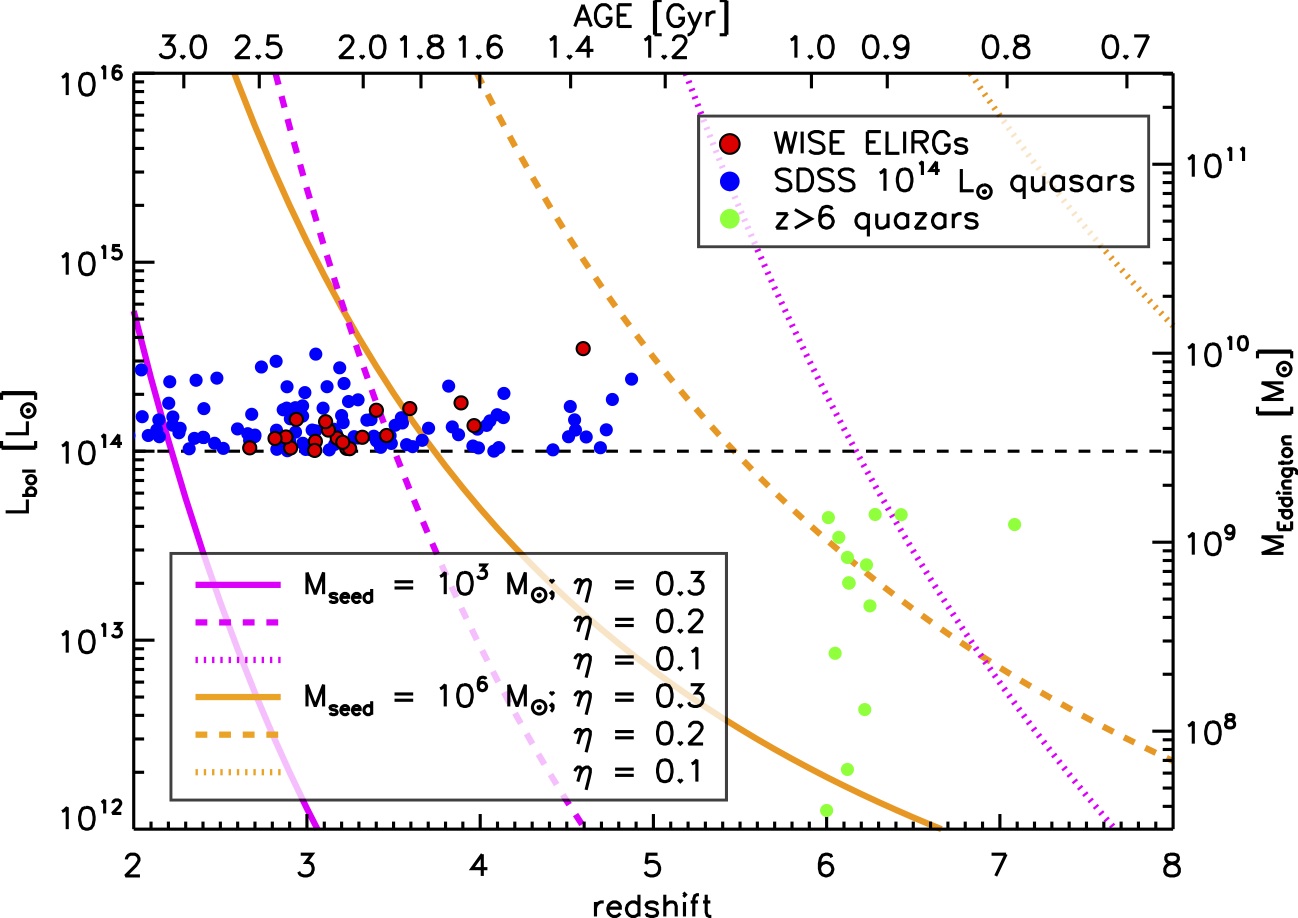}
\end{center}
\caption{Bolometric luminosity $L_{\rm bol}$ vs. redshift. The age of the SMBH since formation at $z \sim 20$ is plotted on the top. The black hole mass for an Eddington ratio of one ($\bar{\epsilon} = 1$) is plotted on the right. The dotted, dashed, and solid curves are the black hole accretion history between $z=8$ and $z=2$ for $\bar{\epsilon} = 1$ and radiative efficiency $\eta = 0.1$, 0.2, and 0.3, respectively. The magenta lines show predictions with initial black hole mass $M_{\rm seed} = 10^{3}\,M_{\sun}$, while yellow lines show the case of $M_{\rm seed} = 10^{6}\,M_{\sun}$. The regions to the top and right of each curve require significant periods of super-Eddington accretion for the curve's seed mass and radiative efficiency. The red dots show \ehylirg\ \hotdogs, the blue dots show \ehylirg\ quasars, and the green dots show resulting quasars at $z>6$ based on \citet[][and references therein]{2013ApJ...778..113B}.
}\label{fig:BH_accretion}
\ifx \apjloutput \undefined
\end{figure}
\else
\end{figure*}
\fi
\ifx \apjloutput \undefined
\begin{figure}
\else
\begin{figure*}
\fi
\epsscale{0.63}
\begin{center}
\plotone{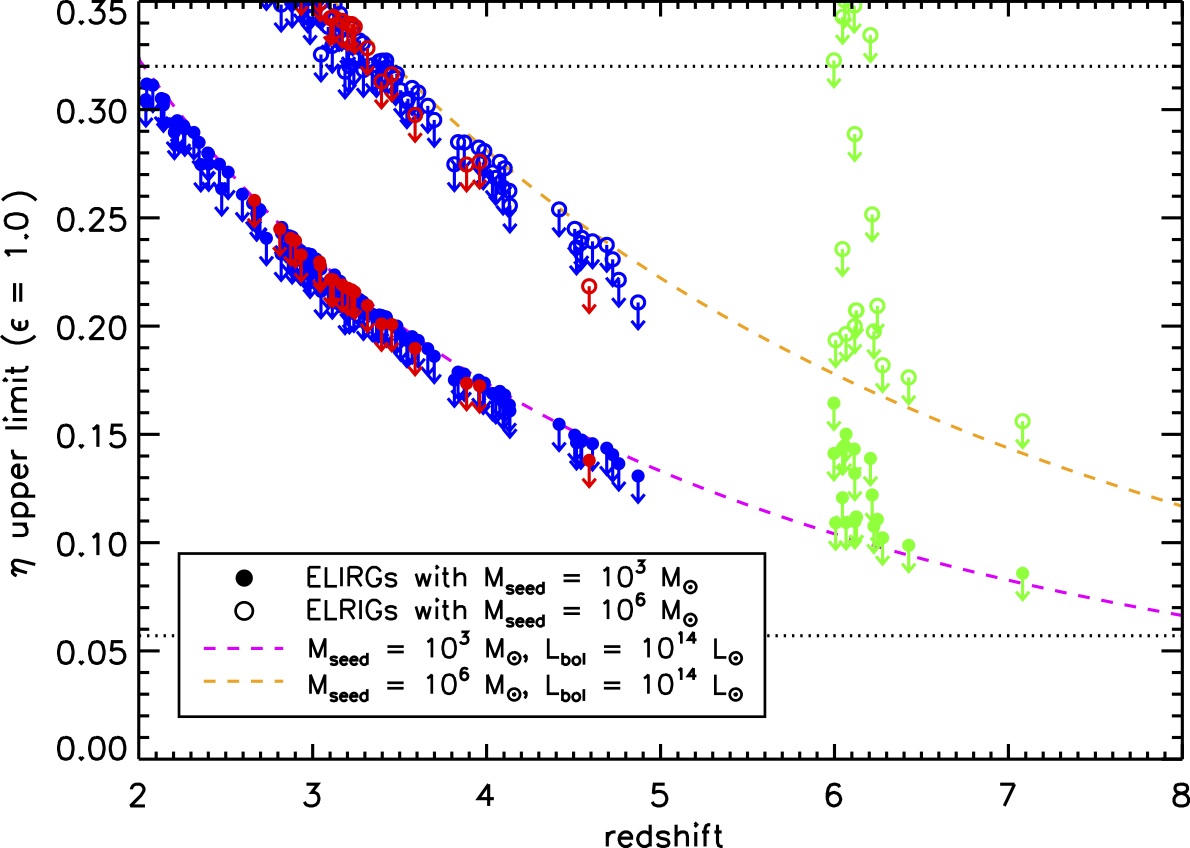}
\end{center}
\caption{Upper limits on radiative efficiency implied by hyperluminous \hotdogs, quasars, and high redshift ($z>6$) quasars (color coded as in Figure \ref{fig:BH_accretion}). Black holes radiating too high a fraction of their accreted mass cannot grow large enough to produce the observed luminosities at the Eddington limit. The filled circles and magenta lines show cases with initial black hole seed mass $M_{\rm seed} = 10^{3}\,M_{\sun}$, and open circles and yellow lines show $M_{\rm seed} = 10^{6}\,M_{\sun}$. The dashed lines outline the luminosity threshold of $L_{\rm bol} > 10^{14}\,L_{\sun}$ used in this paper. The horizontal dotted lines are the theoretically predicted radiative efficiency of a non-spinning black hole ($\eta = 0.057$) and a highly spinning black hole ($\eta = 0.32$). All the lines and dots are plotted assuming a constant Eddington ratio $\epsilon = 1$.
} \label{fig:BH_radiation_efficiency}
\ifx \apjloutput \undefined
\end{figure}
\else
\end{figure*}
\fi
\fi

Figure \ref{fig:BH_accretion} shows $L_{\rm bol}$ vs. $z$ for hyperluminous \hotdogs\ and quasars, and for quasars at $z>6$. 
The curves in the figure show the luminosity vs. redshift tracks followed by black holes as they grow for various choices of $M_{\rm seed}$ and $\eta$. Although in reality the Eddington ratio $\epsilon$ is likely to vary, the overall averaged Eddington ratio $\bar{\epsilon}$ should be lower than unity. With the assumption of $\epsilon \sim 1$ constantly, we consider cases with $\eta \sim 0.1$, a commonly adopted value for a slowly spinning or non-spinning black hole, $\eta \sim 0.2$ for an intermediate spinning black hole, and $\eta \sim 0.3$ for a rapidly spinning black hole.  The radiative efficiency is a factor of 3 higher for a Kerr black hole because the material can still radiate gravitational energy to the last stable circular orbit, which is 3 times smaller compared to a non-spinning Schwarzschild black hole or a slowly spinning black hole \citep{1974ApJ...191..507T}. The curves on Figure \ref{fig:BH_accretion} represent the theoretical limits of black hole mass growth, and the objects toward the upper right of the curves can not be produced with the given initial seed mass and radiative efficiency unless the SMBH is in a super-Eddington state for a significant period of its history. 

If $\eta \sim$ 0.2--0.3 is adopted, the higher radiative efficiency means that the SMBH accumulates less mass to produce the same luminosity. Assuming black hole merging is not important, black holes with $\eta = 0.3$ can not produce the observed \elirgs\ and $z>6$ quasars unless their seed mass $M_{\rm seed} \gg 10^{4}\,M_{\sun}$ or/and they have been accreting at super-Eddington rates since their formation in the early Universe. For the most luminous \hotdogs\ and quasars, the required seed mass is $M_{\rm seed}>10^{7}\,M_{\sun}$, higher than the most massive seed masses predicted by current models \citep{2008MNRAS.387.1649B,2012MNRAS.425.2854A,2013ApJ...778..178H}. On the other hand, for $\eta = 0.1$ (dotted lines in Figure \ref{fig:BH_accretion}), relatively modest seed black holes can grow to \hotdogs\ if they are accreting at $\epsilon({z})>0.4$. If the seed mass exceeds $10^{4}\,M_{\sun}$, even accretion rates $\epsilon < 0.1$ can appropriately create hyperluminous quasars at $z>6$, and \ehylirgs\ at $z< 5$. 

Under the conservative assumption of constant accretion at the Eddington limit since $z=20$, the existence of hyperluminous \hotdogs\ and quasars at $z>3.5$ implies a constraint on the upper limit to the radiative efficiency for given seed black hole mass values. In Figure \ref{fig:BH_radiation_efficiency}, we show the upper limits of radiative efficiency in the cases of $M_{\rm seed} = 10^{3}\,M_{\sun}$ and $M_{\rm seed} = 10^{6}\,M_{\sun}$. The first case represents the upper bound on the predicted black hole seed mass from the first stars \citep[][and references therein]{2014ApJ...781...60H}. The $10^{6}\,M_{\sun}$ case is upper bound of the seed mass from direct collapse of pristine gas clouds \citep{2012MNRAS.425.2854A}, or from rapid mass accretion onto primordial massive stars \citep{2013ApJ...778..178H}. The hyperluminous \hotdogs\ and quasars at $z>3.5$ place similar constraints on black hole radiative efficiency $\eta$ as $z>6$ quasars. If the black hole seeds have masses of $\sim 10^{3}\,M_{\sun}$, we expect that SMBHs have radiative efficiency on average lower than 15\% to form the \elirgs\ and hyperluminous quasars at $z>4$. If a higher seed mass is adopted, the upper limits of $\eta$ are still $<$ 25\%, which corresponds to the radiative efficiency of a mildly rotating black hole. This suggests that the SMBHs in these most luminous systems are either (1) born with high mass \citep[as discussed by][for quasars at $z>7$]{2013ApJ...771..116J}, (2) experience substantial super-Eddington accretion episodes in their growth history \citep{2013ApJ...764...45K}, or (3) have sustained lower radiation feedback to accretion due to lower radiation efficiency for slowly spinning black holes. In the latter case, some mechanism, such as accretion with randomized directions \citep{2006MNRAS.373L..90K,2008MNRAS.385.1621K,2012MNRAS.419.2797F}, has to interrupt the increase of black hole spin by angular momentum transported from a regulated accretion disk, otherwise a black hole can be spun up close to the theoretical limit over a few Salpeter timescales. 

\subsection{High Luminosity State Time Scale}

Analysis of the black hole mass function at $z=0$ suggests that SMBHs spend $\sim$ 1\% of their lifetime in a luminous, high accretion rate mode, and 99\% in a dim, low accretion phase \citep{2006ApJ...643..641H}. The high accretion phase dominates the BH mass growth. For \elirgs, the accretion rate $\dot{M}$ required to radiate at the \elirg\ luminosity level is $\dot{M} = L/(\eta c^{2}) \sim (7/\eta)\,M_{\sun}\,{\rm yr}^{-1}$. For the typical $\eta = 0.1$ radiative efficiency of a slow-spinning black hole, $\dot{M} \sim 70 M_{\sun}\,{\rm yr}^{-1}$.

The minimum lifetime of the high luminosity phase in hyperluminous \hotdogs\ can be estimated by considering the depletion time of the observed dusty material. The peak of the \elirg\ SEDs suggests that the luminosity is dominated by radiation from hot dust at $T_{d} \sim 450 $\,K. At that temperature, the required dust mass to produce the observed $L_{\rm MIR} \sim 8 \times 10^{13}\,L_{\sun}$ is on the order of 2700\,$M_{\sun}$. This dust mass is just $\sim$ 40 times the mass annually accreted by the SMBH in \hotdogs. This timescale is shorter than the light-crossing time scale of 200\,yr discussed in Section \ref{sec:dust_distribution}.

On the other hand, \cite{assef_opt} have studied the luminosity functions of \hotdogs\ and luminous quasars, finding they have comparable number density. This suggests similar lifetimes for both the obscured and the unobscured phases of hyperluminous black hole accretion. Using the life cycle of broad line quasars at $z=1$ \citep{2010ApJ...719.1315K} as an analog, this time scale is $\sim$ 100\,Myr, likely to be the upper bound for \hotdog\ phase. This timescale is a few times the Salpeter timescale. The accreted mass onto the SMBH exceeds $7\times 10^9\,M_{\sun}$. 

\subsection{A Luminosity Limit at $L_{\rm bol} \sim 10^{14.5}\,L_{\sun}$?}

Our systematic search for \elirgs\ using the ``$W1$$W2$-dropout'' selection criteria has identified about 1000 candidates in the \textit{WISE} database \citep{2012ApJ...755..173E}. Among the 150 of these candidates from which we have redshift information, a total of 20 \elirgs\ at $z>2.5$ have been discovered as reported in this paper, including the most luminous system, W2246$-$0526 at $z=4.593$ with $L_{\rm bol} \sim 3.5 \times 10^{14}\,L_{\sun}$. 

Are there other infrared objects, either starbursts or obscured AGNs, with similarly high luminosities? Recently, 1HERMES X24 J161506.65+543846.9 at $z=4.952$ was discovered by \Herschel\ HerMES survey \citep{2012ApJ...761..139C}. Its $L_{\rm IR}$ is reported to be $\sim 1.2 \times 10^{14}\,L_{\sun}$ based on an SED model with $T_{\rm dust} = 98$\,K. It is the brightest object in their sample, and tentatively classified as a starburst system based on its optical spectrum \citep{2012ApJ...761..139C}. Our conservative estimate using \wise\ and \Herschel\ photometry would imply its $L_{\rm bol}$ is $\sim 8 \times 10^{13}\,L_{\sun}$, slightly shy of our $L_{\rm bol}$ threshold, but likely making it one of the most powerful starburst systems known. In addition to \Herschel\ studies, a few systems discovered by \wise\ using different selection criteria have $L_{\rm bol}$ at the $\sim 10^{14}\,L_{\sun}$ level \citep[][]{lonsdale_NVSS_WISE,stern_w1819}, but they all show obvious AGN features in their spectra or have radio emission.

The $L_{\rm bol} \sim 10^{14.5}\,L_{\sun}$ of W2246$-$0526 exceeds the most luminous quasars listed in Table \ref{table:properties_qso} (see Figure \ref{fig:lum_distribution}). 
Could there be even more luminous \hotdogs\ ? 

In Figure \ref{fig:BH_accretion}, a few quasars at $z>6$ have implied black hole mass $\gtrsim 10^{9}\,M_{\sun}$ at an age of $<$ 1.0 Gyr. If they accrete at the Eddington limit, they could potentially reach $> 10^{15}\,L_{\sun}$ by $z\sim 5$. Thus far, we have not observed any system at that $L_{\rm bol}$ level. Such systems are expected to be rare: no more than a few over the whole sky based on the luminosity function of \hotdogs\ and optical quasars. At $z>5$, the hot dust emission peak at 6 \micron\ will shift to $\sim$ 40 \micron, well beyond the \wise\ 22 \micron\ ($W4$) filter. To identify such sources would require deeper imaging at 30--70 \micron\ from future missions such as \textit{SPICA}, or submillimeter photometry at 200 \micron\ -- 1 mm with ALMA or CCAT. As noted by \cite{assef_opt}, such objects may also have been detected by \wise\ but failed to meet the $W1$$W2$-dropout selection, because they will be detectable by \wise\ at 3.4 and 4.6 \micron. If we relax the $W1$ and $W2$ flux limits for \hotdogs, many contaminants fall into the selection criteria, making it much more difficult to identify the \elirg\ systems. It is also possible that AGN at $\sim 10^{14.5}\,L_{\sun}$ have reached a physical limit for BH accretion, or that the accreting material is depleted after $z \sim 5$. If so, this might explain the upper luminosity bound at $10^{14.5}\,L_{\sun}$ for hyperluminous \hotdogs\ and quasars in Figure \ref{fig:BH_accretion}.

\section{Summary}\label{sec:summary}
We report 20 highly obscured, \ehylirg\ AGNs discovered by \wise. These sources, because of their similarity to DOGs with steeply rising SEDs towards the mid-IR albeit with hotter dust components, have been dubbed \hotdogs\ by \cite{2012ApJ...756...96W}. The luminosities of these objects exceed $10^{14}\,L_{\sun}$, making them among the most luminous (non-transient) systems in the universe. They are not likely to be powered by starbursts, but rather by highly obscured and actively accreting AGNs. Based on their lack of variability and the absence of evidence for foreground lensing systems in their spectra and images, we conclude that the high luminosity is generally not a result of relativistic beaming or gravitational lensing.

We present the full SED of these objects from observed optical to far-IR wavelengths. For both hyperluminous \hotdogs\ and unobscured quasars of similarly high luminosity, we show that the conversion from monochromatic mid-IR luminosity to total $L_{\rm bol}$\ can vary by a factor of $\sim$ 3. Thus, the use of single-wavelength $\nu L_{\nu}$ values such as  $L_{\rm 5.8\,\mu \rm m}$ and $L_{\rm 7.8\,\mu \rm m}$\ to represent \Lbol\ needs to be reevaluated considering dust obscuration and thermal dust emission. 

Based on the \hotdog\ SEDs, we suggest that emission from a dust component at $T_{d} \sim 450\,K$ contributes the majority of the luminosity. This dust component, with a characteristic radius of $\sim$ 40 pc, contains 2700\,$M_{\sun}$. Because of its physical size, we expect that large flux variations in these \hotdogs\ should not occur on timescales $\lesssim 200$ yr.  

The high luminosities in the \elirg\ \hotdogs\ are likely maintained by $M_{\rm BH} \sim 3 \times 10^{9}\,M_{\sun}$ SMBHs accreting near the Eddington limit. Their existence at redshift $2.8 < z < 4.6$ implies a time-averaged Eddington ratio of $>$ 25\% up to their observed epochs. It would be difficult for these SMBHs to grow to $10^{9}\,M_{\sun}$ with the relatively high radiative efficiency $\epsilon \sim 0.3$ expected from a Kerr black hole. This suggests that the spin of SMBHs in \hotdogs\ may be low, perhaps as a result of chaotic accretion due to galaxy merger events.

\textbf{\textit{Note added in proof.--}} After this paper was submitted, we were alerted to the discovery of SDSS J0010$+$2802 at $z=6.3$ with $M_{\rm BH} \sim 1.2 \times 10^{10} M_{\sun}$ \citep{2015Natur.518..512W}. Using our methodology, the luminosity of this source is $1.6 \times 10^{14}\,L_{\sun}$.

\acknowledgments
The authors thank the anonymous referee for the constructive comments and for encouraging a more thorough discussion of gravitational lensing in this paper. This publication makes use of data products from the {\it Wide-field Infrared Survey Explorer}, which is a joint project of the University of California, Los Angeles, and the Jet Propulsion Laboratory/California Institute of Technology, and {\it NEOWISE}, which is a project of the Jet Propulsion Laboratory/California Institute of Technology. {\it WISE} and {\it NEOWISE} are funded by the National Aeronautics and Space Administration. This work is also based in part on observations made with the \textit{Spitzer Space Telescope}, which is operated by the Jet Propulsion Laboratory, California Institute of Technology under a contract with NASA. Some of the data presented herein were obtained at the W.M. Keck Observatory, which is operated as a scientific partnership among the California Institute of Technology, the University of California and the National Aeronautics and Space Administration. The Observatory was made possible by the generous financial support of the W.M. Keck Foundation. Part of this research has made use of the Keck Observatory Archive (KOA), which is operated by the W. M. Keck Observatory and the NASA Exoplanet Science Institute (NExScI), under contract with the National Aeronautics and Space Administration. This research has made use of the NASA/IPAC Infrared Science Archive and the NASA/IPAC Extragalactic Database (NED), which are operated by the Jet Propulsion Laboratory, California Institute of Technology, under contracts with the National Aeronautics and Space Administration. This material is based upon work supported by the National Aeronautics and Space Administration under Proposal No. 13-ADAP13-0092 issued through the Astrophysics Data Analysis Program. C.-W. T. was supported by an appointment to the NASA Postdoctoral Program at the Jet Propulsion Laboratory, administered by Oak Ridge Associated Universities through a contract with NASA. R. J. A. was supported by Gemini-CONICYT grant number 32120009.

{\it Facilities:} \facility{{\it WISE}}, \facility{{\it Spitzer} (IRAC)}, \facility{{\it Herschel} (PACS, SPIRE)}, \facility{{Keck:II} (NIRC2)}.


\def\nar{NewAR}
\def\arxiv{arXiv}


\ifx \apjloutput \undefined

\begin{figure}
\epsscale{1.0}
\begin{center}
\plotone{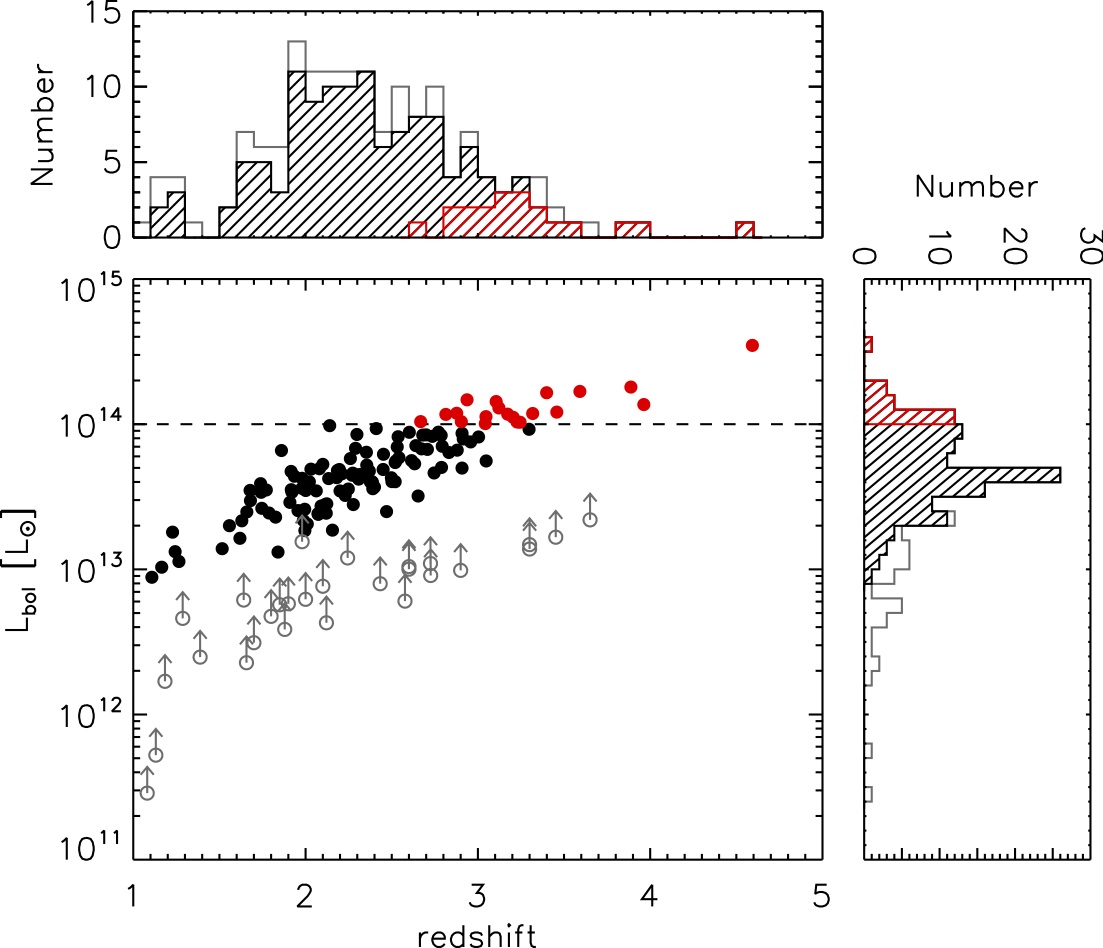}
\end{center}
\caption{The redshift and bolometric luminosity, $L_{\rm bol}$,
of \wise\ Hot DOGs at $1 < z < 5$. The luminosity distribution of the confirmed \hotdogs\ is on the right, and the redshift distribution is on the top. The black dots and black shaded regions represent \hotdogs\ with \Herschel\ measurements, while the gray open circles and open histograms indicate the lower luminosity limits for \hotdogs\ without far-IR data. The red points and histogram are for the sample of objects presented in this paper which exceed the $L_{\rm bol} > 10^{14}\,L_{\sun}$ threshold, shown by the horizontal dashed line.}\label{fig:hotdog_z_nuLnu}
\end{figure}

\clearpage
\begin{figure}
\epsscale{1.0}
\begin{center}
\plotone{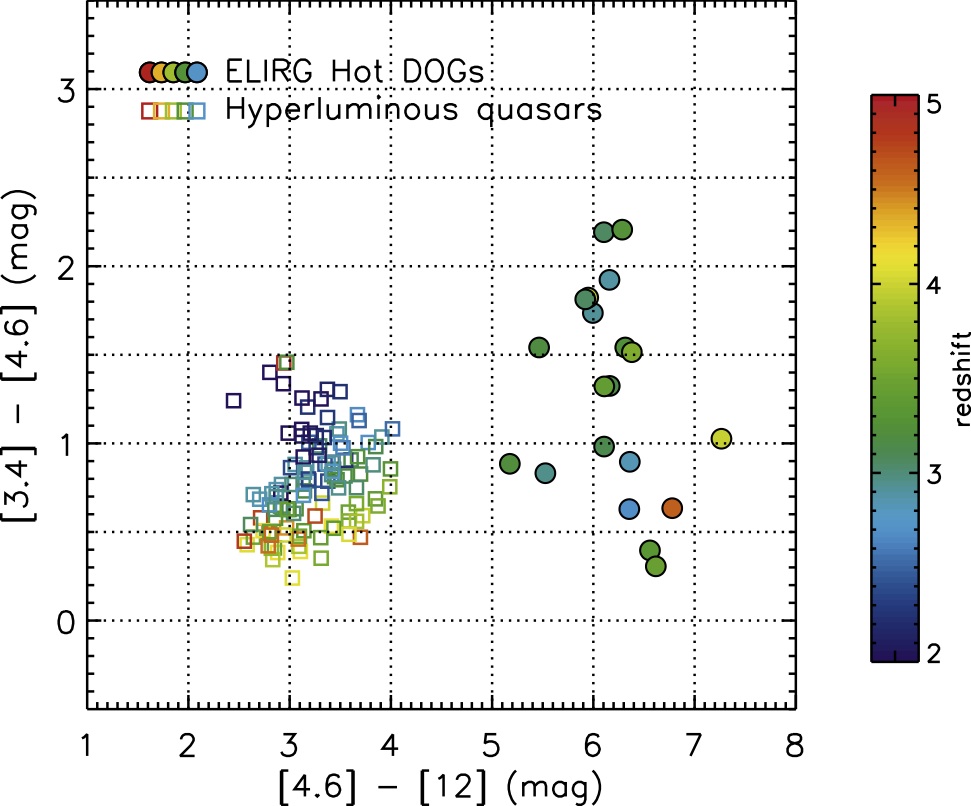}
\end{center}
\caption{Mid-IR colors of the hyperluminous ($L_{\rm bol} > 10^{14}\,L_{\sun}$) \hotdogs\ and optically selected hyperluminous quasars. The open squares show the 116 selected quasars; the filled circles show the 20 \wise-selected \hotdogs. For \hotdogs\ not detected in \WISE\ [3.4] and [4.6], their IRAC measurements are plotted with the color correction from IRAC [3.6] to \WISE\ [3.4] and IRAC [4.5] to \WISE\ [4.6] applied.
}\label{fig:color_color}
\end{figure}

\clearpage
\ifx \apjloutput \undefined
\begin{figure}
\else
\begin{figure*}
\fi
\ifx \apjloutput \undefined
\epsscale{0.75}
\else
\epsscale{0.75}
\fi
\begin{center}
\plotone{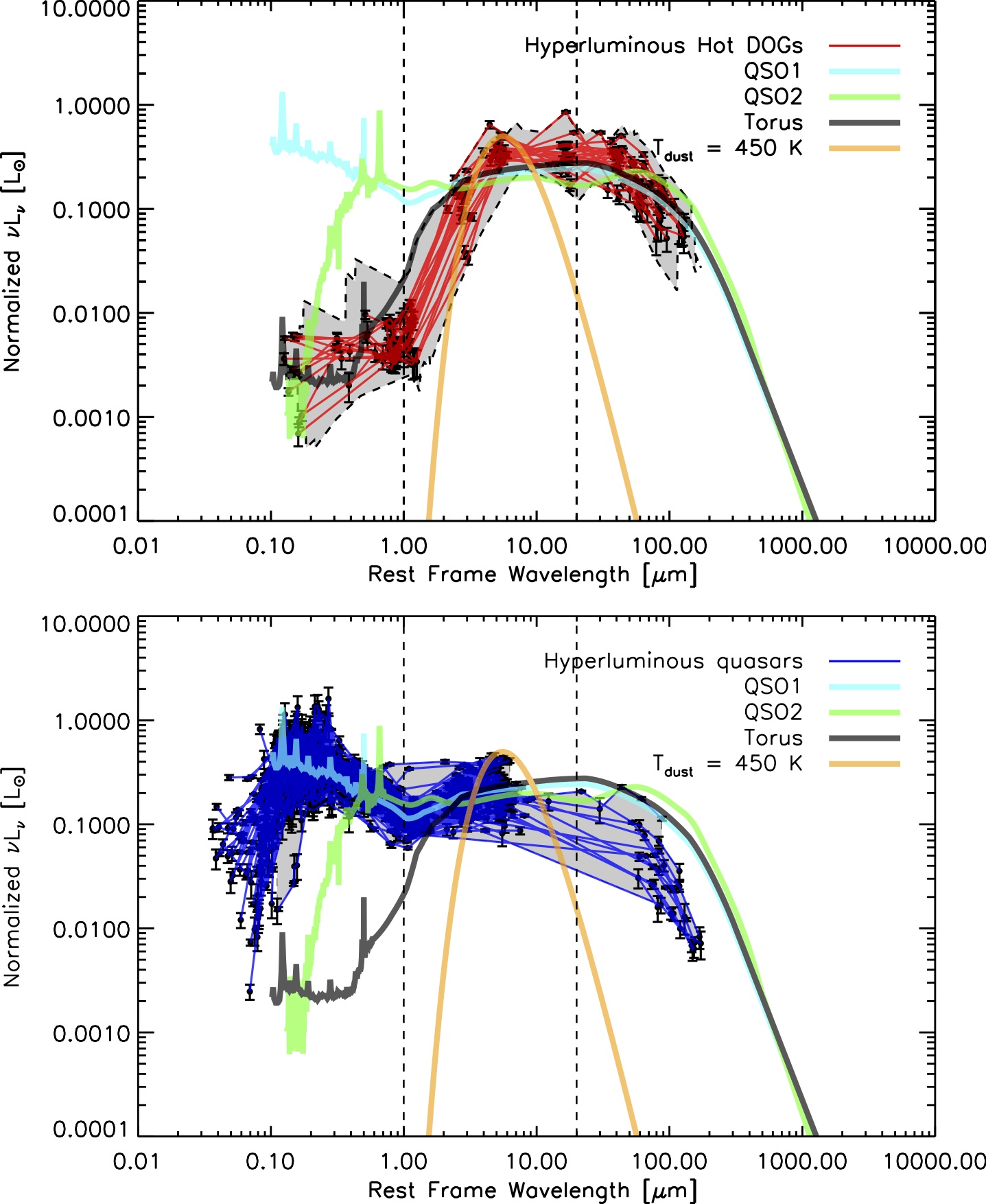}
\end{center}
\caption{Normalized rest-frame SEDs of the hyperluminous ($L_{\rm bol} > 10^{14}\,L_{\sun}$) 
\ehylirg\ \hotdogs\ and optically-selected hyperluminous quasars. The SEDs are normalized to the total bolometric luminosity $L_{\rm bol}$. The shaded region in gray in the upper panel represents the scatter of SEDs for all \hotdogs\ with $z > 1.6$ and $L_{\rm bol} > 10^{13}\,L_{\sun}$, while the gray region in the lower panel covers the scatter of SEDs for the hyperluminous quasar sample. The plotted QSO and torus SED models are adopted from \cite{2006ApJ...642..673P,2007ApJ...663...81P}. The dust model line 
assumes a dust temperature $T_d = 450$ K with emissivity index $\beta = 2.0$. The vertical dashed lines indicate rest-frame wavelengths of 1 and 20\micron.}\label{fig:normalized_SEDs}
\ifx \apjloutput \undefined
\end{figure}
\else
\end{figure*}
\fi
\clearpage
\begin{figure}
\ifx \apjloutput \undefined
\epsscale{0.65}
\else
\epsscale{1.0}
\fi
\begin{center}
\plotone{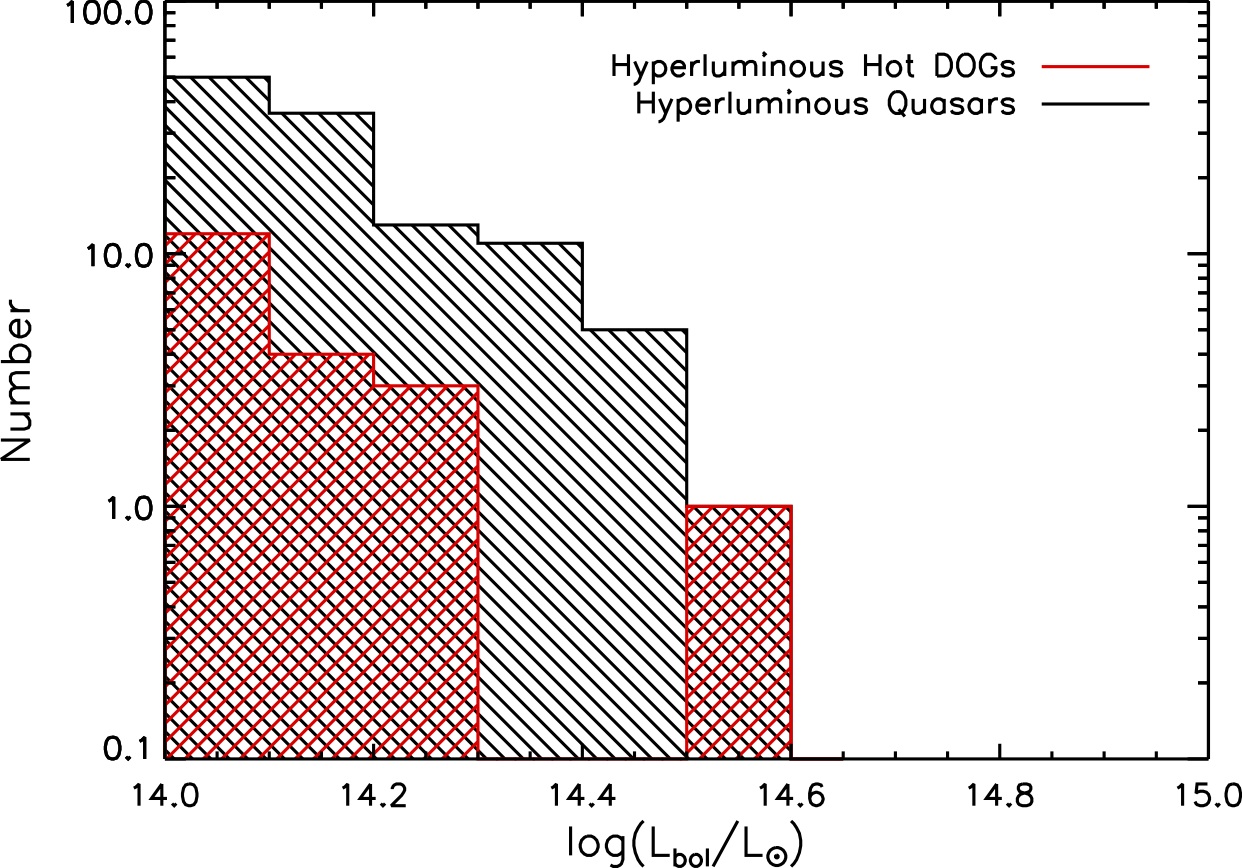}
\end{center}
\caption{Luminosity histogram of $L_{\rm bol} > 10^{14}\,L_{\sun}$  \wise\ \hotdogs\ (in red) and optically selected quasars (in black). There are 20 \hotdogs\ and 116 quasars in this plot. 
}\label{fig:lum_distribution}
\end{figure}

\clearpage
\ifx \apjloutput \undefined
\begin{figure}
\epsscale{0.8}
\else
\begin{figure*}
\epsscale{0.7}
\fi
\begin{center}
\plotone{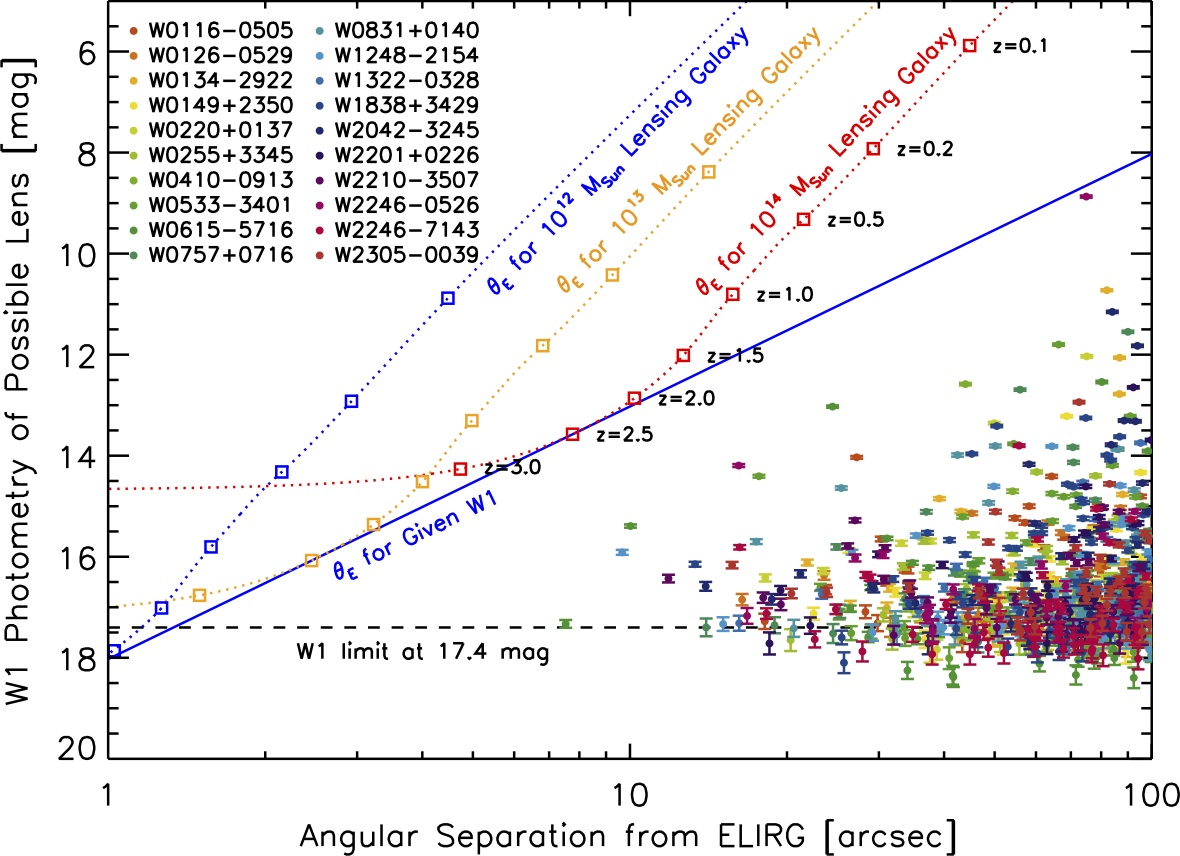}
\end{center}
\caption{
Photometry and angular separation of \WISE\ $W1$ sources within 100\arcsec of the \elirgs. Data points are color-coded by individual \elirg\ as shown in the legend. The dashed black line at $W1 = 17.4$ mag shows the original selection limit for WISE Hot DOGs. The dotted lines show the Einstein radius (\thetaE) and $W1$ magnitude \citep[assuming $M/L_B \sim 5$ and using the elliptical galaxy SED template from][]{2007ApJ...663...81P} for lensing elliptical galaxies with different masses as labeled. Open squares along each dotted line mark representative lensing galaxy redshifts. The solid blue line represents the maximum \thetaE\ vs. $W1$ for a source at $z = 3.2$, the median redshift of the 20 \WISE\ \elirgs, which occurs for a lens redshift of $z \sim 2.5$. \WISE\ sources below the solid blue line are too faint and have too large an angular separation to be lenses. 
}\label{fig:GLens_resolved}
\ifx \apjloutput \undefined
\end{figure}
\else
\end{figure*}
\fi
\clearpage
\begin{figure}
\ifx \apjloutput \undefined
\epsscale{0.7}
\else
\epsscale{1.1}
\fi
\begin{center}
\plotone{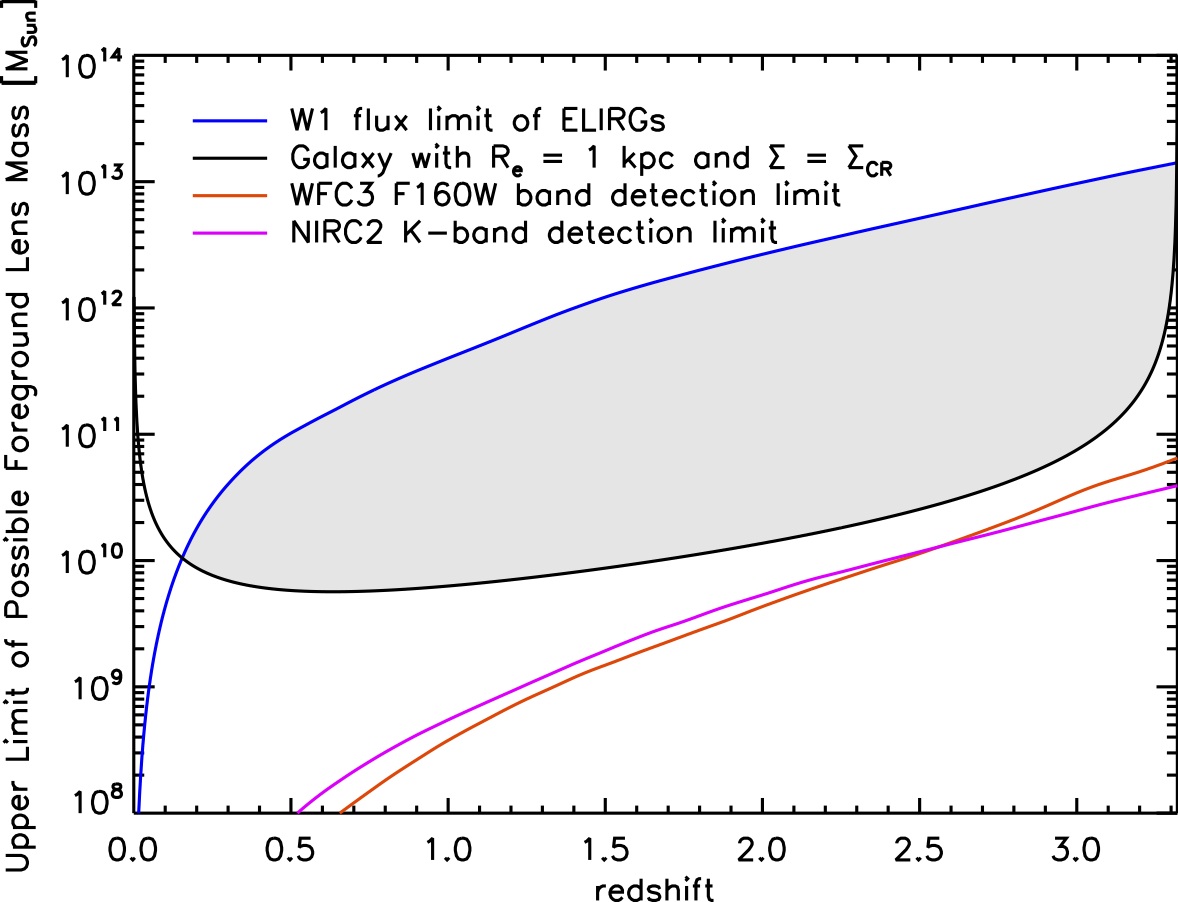}
\end{center}
\caption{
Mass limit of unresolved lenses vs. redshift, assuming a source at $z = 3.2$. The blue curve shows the upper limit due to the brightest $W1$ photometry ($W1 =16.8$) observed for \elirgs. The red-orange and magenta curves show the 3-$\sigma$ detection limit of our high resolution imaging with \HST\ WFC3 in F160W, and Keck NIRC2 with AO in the $K$-band. These curves are calculated using elliptical galaxy SED template from \cite{2007ApJ...663...81P} with assumption of $M/L_{B} \sim 5$. The solid black line shows the mass of a lensing galaxy with size $R_e \sim 1$ kpc and surface mass density $\Sigma$ equals to $\Sigma_{\rm CR}$, the critical surface mass density necessary for gravitational lensing. The shaded region indicates the remaining lens galaxy mass parameter space.
}\label{fig:GLens_unresolved}
\end{figure}

\clearpage
\begin{figure}
\ifx \apjloutput \undefined
\epsscale{0.5}
\else
\epsscale{0.9}
\fi
\begin{center}
\plotone{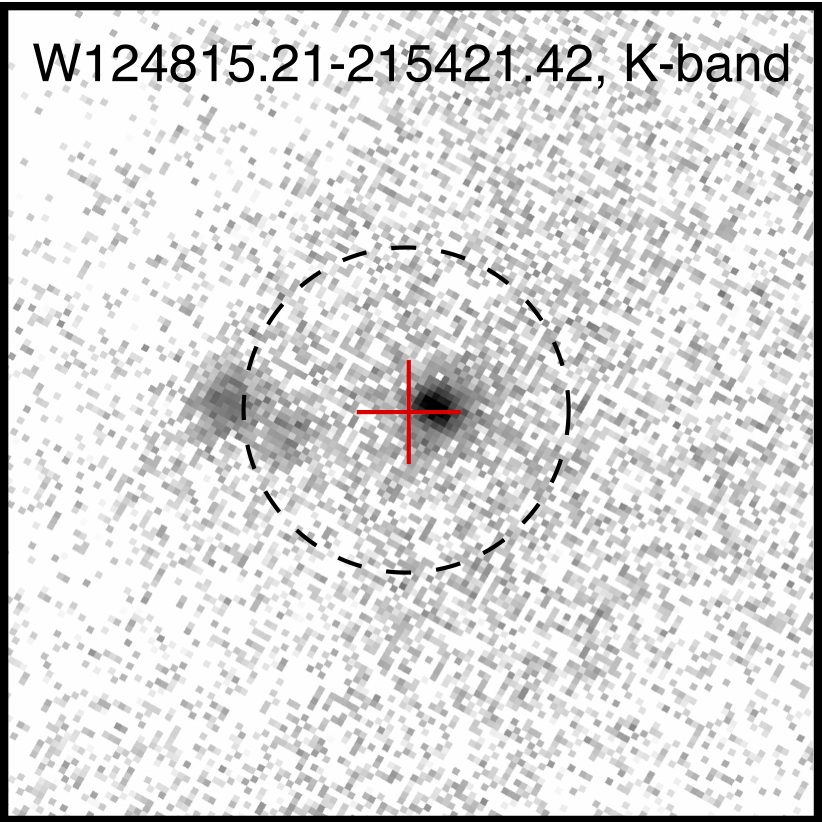}
\end{center}
\caption{
A 5\arcsec $\times$ 5\arcsec\ subsection of the Keck NIRC2 image of W1248$-$2154 in the $K$-band. The PSF FWHM is $\sim$ 0\farcs14. The red cross shows the position of W1248$-$2154 from the AllWISE Catalog and the 0\farcs32 uncertainty, which includes 0\farcs2 registration uncertainty (1\,$\sigma$) of the $K$-band image and the $\sim$ 0\farcs25 position uncertainty of W1248$-$2154 in AllWISE images. The black dashed circle indicates a $r=1\arcsec$ region. The object on the right of the red cross is at $z=3.326$, and the source on the left lies just outside the $r=1\arcsec$ circle is at $z=0.339$. 
}\label{fig:w1248_nirc2}
\end{figure}

\clearpage
\begin{figure}
\ifx \apjloutput \undefined
\epsscale{0.65}
\else
\epsscale{1.0}
\fi
\begin{center}
\plotone{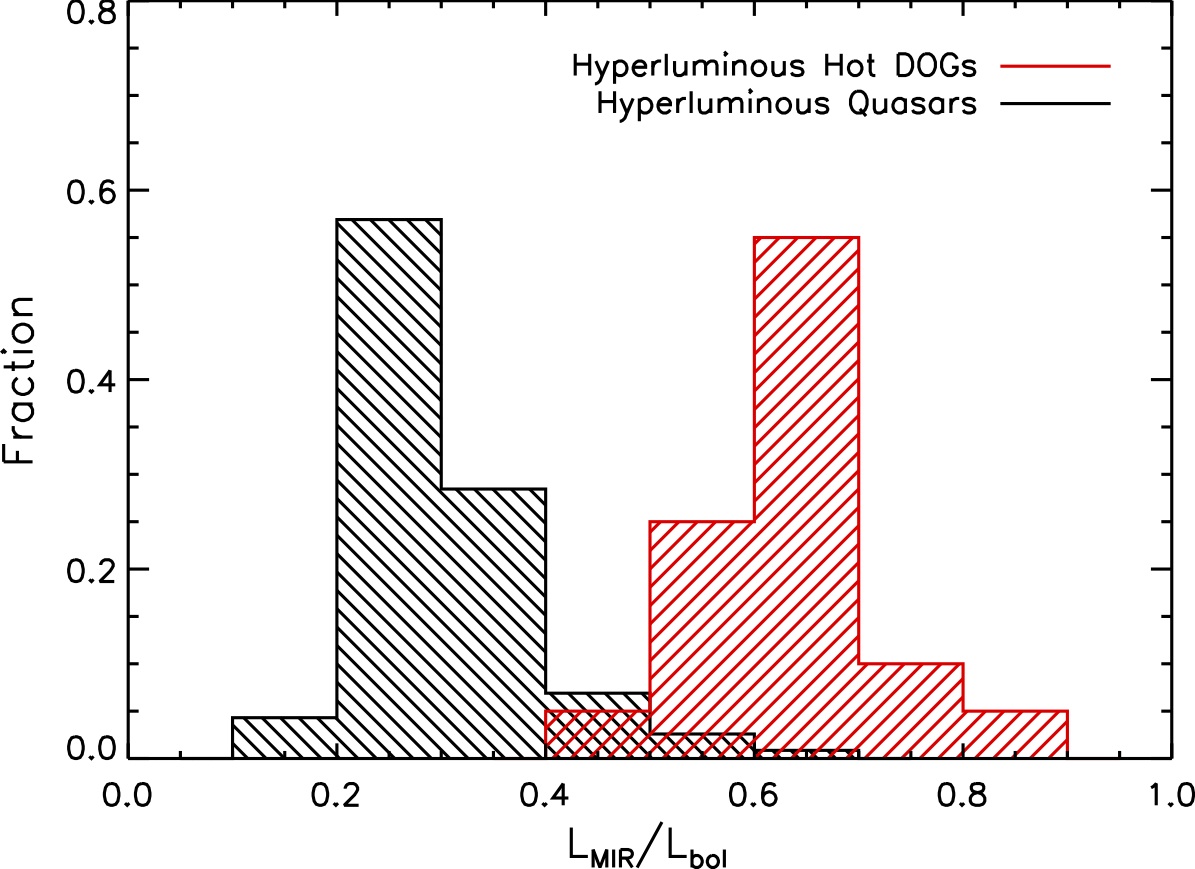}
\end{center}
\caption{Normalized histogram of $L_{\rm MIR} / L_{\rm bol}$ for \wise\ \hotdogs\ and optically selected quasars with $L_{\rm bol} > 10^{14}\,L_{\sun}$, where $L_{\rm bol}$ is the total integrated luminosity calculated using power-law-connected SEDs and $L_{\rm MIR}$ is the integrated luminosity from 1 to 20\,\micron.}\label{fig:Rmir}
\end{figure}

\clearpage
\ifx \apjloutput \undefined
\begin{figure}
\else
\begin{figure*}
\fi
\epsscale{1.0}
\begin{center}
\plottwo{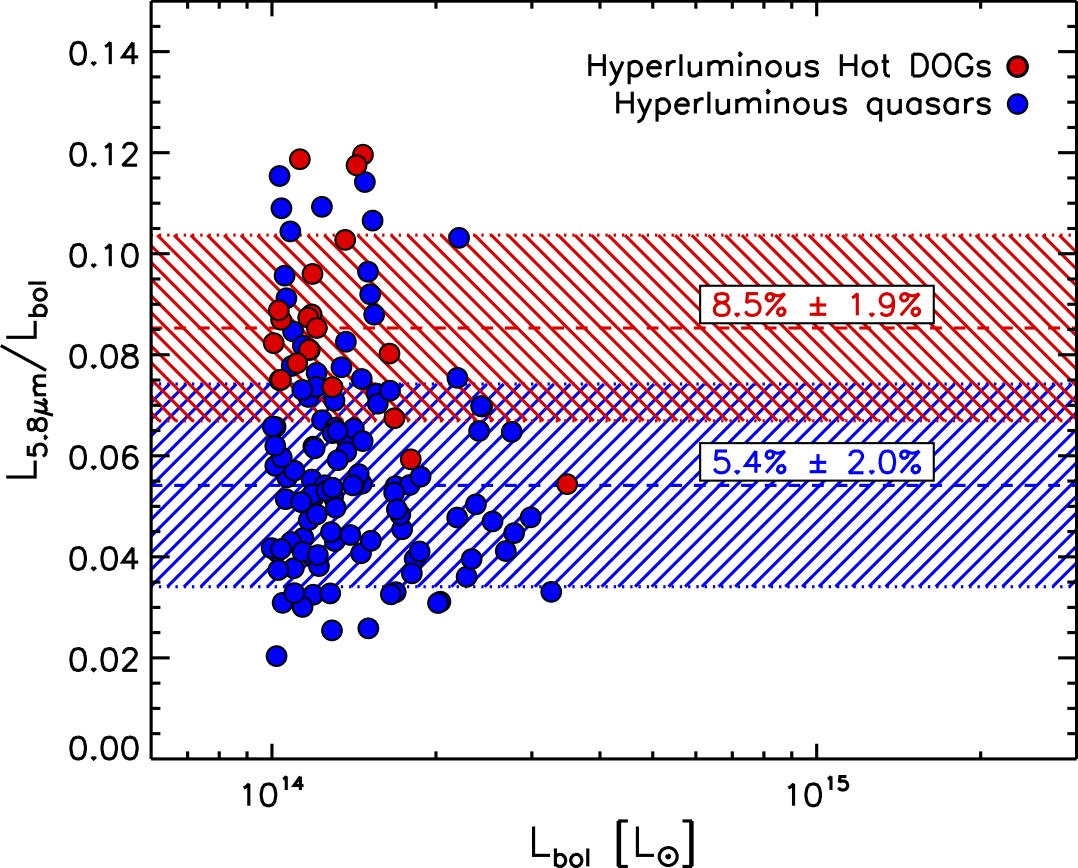}{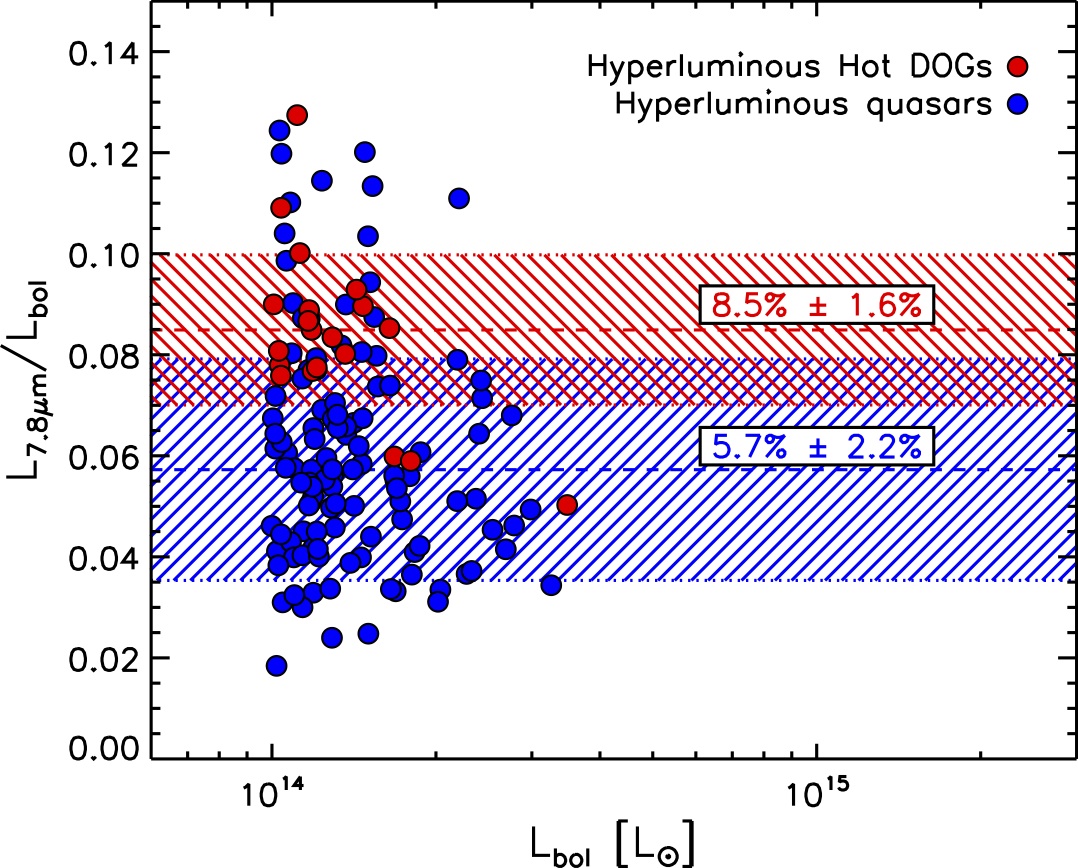}
\end{center}
\caption{Ratio of $L_{\rm 5.8\micron}$ to $L_{\rm bol}$ (\textit{\textbf{left}}) and $L_{\rm 7.8\micron}$ to $L_{\rm bol}$ (\textit{\textbf{right}}) of hyperluminous \hotdogs\ and quasars. The dashed lines represent the median values of luminosity ratios for \hotdogs\ (red), and quasars (blue). The hatched regions show a $1\,\sigma$ range from the median value. 
}\label{fig:Lbol_ratio}
\ifx \apjloutput \undefined
\end{figure}
\else
\end{figure*}
\fi

\clearpage
\ifx \apjloutput \undefined
\begin{figure}
\else
\begin{figure*}
\fi
\epsscale{0.7}
\begin{center}
\plotone{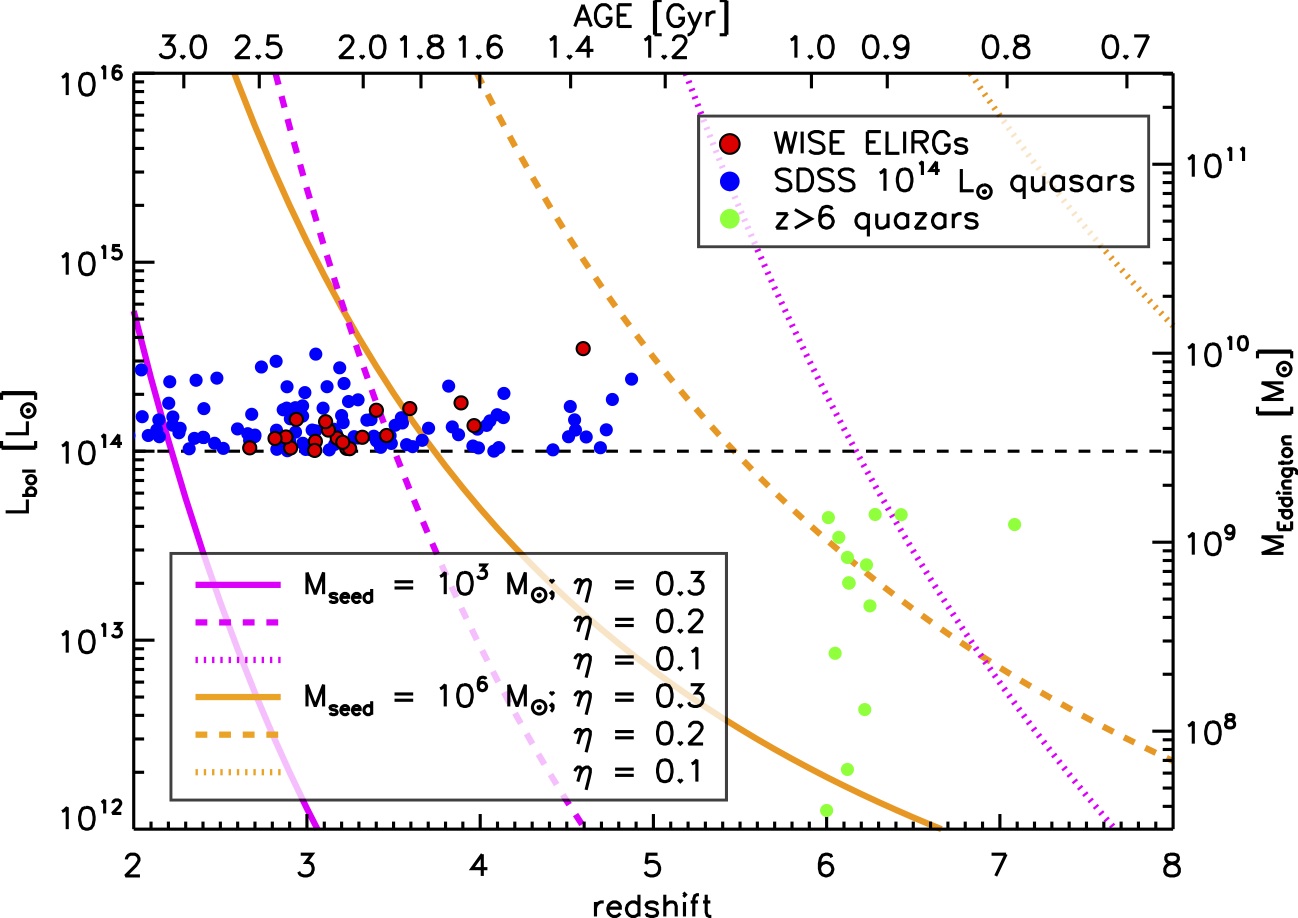}
\end{center}
\caption{Bolometric luminosity $L_{\rm bol}$ vs. redshift. The age of the SMBH since formation at $z \sim 20$ is plotted on the top. The black hole mass for an Eddington ratio of one ($\bar{\epsilon} = 1$) is plotted on the right. The dotted, dashed, and solid curves are the black hole accretion history between $z=8$ and $z=2$ for $\bar{\epsilon} = 1$ and radiative efficiency $\eta = 0.1$, 0.2, and 0.3, respectively. The magenta lines show predictions with initial black hole mass $M_{\rm seed} = 10^{3}\,M_{\sun}$, while yellow lines show the case of $M_{\rm seed} = 10^{6}\,M_{\sun}$. The regions to the top and right of each curve require significant periods of super-Eddington accretion for the curve's seed mass and radiative efficiency. The red dots show \ehylirg\ \hotdogs, the blue dots show \ehylirg\ quasars, and the green dots show resulting quasars at $z>6$ based on \citet[][and references therein]{2013ApJ...778..113B}.
}\label{fig:BH_accretion}
\ifx \apjloutput \undefined
\end{figure}
\else
\end{figure*}
\fi
\clearpage
\ifx \apjloutput \undefined
\begin{figure}
\else
\begin{figure*}
\fi
\epsscale{0.63}
\begin{center}
\plotone{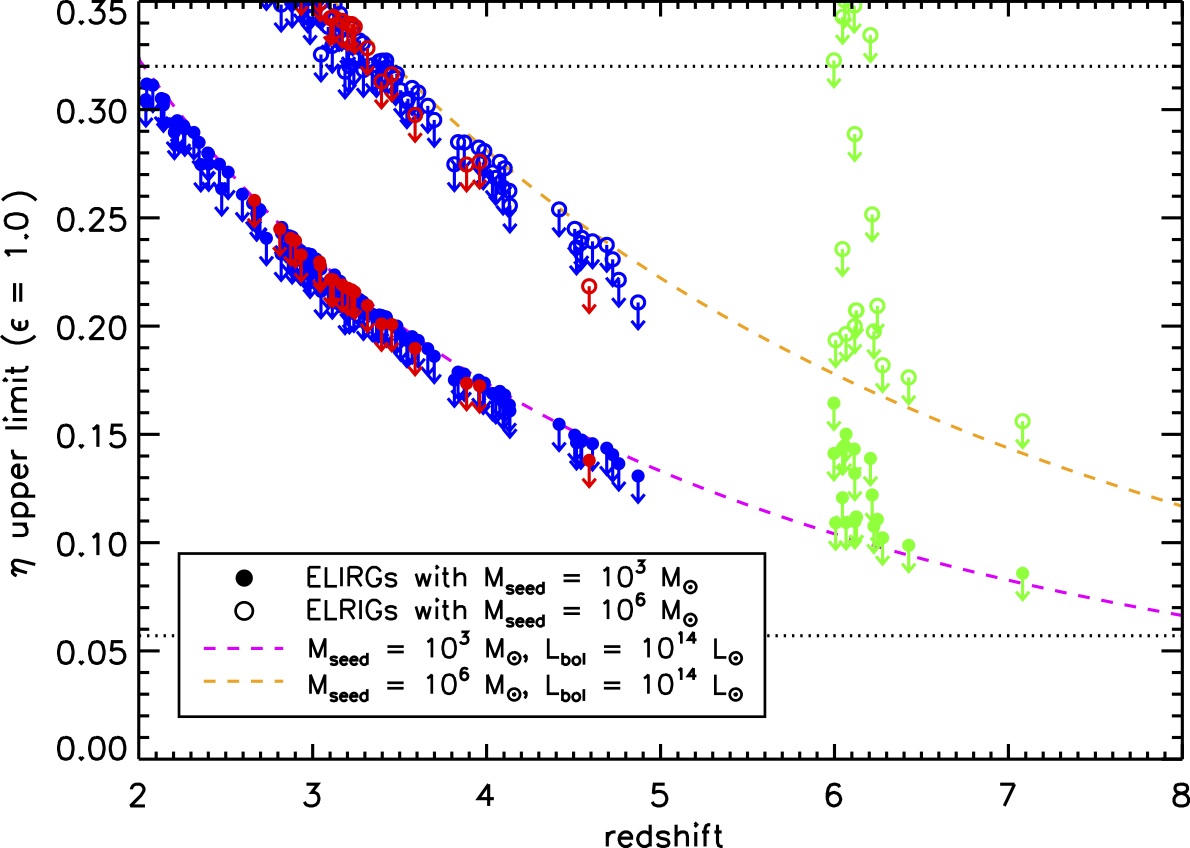}
\end{center}
\caption{Upper limits on radiative efficiency implied by hyperluminous \hotdogs, quasars, and high redshift ($z>6$) quasars (color coded as in Figure \ref{fig:BH_accretion}). Black holes radiating too high a fraction of their accreted mass cannot grow large enough to produce the observed luminosities at the Eddington limit. The filled circles and magenta lines show cases with initial black hole seed mass $M_{\rm seed} = 10^{3}\,M_{\sun}$, and open circles and yellow lines show $M_{\rm seed} = 10^{6}\,M_{\sun}$. The dashed lines outline the luminosity threshold of $L_{\rm bol} > 10^{14}\,L_{\sun}$ used in this paper. The horizontal dotted lines are the theoretically predicted radiative efficiency of a non-spinning black hole ($\eta = 0.057$) and a highly spinning black hole ($\eta = 0.32$). All the lines and dots are plotted assuming a constant Eddington ratio $\epsilon = 1$.
} \label{fig:BH_radiation_efficiency}
\ifx \apjloutput \undefined
\end{figure}
\else
\end{figure*}
\fi

\else
\fi
\clearpage


\ifx \apjloutput \undefined

\ifx \apjloutput \undefined
\begin{deluxetable}{llllcl}  
\else
\begin{deluxetable*}{llllcl}  
\fi
\tabletypesize{\scriptsize}
\tablenum{3}
\tablewidth{0in}
\tablecaption{Properties of Optically Selected Quasars with $L_{\rm bol} > 10^{14} L_{\sun}$ \textbf{(Full Version)} \label{table:properties_qso}}
\tablehead{
\colhead{Source} &  
\multicolumn{2}{c}{\WISE\ Coordinate} &
 \colhead{Redshift} & 
 \colhead{$L_{\rm bol}$\tablenotemark{a}} &
 \colhead{Redshift Ref.}
\\
\colhead{} & 
  \colhead{R.A. (J2000)} & 
  \colhead{Decl. (J2000)} & 
  \colhead{} & 
  \colhead{($10^{14}\,L_{\sun}$)} & 
  \colhead{}
}
\startdata

\ifx \apjloutput \undefined
J000322.91$-$260316.8 &  00:03:22.91 & $-$26:03:16.8 &  4.098   &   1.6   & NED, V10 \\
J001527.40$+$064012.0 &  00:15:27.40 & $+$06:40:12.0 &  3.17    &   1.2   & V10 \\
J004131.50$-$493612.0 &  00:41:31.50 & $-$49:36:12.0 &  3.24    &   1.8   & V10 \\
J010311.30$+$131618.0 &  01:03:11.30 & $+$13:16:18.0 &  2.681   &   1.6   & NED, V10 \\
J012156.04$+$144823.9 &  01:21:56.03 & $+$14:48:23.9 &  2.870    &   1.1   & S10, V10 \\
J012412.47$-$010049.8 &  01:24:12.47 & $-$01:00:49.7 &  2.826   &   1.0   & S10, P12, V10 \\
J013301.90$-$400628.0 &  01:33:01.90 & $-$40:06:28.0 &  3.023   &   1.0   & V10 \\
J015636.00$+$044528.0 &  01:56:36.00 & $+$04:45:28.0 &  2.993   &   1.0   & V10 \\
J020727.20$-$374156.0 &  02:07:27.20 & $-$37:41:56.0 &  2.404   &   1.2   & V10 \\
J020950.70$-$000506.0 &  02:09:50.71 & $-$00:05:06.4 &  2.850    &   1.2   & V10, S10, P12 \\
J024008.10$-$230915.0 &  02:40:08.10 & $-$23:09:15.0 &  2.225   &   1.4   & V10 \\
J024854.30$+$180250.0 &  02:48:54.30 & $+$18:02:50.0 &  4.42    &   1.0   & V10 \\
J025240.10$-$553832.0 &  02:52:40.10 & $-$55:38:32.0 &  2.35    &   1.2   & V10 \\
J030722.80$-$494548.0 &  03:07:22.80 & $-$49:45:48.0 &  4.728   &   1.3   & V10 \\
J032108.45$+$413220.9 &  03:21:08.45 & $+$41:32:20.8 &  2.467   &   1.1   & S10 \\
J035504.90$-$381142.0 &  03:55:04.90 & $-$38:11:42.0 &  4.545   &   1.5   & V10 \\
J051707.60$-$441056.0 &  05:17:07.60 & $-$44:10:56.0 &  1.713   &   2.5   & V10 \\
J055200.40$-$531244.0 &  05:52:00.40 & $-$53:12:44.0 &  1.59    &   1.3   & V10 \\
J055445.70$-$330517.0 &  05:54:45.70 & $-$33:05:17.0 &  2.36    &   2.4   & V10 \\
J073502.30$+$265911.0 &  07:35:02.30 & $+$26:59:11.5 &  1.972   &   1.2   & V10, S10 \\
J074521.70$+$473436.0 &  07:45:21.78 & $+$47:34:36.1 &  3.214   &   2.3   & V10, DR10, P12, S10 \\
J075054.64$+$425219.2 &  07:50:54.64 & $+$42:52:19.2 &  1.905   &   1.4   & S10, P12, V10 \\
J080117.80$+$521034.6 &  08:01:17.79 & $+$52:10:34.5 &  3.209   &   1.4   & P12, S10, V10 \\
J084323.70$+$165655.0 &  08:43:23.70 & $+$16:56:55.0 &  4.874   &   2.4   & V10 \\
J084401.90$+$050358.0 &  08:44:01.95 & $+$05:03:57.9 &  3.346   &   1.5   & V10, S10 \\
J090033.50$+$421547.0 &  09:00:33.50 & $+$42:15:47.0 &  3.295   &   1.9   & S10, DR10 \\
J090423.37$+$130920.7 &  09:04:23.37 & $+$13:09:20.7 &  2.975   &   1.6   & S10, V10 \\
J094202.00$+$042244.0 &  09:42:02.04 & $+$04:22:44.5 &  3.272   &   1.2   & V10, P12, S10 \\
J094253.60$-$110426.0 &  09:42:53.60 & $-$11:04:26.0 &  3.093   &   1.7   & V10 \\
J094734.20$+$142117.0 &  09:47:34.19 & $+$14:21:16.9 &  3.030    &   1.3   & S10, V10 \\
J094932.27$+$033531.8 &  09:49:32.26 & $+$03:35:31.7 &  4.107   &   1.1   & P12, S10, V10 \\
J095937.12$+$131215.4 &  09:59:37.11 & $+$13:12:15.4 &  4.056   &   1.5   & S10, V10 \\
J100711.80$+$053209.0 &  10:07:11.80 & $+$05:32:08.8 &  2.148   &   1.2   & V10, S10, P12 \\
J100716.50$+$742146.0 &  10:07:16.50 & $+$74:21:46.0 &  2.88    &   1.1   & V10 \\
J101447.18$+$430030.1 &  10:14:47.18 & $+$43:00:30.1 &  2.988   &   2.0   & DR10, P12, S10, V10 \\
J101956.60$+$274401.7 &  10:19:56.59 & $+$27:44:01.7 &  1.925   &   1.1   & S10, V10 \\
J102714.70$+$354317.0 &  10:27:14.77 & $+$35:43:17.4 &  3.116   &   2.2   & V10, DR10, P12, S10 \\
J103419.70$+$191221.0 &  10:34:19.71 & $+$19:12:22.1 &  3.161    &   1.1   & V10, S10 \\
J104856.70$-$163710.0 &  10:48:56.70 & $-$16:37:10.0 &  3.37    &   1.5   & V10 \\
J105756.26$+$455553.0 &  10:57:56.25 & $+$45:55:53.0 &  4.137   &   2.0   & S10, V10 \\
J110325.30$-$264515.0 &  11:03:25.30 & $-$26:45:15.0 &  2.145   &   1.5   & V10 \\
J110610.70$+$640009.0 &  11:06:10.73 & $+$64:00:09.6 &  2.202   &   1.9   & V10, S10 \\
J111038.50$+$483116.0 &  11:10:38.63 & $+$48:31:15.6 &  2.954   &   1.7   & V10, S10 \\
J111055.20$+$430509.0 &  11:10:55.21 & $+$43:05:10.1 &  3.839   &   1.3   & V10, P12, S10 \\
J111119.10$+$133604.0 &  11:11:19.10 & $+$13:36:03.9 &  3.48    &   1.2   & V10, S10 \\
J112010.30$-$134625.0 &  11:20:10.30 & $-$13:46:25.0 &  3.958   &   1.1   & NED, V10 \\
J112442.80$-$170518.0 &  11:24:42.80 & $-$17:05:18.0 &  2.40    &   1.2   & V10 \\
J113017.37$+$073212.9 &  11:30:17.37 & $+$07:32:12.9 &  2.658   &   1.2   & S10 \\
J114308.88$+$345222.3 &  11:43:08.88 & $+$34:52:22.2 &  3.155   &   1.3   & P12, S10, V10 \\
J115023.58$+$281907.5 &  11:50:23.58 & $+$28:19:07.5 &  3.132   &   1.3   & S10, V10 \\
J115906.50$+$133737.0 &  11:59:06.52 & $+$13:37:37.7 &  3.984   &   1.3   & V10, S10 \\
J115954.30$+$201920.0 &  11:59:54.33 & $+$20:19:21.1 &  3.426   &   1.0   & V10, S10 \\
J120006.25$+$312630.8 &  12:00:06.24 & $+$31:26:30.8 &  2.975   &   1.7   & DR10, P12, S10, V10 \\
J120144.37$+$011611.7 &  12:01:44.36 & $+$01:16:11.6 &  3.233   &   1.1   & S10, P12, V10 \\
J120147.80$+$120630.0 &  12:01:47.90 & $+$12:06:30.2 &  3.509   &   1.3   & V10, S10 \\
J120331.20$+$152256.0 &  12:03:31.29 & $+$15:22:54.7 &  2.976   &   1.1   & V10, S10 \\
J120523.10$-$074232.0 &  12:05:23.13 & $-$07:42:32.7 &  4.694   &   1.0   & V10, NED \\
J121027.60$+$174109.0 &  12:10:27.62 & $+$17:41:08.9 &  3.610    &   1.1   & V10, S10 \\
J121843.39$+$153617.2 &  12:18:43.39 & $+$15:36:17.2 &  2.261   &   1.3   & S10, V10 \\
J121930.77$+$494052.2 &  12:19:30.77 & $+$49:40:52.2 &  2.698   &   1.2   & S10, V10 \\
J122343.10$+$503753.0 &  12:23:43.15 & $+$50:37:53.4 &  3.487   &   1.1   & V10, S10 \\
J122527.40$+$223513.0 &  12:25:27.40 & $+$22:35:12.9 &  2.049   &   1.5   & V10, S10 \\
J123549.40$+$591028.0 &  12:35:49.46 & $+$59:10:27.0 &  2.824   &   1.3   & V10, S10 \\
J123628.27$+$342003.8 &  12:36:28.26 & $+$34:20:03.8 &  4.079   &   1.0   & P12 \\
J123641.46$+$655442.2 &  12:36:41.45 & $+$65:54:42.1 &  3.386   &   1.2   & S10, V10 \\
J123754.80$+$084106.0 &  12:37:54.80 & $+$08:41:06.0 &  2.887   &   1.0   & V10 \\
J124637.06$+$262500.3 &  12:46:37.06 & $+$26:25:00.3 &  3.129   &   1.0   & S10, V10 \\
J124913.90$-$055918.0 &  12:49:13.90 & $-$05:59:18.0 &  2.226   &   1.5   & V10 \\
J124957.20$-$015929.0 &  12:49:57.23 & $-$01:59:28.8 &  3.665   &   1.1   & V10, S10, P12 \\
J125005.70$+$263107.0 &  12:50:05.72 & $+$26:31:07.5 &  2.044   &   2.7   & V10, NED, DR10, S10 \\
J131011.60$+$460124.0 &  13:10:11.60 & $+$46:01:24.4 &  2.133   &   1.3   & V10, S10 \\
J131048.10$+$361557.0 &  13:10:48.12 & $+$36:15:57.3 &  3.429   &   1.2   & V10, P12, S10 \\
J132611.80$+$074358.0 &  13:26:11.85 & $+$07:43:58.4 &  4.134   &   1.5   & V10, S10 \\
J134743.20$+$495621.0 &  13:47:43.29 & $+$49:56:21.3 &  4.510    &   1.2   & V10, S10 \\
J135038.90$-$251216.0 &  13:50:38.90 & $-$25:12:16.0 &  2.599   &   1.3   & V10 \\
J142243.00$+$441721.0 &  14:22:43.02 & $+$44:17:21.3 &  3.545   &   1.5   & V10, S10 \\
J142656.10$+$602550.0 &  14:26:56.19 & $+$60:25:50.8 &  3.189   &   2.8   & V10, S10, DR10 \\
J144453.70$+$291905.0 &  14:44:53.70 & $+$29:19:05.0 &  2.669   &   1.2   & V10 \\
J144542.76$+$490248.9 &  14:45:42.76 & $+$49:02:48.9 &  3.875   &   1.2   & S10 \\
J145147.10$-$151220.0 &  14:51:47.10 & $-$15:12:20.0 &  4.763   &   1.9   & V10 \\
J151336.70$+$430708.0 &  15:13:36.70 & $+$43:07:08.3 &  3.292   &   1.2   & V10, S10 \\
J151352.53$+$085555.7 &  15:13:52.52 & $+$08:55:55.7 &  2.904   &   1.5   & S10, V10 \\
J152156.48$+$520238.6 &  15:21:56.48 & $+$52:02:38.5 &  2.208   &   2.3   & S10, V10 \\
J152553.89$+$513649.1 &  15:25:53.89 & $+$51:36:49.1 &  2.883   &   1.7   & S10, V10 \\
J153830.55$+$085517.1 &  15:38:30.55 & $+$08:55:17.1 &  3.551   &   1.4   & S10, V10 \\
J155152.46$+$191104.1 &  15:51:52.46 & $+$19:11:04.0 &  2.822   &   3.0   & DR10, P12 \\
J155426.10$+$193703.0 &  15:54:26.16 & $+$19:37:03.0 &  4.612   &   1.2   & V10, P12, S10 \\
J155633.78$+$351757.4 &  15:56:33.78 & $+$35:17:57.3 &  3.20    &   1.5   & P12 \\
J160441.40$+$164538.0 &  16:04:41.47 & $+$16:45:38.3 &  2.934   &   1.3   & V10, P12, S10 \\
J160455.40$+$381201.6 &  16:04:55.39 & $+$38:12:01.6 &  2.480    &   2.5   & DR10, V10 \\
J160843.90$+$071508.0 &  16:08:43.90 & $+$07:15:08.6 &  2.863   &   1.7   & V10, DR10, P12, S10 \\
J161434.67$+$470420.1 &  16:14:34.67 & $+$47:04:20.1 &  1.860    &   1.1   & S10, V10 \\
J162116.90$-$004250.0 &  16:21:16.92 & $-$00:42:50.8 &  3.702   &   1.3   & V10, S10 \\
J162645.60$+$642655.0 &  16:26:45.60 & $+$64:26:55.0 &  2.32    &   1.0   & V10 \\
J163300.10$+$362904.0 &  16:33:00.13 & $+$36:29:04.8 &  3.575   &   1.1   & V10, S10 \\
J163429.00$+$703132.4 &  16:34:29.00 & $+$70:31:32.4 &  1.334   &   2.2   & NED, V10 \\
J163909.10$+$282447.0 &  16:39:09.11 & $+$28:24:47.1 &  3.818   &   2.2   & V10, S10 \\
J163950.50$+$434004.0 &  16:39:50.52 & $+$43:40:03.6 &  3.990    &   1.0   & V10, S10 \\
J164656.30$+$551446.0 &  16:46:56.30 & $+$55:14:46.0 &  4.037   &   1.4   & V10 \\
J170100.60$+$641208.0 &  17:01:00.61 & $+$64:12:09.1 &  2.737   &   2.8   & V10, S10, DR10 \\
J171635.40$+$532815.0 &  17:16:35.40 & $+$53:28:15.0 &  1.94    &   1.8   & V10 \\
J172323.10$+$224357.0 &  17:23:23.10 & $+$22:43:57.0 &  4.52    &   1.7   & V10 \\
J173352.20$+$540030.0 &  17:33:52.23 & $+$54:00:30.4 &  3.424   &   1.1   & V10, S10 \\
J175746.40$+$753916.0 &  17:57:46.40 & $+$75:39:16.0 &  3.05    &   1.3   & V10 \\
J194454.90$+$770552.0 &  19:44:54.90 & $+$77:05:52.0 &  3.051   &   3.2   & V10 \\
J212329.47$-$005052.9 &  21:23:29.46 & $-$00:50:52.9 &  2.268   &   1.3   & S10, P12, V10 \\
J215647.30$+$224250.0 &  21:56:47.30 & $+$22:42:50.0 &  1.29    &   1.4   & V10 \\
J222006.77$-$280323.9 &  22:20:06.77 & $-$28:03:23.9 &  2.405   &   1.7   & C04, V10 \\
J223953.60$-$055219.0 &  22:39:53.60 & $-$05:52:19.0 &  4.55    &   1.3   & V10 \\
J225244.00$-$502137.0 &  22:52:44.00 & $-$50:21:37.0 &  2.90    &   1.1   & V10 \\
J231324.46$+$003444.5 &  23:13:24.45 & $+$00:34:44.5 &  2.084   &   1.2   & S10, V10 \\
J234235.50$-$553439.0 &  23:42:35.50 & $-$55:34:39.0 &  2.70    &   1.2   & V10 \\
J234628.20$+$124860.0 &  23:46:28.20 & $+$12:48:60.0 &  2.517   &   1.0   & V10 \\
J235034.30$-$432560.0 &  23:50:34.30 & $-$43:25:60.0 &  2.885   &   2.2   & V10 \\
J235129.80$-$142757.0 &  23:51:29.80 & $-$14:27:57.0 &  2.933   &   1.7   & V10 \\
J235808.54$+$012507.2 &  23:58:08.54 & $+$01:25:07.2 &  3.401   &   1.1   & P12, V10 
\else
J000322.91$-$260316.8 &  00:03:22.91 & $-$26:03:16.8 &  4.098   &   1.6   & NED, V10 \\
J001527.40$+$064012.0 &  00:15:27.40 & $+$06:40:12.0 &  3.17    &   1.2   & V10 \\
J004131.50$-$493612.0 &  00:41:31.50 & $-$49:36:12.0 &  3.24    &   1.8   & V10 \\
J010311.30$+$131618.0 &  01:03:11.30 & $+$13:16:18.0 &  2.681   &   1.6   & NED, V10 \\
J012156.04$+$144823.9 &  01:21:56.03 & $+$14:48:23.9 &  2.870    &   1.1   & S10, V10 \\
J012412.47$-$010049.8 &  01:24:12.47 & $-$01:00:49.7 &  2.826   &   1.0   & S10, P12, V10 \\
J013301.90$-$400628.0 &  01:33:01.90 & $-$40:06:28.0 &  3.023   &   1.0   & V10 \\
J015636.00$+$044528.0 &  01:56:36.00 & $+$04:45:28.0 &  2.993   &   1.0   & V10 \\
J020727.20$-$374156.0 &  02:07:27.20 & $-$37:41:56.0 &  2.404   &   1.2   & V10 \\
J020950.70$-$000506.0 &  02:09:50.71 & $-$00:05:06.4 &  2.850    &   1.2   & V10, S10, P12 \\
J024008.10$-$230915.0 &  02:40:08.10 & $-$23:09:15.0 &  2.225   &   1.4   & V10 \\
J024854.30$+$180250.0 &  02:48:54.30 & $+$18:02:50.0 &  4.42    &   1.0   & V10 \\
J025240.10$-$553832.0 &  02:52:40.10 & $-$55:38:32.0 &  2.35    &   1.2   & V10 \\
J030722.80$-$494548.0 &  03:07:22.80 & $-$49:45:48.0 &  4.728   &   1.3   & V10 \\
J032108.45$+$413220.9 &  03:21:08.45 & $+$41:32:20.8 &  2.467   &   1.1   & S10
\fi

\enddata
\tablecomments{Redshifts from: V10 \citep{2010A&A...518A..10V}; C04 \citep{2004MNRAS.349.1397C}, C09 \citep{2009MNRAS.392...19C}; S10 \citep{2010AJ....139.2360S}; P12 \citep{2012A&A...548A..66P}, DR10 \citep{2013arXiv1307.7735A}, and NED (2013 April version of HyLIRG list from NASA/IPAC Extragalactic Database). 
}
\tablenotetext{a}{See Section \ref{sec:luminosity_cal} for definition.}
\ifx \apjloutput \undefined
\end{deluxetable}  
\else
\end{deluxetable*}  
\fi

\ifx \apjloutput \undefined
\begin{deluxetable}{lllllllllll}  
\else
\begin{deluxetable*}{lllllllllll}  
\fi
\tabletypesize{\scriptsize}
\tablenum{4}
\tablewidth{0in}
\tablecaption{Photometry of Optically Selected Quasars with $L_{\rm bol} > 10^{14}\,L_{\sun}$ \textbf{(Full Version)}\label{table:photometry_qso}}
\tablehead{
\colhead{Source} &  
 \colhead{$R$-band} & 
 \colhead{3.4$\mu$m} & 
 \colhead{4.6$\mu$m} &
 \colhead{12$\mu$m} & 
 \colhead{22$\mu$m} &
 \colhead{70$\mu$m} &
 \colhead{160$\mu$m} &
 \colhead{250$\mu$m} &
 \colhead{350$\mu$m} &
 \colhead{500$\mu$m} 
\\
\colhead{} & 
  \colhead{(mJy)} & 
  \colhead{(mJy)} & 
  \colhead{(mJy)} &
  \colhead{(mJy)} & 
  \colhead{(mJy)} & 
  \colhead{(mJy)} &
  \colhead{(mJy)} & 
  \colhead{(mJy)} & 
  \colhead{(mJy)} & 
  \colhead{(mJy)}
}
\startdata
\ifx \apjloutput \undefined
J000322.910$-$260316.80 & 0.40(0.11) & 0.86(0.02) & 0.69(0.02) & 2.2(0.1) & 8.5(0.9) &  \nodata &  \nodata &  \nodata &  \nodata &  \nodata \\ 
J001527.400$+$064012.00 & 0.43(0.12) & 0.93(0.02) & 1.09(0.03) & 5.3(0.2) & 10.9(1.1) &  \nodata &  \nodata &  \nodata &  \nodata &  \nodata \\ 
J004131.500$-$493612.00 & 0.79(0.22) & 1.61(0.04) & 1.62(0.04) & 4.3(0.1) & 9.1(0.8) &  \nodata &  \nodata &  \nodata &  \nodata &  \nodata \\ 
J010311.300$+$131618.00 & 1.09(0.30) & 1.08(0.03) & 1.51(0.04) & 9.0(0.2) & 19.4(1.0) &  \nodata &  \nodata &  \nodata &  \nodata &  \nodata \\ 
J012156.038$+$144823.94 & 0.31(0.09) & 0.93(0.03) & 1.16(0.03) & 3.6(0.2) & 9.2(1.0) &  \nodata &  \nodata &  \nodata &  \nodata &  \nodata \\ 
J012412.470$-$010049.76 & 0.45(0.12) & 1.28(0.03) & 1.40(0.03) & 3.6(0.1) & 6.7(1.0) &  \nodata &  \nodata &  \nodata &  \nodata &  \nodata \\ 
J013301.900$-$400628.00 & 0.49(0.13) & 0.65(0.02) & 0.89(0.03) & 3.4(0.1) & 9.4(0.9) &  \nodata &  \nodata &  \nodata &  \nodata &  \nodata \\ 
J015636.000$+$044528.00 & 0.43(0.12) & 0.60(0.02) & 0.67(0.02) & 2.2(0.1) & 3.1(0.8) &  \nodata &  \nodata &  \nodata &  \nodata &  \nodata \\ 
J020727.200$-$374156.00 & 0.85(0.23) & 1.49(0.03) & 1.71(0.04) & 6.0(0.1) & 12.3(0.8) &  \nodata &  \nodata &  \nodata &  \nodata &  \nodata \\ 
J020950.712$-$000506.49 & 0.63(0.17) & 0.93(0.02) & 1.36(0.03) & 6.1(0.2) & 15.0(0.8) &  \nodata &  \nodata & 66(6) & 48(6) & 22(7) \\ 
J024008.100$-$230915.00 & 1.01(0.28) & 1.92(0.04) & 3.23(0.07) & 11.1(0.2) & 21.1(1.0) &  \nodata &  \nodata &  \nodata &  \nodata &  \nodata \\ 
J024854.300$+$180250.00 & 0.17(0.05) & 0.62(0.02) & 0.55(0.02) & 1.6(0.1) & 4.1(1.1) &  \nodata &  \nodata &  \nodata &  \nodata &  \nodata \\ 
J025240.100$-$553832.00 & 0.88(0.24) & 1.73(0.04) & 1.99(0.04) & 8.2(0.2) & 19.0(0.9) &  \nodata &  \nodata & 46(2) & 39(3) & 14(3) \\ 
J030722.800$-$494548.00 & 0.06(0.02) & 0.54(0.01) & 0.51(0.02) & 1.1(0.1) & 4.0(0.7) &  \nodata &  \nodata &  \nodata &  \nodata &  \nodata \\ 
J032108.450$+$413220.87 & 0.56(0.15) & 1.39(0.03) & 1.94(0.05) & 6.8(0.2) & 12.7(1.1) &  \nodata &  \nodata &  \nodata &  \nodata &  \nodata \\ 
J035504.900$-$381142.00 & 0.22(0.06) & 0.81(0.02) & 0.71(0.02) & 1.7(0.1) & 5.2(0.5) &  \nodata &  \nodata &  \nodata &  \nodata &  \nodata \\ 
J051707.600$-$441056.00 & 4.17(1.15) & 6.19(0.13) & 9.10(0.17) & 26.3(0.4) & 44.3(1.1) &  \nodata &  \nodata &  \nodata &  \nodata &  \nodata \\ 
J055200.400$-$531244.00 & 2.46(0.68) & 4.78(0.10) & 9.63(0.18) & 23.5(0.3) & 38.1(1.1) &  \nodata &  \nodata &  \nodata &  \nodata &  \nodata \\ 
J055445.700$-$330517.00 & 2.04(0.56) & 3.72(0.08) & 4.57(0.09) & 13.5(0.2) & 26.3(1.1) &  \nodata &  \nodata &  \nodata &  \nodata &  \nodata \\ 
J073502.309$+$265911.58 & 1.34(0.37) & 2.18(0.05) & 3.83(0.08) & 14.9(0.3) & 27.8(1.4) &  \nodata &  \nodata & 87(5) & 56(5) & 26(6) \\ 
J074521.787$+$473436.19 & 1.08(0.30) & 1.78(0.04) & 1.76(0.04) & 5.5(0.2) & 10.6(1.0) &  \nodata &  \nodata & 50(5) & 52(5) & 52(6) \\ 
J075054.644$+$425219.26 & 0.82(0.23) & 2.61(0.06) & 4.60(0.10) & 15.0(0.3) & 22.8(1.2) &  \nodata &  \nodata & 30(5) & 25(5) & 26(8) \\ 
J080117.796$+$521034.55 & 0.31(0.08) & 0.92(0.02) & 1.10(0.03) & 6.1(0.2) & 11.9(0.9) &  \nodata &  \nodata & 78(9) & 76(8) & 56(10) \\ 
J084323.700$+$165655.00 & 0.08(0.02) & 1.03(0.03) & 2.19(0.05) & 6.1(0.2) & 10.7(1.4) &  \nodata &  \nodata &  \nodata &  \nodata &  \nodata \\ 
J084401.954$+$050357.96 & 0.35(0.10) & 1.45(0.03) & 1.29(0.04) & 4.3(0.2) & 7.3(1.1) &  \nodata &  \nodata &  \nodata &  \nodata &  \nodata \\ 
J090033.509$+$421547.06 & 0.70(0.19) & 1.70(0.04) & 1.69(0.04) & 4.6(0.2) & 9.4(0.9) &  \nodata &  \nodata & 23(5) & 17(5) &  $<21$ \\ 
J090423.374$+$130920.74 & 0.79(0.22) & 1.30(0.03) & 1.62(0.04) & 10.1(0.2) & 23.8(1.4) &  \nodata &  \nodata &  \nodata &  \nodata &  \nodata \\ 
J094202.044$+$042244.54 & 0.55(0.15) & 1.04(0.03) & 0.98(0.03) & 2.5(0.1) & 4.9(0.8) &  \nodata &  \nodata &  \nodata &  \nodata &  \nodata \\ 
J094253.600$-$110426.00 & 0.52(0.14) & 0.97(0.02) & 1.24(0.03) & 6.3(0.2) & 12.5(0.9) &  \nodata &  \nodata &  \nodata &  \nodata &  \nodata \\ 
J094734.198$+$142116.96 & 0.66(0.18) & 0.89(0.02) & 0.99(0.03) & 5.3(0.2) & 9.9(1.0) &  \nodata &  \nodata &  \nodata &  \nodata &  \nodata \\ 
J094932.269$+$033531.79 & 0.33(0.09) & 0.54(0.02) & 0.49(0.02) & 2.1(0.1) & 5.1(0.9) &  \nodata &  \nodata &  \nodata &  \nodata &  \nodata \\ 
J095937.116$+$131215.44 & 0.53(0.15) & 1.09(0.03) & 0.89(0.03) & 1.8(0.1) & 6.4(1.0) &  \nodata &  \nodata &  \nodata &  \nodata &  \nodata \\ 
J100711.809$+$053208.89 & 1.07(0.30) & 1.57(0.04) & 2.28(0.05) & 8.5(0.2) & 15.9(1.1) &  \nodata &  \nodata &  \nodata &  \nodata &  \nodata \\ 
J100716.500$+$742146.00 & 0.63(0.18) & 1.25(0.03) & 1.41(0.03) & 3.8(0.1) & 7.8(0.8) &  \nodata &  \nodata &  \nodata &  \nodata &  \nodata \\ 
J101447.182$+$430030.10 & 1.05(0.29) & 1.25(0.03) & 1.21(0.03) & 3.7(0.1) & 9.1(1.0) &  \nodata &  \nodata &  \nodata &  \nodata &  \nodata \\ 
J101956.597$+$274401.72 & 2.97(0.82) & 2.02(0.05) & 2.17(0.05) & 5.9(0.2) & 10.7(0.9) &  \nodata &  \nodata &  \nodata &  \nodata &  \nodata \\ 
J102714.777$+$354317.43 & 0.65(0.18) & 1.89(0.04) & 2.31(0.05) & 10.1(0.2) & 22.2(1.0) &  \nodata &  \nodata &  \nodata &  \nodata &  \nodata \\ 
J103419.716$+$191222.18 & 0.69(0.19) & 0.66(0.02) & 0.77(0.02) & 3.4(0.2) & 6.2(1.0) &  \nodata &  \nodata &  \nodata &  \nodata &  \nodata \\ 
J104856.700$-$163710.00 & 0.44(0.12) & 1.06(0.03) & 0.95(0.03) & 4.2(0.1) & 10.6(1.0) &  \nodata &  \nodata &  \nodata &  \nodata &  \nodata \\ 
J105756.258$+$455553.04 & 0.70(0.19) & 1.35(0.03) & 1.07(0.03) & 2.8(0.1) & 5.3(0.9) &  \nodata &  \nodata &  \nodata &  \nodata &  \nodata \\ 
J110325.300$-$264515.00 & 1.49(0.41) & 2.43(0.06) & 3.07(0.07) & 13.2(0.2) & 29.7(1.1) &  \nodata &  \nodata &  \nodata &  \nodata &  \nodata \\ 
J110610.730$+$640009.65 & 2.66(0.73) & 1.96(0.05) & 2.50(0.05) & 12.2(0.2) & 24.4(1.0) &  \nodata &  \nodata &  \nodata &  \nodata &  \nodata \\ 
J111038.630$+$483115.67 & 0.37(0.10) & 1.35(0.03) & 2.03(0.04) & 9.3(0.2) & 17.9(0.8) &  \nodata &  \nodata &  \nodata &  \nodata &  \nodata \\ 
J111055.215$+$430510.11 & 0.12(0.03) & 1.51(0.03) & 1.24(0.03) & 3.9(0.1) & 9.5(0.9) &  \nodata &  \nodata &  \nodata &  \nodata &  \nodata \\ 
J111119.106$+$133603.92 & 0.33(0.09) & 0.74(0.02) & 0.72(0.02) & 3.6(0.1) & 9.6(0.9) &  \nodata &  \nodata &  \nodata &  \nodata &  \nodata \\ 
J112010.300$-$134625.00 & 0.24(0.07) & 0.50(0.02) & 0.48(0.02) & 2.7(0.1) & 8.1(0.9) &  \nodata &  \nodata &  \nodata &  \nodata &  \nodata \\ 
J112442.800$-$170518.00 & 0.98(0.27) & 1.37(0.03) & 1.71(0.04) & 6.9(0.2) & 14.1(1.0) &  \nodata &  \nodata &  \nodata &  \nodata &  \nodata \\ 
J113017.374$+$073212.95 & 0.51(0.14) & 1.28(0.03) & 2.08(0.05) & 11.3(0.3) & 24.2(1.4) &  \nodata &  \nodata &  \nodata &  \nodata &  \nodata \\ 
J114308.880$+$345222.29 & 0.83(0.23) & 0.85(0.02) & 0.93(0.03) & 3.1(0.1) & 7.4(1.0) &  \nodata &  \nodata &  \nodata &  \nodata &  \nodata \\ 
J115023.580$+$281907.50 & 0.55(0.15) & 0.86(0.02) & 0.85(0.03) & 2.5(0.1) & 7.5(0.9) &  \nodata &  \nodata &  \nodata &  \nodata &  \nodata \\ 
J115906.526$+$133737.74 & 0.38(0.10) & 0.63(0.02) & 0.64(0.02) & 2.5(0.1) & 7.2(0.9) &  \nodata &  \nodata &  \nodata &  \nodata &  \nodata \\ 
J115954.334$+$201921.11 & 0.31(0.09) & 0.62(0.02) & 0.53(0.02) & 2.0(0.1) & 3.8(0.9) &  \nodata &  \nodata &  \nodata &  \nodata &  \nodata \\ 
J120006.247$+$312630.85 & 0.77(0.21) & 1.06(0.03) & 1.25(0.03) & 5.4(0.2) & 11.5(0.9) &  \nodata &  \nodata &  \nodata &  \nodata &  \nodata \\ 
J120144.367$+$011611.65 & 0.26(0.07) & 0.67(0.02) & 0.87(0.03) & 4.7(0.2) & 11.5(1.1) &  \nodata &  \nodata &  \nodata &  \nodata &  \nodata \\ 
J120147.909$+$120630.28 & 0.40(0.11) & 1.15(0.03) & 0.99(0.03) & 3.2(0.1) & 8.8(1.0) &  \nodata &  \nodata &  \nodata &  \nodata &  \nodata \\ 
J120331.291$+$152254.77 & 0.63(0.17) & 0.90(0.02) & 1.01(0.03) & 3.0(0.1) & 5.3(0.9) &  \nodata &  \nodata &  \nodata &  \nodata &  \nodata \\ 
J120523.130$-$074232.70 & 0.10(0.03) & 0.41(0.01) & 0.39(0.02) & 1.4(0.2) & 6.5(1.2) & 16(3) & 32(9) & 76(3) & 82(3) & 75(3) \\ 
J121027.626$+$174108.92 & 0.11(0.03) & 0.74(0.02) & 0.69(0.02) & 3.4(0.1) & 10.1(0.9) &  \nodata &  \nodata &  \nodata &  \nodata &  \nodata \\ 
J121843.392$+$153617.24 & 1.50(0.41) & 1.63(0.04) & 2.23(0.05) & 8.3(0.2) & 16.1(1.0) &  \nodata &  \nodata &  \nodata &  \nodata &  \nodata \\ 
J121930.770$+$494052.25 & 1.14(0.31) & 0.88(0.02) & 1.23(0.03) & 5.7(0.2) & 12.9(0.8) &  \nodata &  \nodata &  \nodata &  \nodata &  \nodata \\ 
J122343.157$+$503753.47 & 0.26(0.07) & 1.06(0.03) & 0.91(0.03) & 1.9(0.1) & 4.5(0.8) &  \nodata &  \nodata &  \nodata &  \nodata &  \nodata \\ 
J122527.401$+$223512.98 & 1.91(0.53) & 2.04(0.05) & 2.95(0.06) & 9.7(0.2) & 18.6(1.0) &  \nodata &  \nodata &  \nodata &  \nodata &  \nodata \\ 
J123549.469$+$591027.01 & 0.49(0.14) & 1.48(0.03) & 1.54(0.04) & 3.4(0.1) & 5.3(0.7) &  \nodata &  \nodata &  \nodata &  \nodata &  \nodata \\ 
J123628.268$+$342003.82 & 0.45(0.12) & 0.39(0.01) & 0.27(0.01) & 0.8(0.1) & 3.2(0.8) &  \nodata &  \nodata &  \nodata &  \nodata &  \nodata \\ 
J123641.455$+$655442.19 & 0.56(0.16) & 0.63(0.02) & 0.65(0.02) & 4.2(0.1) & 8.6(0.8) &  \nodata &  \nodata &  \nodata &  \nodata &  \nodata \\ 
J123754.800$+$084106.00 & 0.36(0.10) & 1.22(0.03) & 1.50(0.04) & 5.1(0.2) & 10.2(1.0) &  \nodata &  \nodata &  \nodata &  \nodata &  \nodata \\ 
J124637.063$+$262500.30 & 0.53(0.15) & 0.45(0.01) & 0.62(0.02) & 4.0(0.1) & 8.4(0.9) &  \nodata &  \nodata &  \nodata &  \nodata &  \nodata \\ 
J124913.900$-$055918.00 & 1.08(0.30) & 2.06(0.05) & 3.75(0.08) & 17.3(0.3) & 32.1(1.35) & 128(2) & 156(5) & 111(4) & 72(3) & 32(4) \\ 
J124957.239$-$015928.81 & 0.45(0.12) & 0.44(0.01) & 0.49(0.02) & 3.6(0.2) & 9.2(1.1) &  \nodata &  \nodata &  \nodata &  \nodata &  \nodata \\ 
J125005.724$+$263107.59 & 5.22(1.44) & 3.22(0.07) & 4.73(0.10) & 16.7(0.3) & 31.3(1.4) &  \nodata &  \nodata &  \nodata &  \nodata &  \nodata \\ 
J131011.604$+$460124.46 & 0.65(0.18) & 1.42(0.03) & 2.62(0.05) & 10.8(0.2) & 22.4(1.0) &  \nodata &  \nodata &  \nodata &  \nodata &  \nodata \\ 
J131048.124$+$361557.32 & 0.42(0.12) & 1.24(0.03) & 1.01(0.03) & 2.6(0.1) & 6.2(0.7) &  \nodata &  \nodata &  \nodata &  \nodata &  \nodata \\ 
J132611.851$+$074358.44 & 0.57(0.16) & 1.12(0.03) & 0.99(0.03) & 2.3(0.1) & 3.6(0.9) &  \nodata &  \nodata &  \nodata &  \nodata &  \nodata \\ 
J134743.294$+$495621.30 & 0.14(0.04) & 0.51(0.01) & 0.43(0.02) & 1.4(0.1) & 3.3(0.7) &  \nodata &  \nodata &  \nodata &  \nodata &  \nodata \\ 
J135038.900$-$251216.00 & 0.79(0.22) & 0.97(0.02) & 1.32(0.03) & 6.3(0.2) & 12.1(0.8) &  \nodata &  \nodata &  \nodata &  \nodata &  \nodata \\ 
J142243.025$+$441721.30 & 0.33(0.09) & 1.06(0.03) & 1.08(0.03) & 5.8(0.1) & 15.1(0.9) &  \nodata &  \nodata &  \nodata &  \nodata &  \nodata \\ 
J142656.196$+$602550.89 & 1.32(0.36) & 1.97(0.05) & 2.28(0.05) & 10.2(0.2) & 23.0(0.8) &  \nodata &  \nodata &  \nodata &  \nodata &  \nodata \\ 
J144453.700$+$291905.00 & 1.41(0.39) & 1.92(0.04) & 1.94(0.04) & 4.7(0.1) & 8.2(0.6) &  \nodata &  \nodata &  \nodata &  \nodata &  \nodata \\ 
J144542.761$+$490248.91 & 0.36(0.10) & 0.94(0.02) & 0.72(0.02) & 1.8(0.1) & 4.2(0.7) &  \nodata &  \nodata &  \nodata &  \nodata &  \nodata \\ 
J145147.100$-$151220.00 & 0.18(0.05) & 0.98(0.02) & 0.82(0.02) & 1.6(0.1) & 6.0(0.8) &  \nodata &  \nodata &  \nodata &  \nodata &  \nodata \\ 
J151336.706$+$430708.33 & 0.52(0.14) & 0.35(0.01) & 0.43(0.01) & 3.1(0.1) & 7.3(0.6) &  \nodata &  \nodata &  \nodata &  \nodata &  \nodata \\ 
J151352.526$+$085555.74 & 0.51(0.14) & 1.19(0.03) & 1.72(0.04) & 11.6(0.2) & 25.8(1.0) &  \nodata &  \nodata &  \nodata &  \nodata &  \nodata \\ 
J152156.482$+$520238.58 & 2.36(0.65) & 2.41(0.05) & 3.47(0.07) & 13.9(0.2) & 21.8(0.8) &  \nodata &  \nodata &  \nodata &  \nodata &  \nodata \\ 
J152553.892$+$513649.10 & 0.64(0.18) & 2.07(0.05) & 2.21(0.04) & 4.6(0.1) & 8.7(0.6) &  \nodata &  \nodata &  \nodata &  \nodata &  \nodata \\ 
J153830.554$+$085517.11 & 0.74(0.21) & 1.12(0.03) & 0.86(0.03) & 3.3(0.1) & 8.0(0.8) &  \nodata &  \nodata &  \nodata &  \nodata &  \nodata \\ 
J155152.464$+$191104.08 & 1.10(0.30) & 2.15(0.05) & 2.70(0.06) & 11.2(0.2) & 23.0(1.0) &  \nodata &  \nodata &  \nodata &  \nodata &  \nodata \\ 
J155426.160$+$193703.04 & 0.10(0.03) & 0.42(0.01) & 0.36(0.02) & 2.0(0.1) & 4.3(0.7) &  \nodata &  \nodata &  \nodata &  \nodata &  \nodata \\ 
J155633.782$+$351757.39 & 0.05(0.02) & 1.63(0.04) & 3.47(0.07) & 9.9(0.2) & 17.8(0.8) &  \nodata &  \nodata &  \nodata &  \nodata &  \nodata \\ 
J160441.472$+$164538.34 & 0.45(0.12) & 0.62(0.02) & 0.69(0.02) & 3.1(0.1) & 9.9(1.0) &  \nodata &  \nodata &  \nodata &  \nodata &  \nodata \\ 
J160455.397$+$381201.63 & 2.15(0.60) & 2.01(0.04) & 3.16(0.06) & 17.4(0.3) & 34.3(1.0) &  \nodata &  \nodata &  \nodata &  \nodata &  \nodata \\ 
J160843.901$+$071508.68 & 0.68(0.19) & 1.54(0.04) & 1.66(0.04) & 4.1(0.2) & 8.4(0.9) &  \nodata &  \nodata &  \nodata &  \nodata &  \nodata \\ 
J161434.673$+$470420.10 & 1.94(0.54) & 2.99(0.06) & 4.48(0.08) & 14.6(0.2) & 29.0(0.9) &  \nodata &  \nodata &  \nodata &  \nodata &  \nodata \\ 
J162116.922$-$004250.87 & 0.40(0.11) & 1.03(0.03) & 0.83(0.03) & 2.1(0.1) & 7.3(1.0) &  \nodata &  \nodata &  \nodata &  \nodata &  \nodata \\ 
J162645.600$+$642655.00 & 1.46(0.40) & 1.10(0.02) & 1.28(0.03) & 4.4(0.1) & 8.7(0.6) &  \nodata &  \nodata &  \nodata &  \nodata &  \nodata \\ 
J163300.134$+$362904.88 & 0.09(0.03) & 0.76(0.02) & 0.76(0.02) & 5.0(0.1) & 11.8(0.9) &  \nodata &  \nodata &  \nodata &  \nodata &  \nodata \\ 
J163429.000$+$703132.40 & 5.81(1.61) & 10.43(0.23) & 19.83(0.37) & 54.7(0.7) & 115.4(2.6) &  \nodata &  \nodata &  \nodata &  \nodata &  \nodata \\ 
J163909.110$+$282447.16 & 0.30(0.08) & 1.49(0.04) & 1.38(0.04) & 7.4(0.2) & 20.6(1.3) &  \nodata &  \nodata &  \nodata &  \nodata &  \nodata \\ 
J163950.520$+$434003.68 & 0.32(0.09) & 0.55(0.01) & 0.48(0.02) & 1.3(0.1) & 3.6(0.7) &  \nodata &  \nodata &  \nodata &  \nodata &  \nodata \\ 
J164656.300$+$551446.00 & 0.37(0.10) & 0.64(0.01) & 0.55(0.01) & 2.8(0.1) & 9.0(0.5) &  \nodata &  \nodata &  \nodata &  \nodata &  \nodata \\ 
J170100.619$+$641209.12 & 1.27(0.35) & 1.98(0.04) & 2.38(0.05) & 10.3(0.2) & 21.1(0.7) &  \nodata &  \nodata & 80(1) & 56(1) & 39(1) \\ 
J171635.400$+$532815.00 & 6.97(1.93) & 2.63(0.06) & 3.42(0.07) & 11.3(0.2) & 20.3(0.9) &  \nodata &  \nodata &  \nodata &  \nodata &  \nodata \\ 
J172323.100$+$224357.00 & 0.12(0.03) & 1.13(0.03) & 0.92(0.02) & 2.2(0.1) & 5.7(0.7) &  \nodata &  \nodata &  \nodata &  \nodata &  \nodata \\ 
J173352.231$+$540030.49 & 0.53(0.15) & 0.75(0.02) & 0.66(0.02) & 1.6(0.1) & 5.9(0.5) &  \nodata &  \nodata &  \nodata &  \nodata &  \nodata \\ 
J175746.400$+$753916.00 & 0.44(0.12) & 1.54(0.03) & 1.41(0.03) & 2.9(0.1) & 5.9(0.6) &  \nodata &  \nodata &  \nodata &  \nodata &  \nodata \\ 
J194454.900$+$770552.00 & 1.84(0.51) & 2.80(0.06) & 2.77(0.06) & 7.1(0.1) & 15.1(0.6) &  \nodata &  \nodata &  \nodata &  \nodata &  \nodata \\ 
J212329.468$-$005052.93 & 1.59(0.44) & 1.47(0.04) & 2.35(0.05) & 9.7(0.2) & 20.6(1.1) &  \nodata &  \nodata & 33(3) & 24(4) & 16(3) \\ 
J215647.300$+$224250.00 & 1.52(0.42) & 7.62(0.16) & 13.26(0.24) & 23.3(0.4) & 32.7(1.3) &  \nodata &  \nodata &  \nodata &  \nodata &  \nodata \\ 
J222006.770$-$280323.90 & 1.20(0.33) & 2.03(0.05) & 2.18(0.05) & 8.5(0.2) & 19.2(1.3) &  \nodata &  \nodata & 39(4) & 29(4) & 20(3) \\ 
J223953.600$-$055219.00 & 0.16(0.04) & 0.74(0.02) & 0.64(0.02) & 1.6(0.1) & 4.6(1.0) &  \nodata &  \nodata &  \nodata &  \nodata &  \nodata \\ 
J225244.000$-$502137.00 & 0.50(0.14) & 1.14(0.03) & 1.44(0.04) & 6.2(0.2) & 12.7(0.9) &  \nodata &  \nodata &  \nodata &  \nodata &  \nodata \\ 
J231324.456$+$003444.51 & 1.25(0.34) & 1.96(0.05) & 2.57(0.05) & 9.8(0.2) & 15.1(1.1) &  \nodata &  \nodata &  \nodata &  \nodata &  \nodata \\ 
J234235.500$-$553439.00 & 0.62(0.17) & 1.19(0.03) & 1.27(0.03) & 4.2(0.1) & 8.5(0.8) &  \nodata &  \nodata &  \nodata &  \nodata &  \nodata \\ 
J234628.200$+$124860.00 & 0.68(0.19) & 0.89(0.02) & 1.34(0.04) & 10.0(0.2) & 23.8(1.3) &  \nodata &  \nodata &  \nodata &  \nodata &  \nodata \\ 
J235034.300$-$432560.00 & 0.96(0.27) & 1.69(0.04) & 2.03(0.05) & 6.8(0.2) & 16.1(1.0) &  \nodata &  \nodata &  \nodata &  \nodata &  \nodata \\ 
J235129.800$-$142757.00 & 0.90(0.25) & 1.77(0.04) & 1.87(0.04) & 4.8(0.2) & 12.5(1.0) &  \nodata &  \nodata &  \nodata &  \nodata &  \nodata \\ 
J235808.542$+$012507.20 & 0.43(0.12) & 0.88(0.02) & 0.87(0.03) & 2.5(0.1) & 6.5(0.9) &  \nodata &  \nodata &  \nodata &  \nodata &  \nodata 

\else

\fi
\enddata
\tablecomments{The numbers in parentheses are the 1--$\sigma$ uncertainty in photometry.}
\ifx \apjloutput \undefined
\end{deluxetable}  
\else
\end{deluxetable*}  
\fi

\else
\fi

\end{document}